\documentclass[10pt]{iopart}
\usepackage{iopams}
\usepackage{amsmath_iop}
\usepackage{graphicx}
\usepackage{bm}
\usepackage{dcolumn}
\usepackage{lscape}
\usepackage{CJK}
\usepackage{url}
\def\ket#1{\left|{#1}\right\rangle}

\def\brakket#1#2#3{\left\langle{#1}\middle|{#2}\middle|{#3}\right\rangle}
\def\ve#1{{\bm{#1}}}
\def\nuc#1#2#3{{}^{#2}_{#3}\mathrm{#1}}
\def\urm#1{\scriptstyle{\text{\textrm{\textmd{\textup{#1}}}}}}

\def\avr#1{\left\langle{#1}\right\rangle}
\def\Nabla{\bm{\nabla}}
\def\ca#1{{\mathcal{#1}}}
\def\fr#1{{\mathfrak{#1}}}
\let\temp\epsilon
\let\epsilon\varepsilon
\let\varepsilon\temp
\let\temp\relax
\DeclareMathOperator{\laplace}{\Delta}
\def\defeq{\mathrel{:=}}

\begin{document}
% 
%%%%%%%%%%%%%%%%%%%%%%%%%%%%%%%%%%%%%%%%%%%%%%%%%% 
\begin{CJK*}{UTF8}{}
  \title[On deformability of atoms]
    {On deformability of atoms---comparative study between atoms and atomic nuclei}
  \author{
    Tomoya Naito    (\CJKfamily{min}{内藤智也})$ {}^{1, \, 2} $,
    Shimpei Endo    (\CJKfamily{min}{遠藤晋平})$ {}^{3, \, 4} $,
    Kouichi Hagino  (\CJKfamily{min}{萩野浩一})$ {}^{5} $,
    and
    Yusuke Tanimura (\CJKfamily{min}{谷村雄介})$ {}^{3, \, 6} $}
  \address{
    $ {}^{1} $ Department of Physics, Graduate School of Science, The University of Tokyo,
    Tokyo 113-0033, Japan}
  \address{
    $ {}^{2} $ RIKEN Nishina Center, Wako 351-0198, Japan}
  \address{
    $ {}^{3} $ Department of Physics, Tohoku University,
    Sendai 980-8578, Japan}
  \address{
    $ {}^{4} $ Frontier Research Institute for Interdisciplinary Science, Tohoku University,
    Sendai 980-8578, Japan}
  \address{
    $ {}^{5} $ Department of Physics, Kyoto University,
    Kyoto 606-8502, Japan}
  \address{
    $ {}^{6} $ Graduate Program on Physics for the Universe, Tohoku University,
    Sendai 980-8578, Japan}
  \ead{tomoya.naito@phys.s.u-tokyo.ac.jp,
    shimpei.endo@nucl.phys.tohoku.ac.jp,
    hagino.kouichi.5m@kyoto-u.ac.jp,
    tanimura@nucl.phys.tohoku.ac.jp}
  \date{\today}
  %%%%%%%%%%%%%%%%%%%%%%%%%%%%%%%%%%%%%%%%%%%%%%%%%% 
  \begin{abstract}
    Atomic nuclei can be spontaneously deformed into non-spherical shapes as many-nucleon systems.
    We discuss to what extent a similar deformation takes place in many-electron systems.
    To this end,
    we employ several many-body methods,
    such as the unrestricted Hartree-Fock method, post-Hartree-Fock methods, and the density functional theory,
    to compute the electron density distribution in atoms.
    We show that the electron density distribution of open-shell atoms is
    deformed due solely to the single-particle valence orbitals, while the core part remains spherical.
    This is in contrast to atomic nuclei, which can be deformed collectively.
    We qualitatively discuss the origin for this apparent difference between atoms and nuclei by estimating the energy change due to deformation.
    We find that nature of the interaction plays an essential role for the collective deformation.
  \end{abstract}
  \submitto{\JPB}
  \maketitle
\end{CJK*}
% \ioptwocol
%%%%%%%%%%%%%%%%%%%%%%%%%%%%%%%%%%%%%%%%%%%%%%%%%%
%
% Introduction
% -*- coding: utf-8 -*-
%%%%%%%%%%%%%%%%%%%%%%%%%%%%%%%%%%%%%%%%%%%%%%%%%% 
% 
% Deformation of atoms paper manuscript for J. Phys. B
% Introduction Part
% 
% Begin to write: 2020-07-25
% 
% Tomoya Naito, Shimpei Endo,
% Kouichi Hagino, and Yusuke Tanimura
% 
%%%%%%%%%%%%%%%%%%%%%%%%%%%%%%%%%%%%%%%%%%%%%%%%%% 
% 
\section{Introduction}
\label{sec:intro}
\par
Atoms and atomic nuclei share common features as
quantum many-body systems of interacting fermions~\cite{
  Maruhn2010SimpleModelsofManyFermionSystems_SpringerVerlag}. 
In atoms, the inter-particle interaction is the repulsive Coulomb interaction between electrons,
and there exists the spherical external Coulomb potential due to the nucleus. 
On the other hand, in atomic nuclei,
the inter-particle interaction is the attractive nuclear force between two or more nucleons,
with no external potential. 
Despite the differences in the fundamental interactions, 
many similar properties have been observed in both systems,
such as the shell structure and the associated magic numbers~\cite{
  Hagino2020Found.Chem.22_267,
  Maeno2021Found.Chem.23_201}.
\par
One of the most important properties of atomic nuclei is the collective nuclear deformation, 
which is evidenced by the characteristic rotational spectra~\cite{
  Bohr1975NuclearDeformations_W.A.Benjamin,
  Ring1980TheNuclearManyBodyProblem_SpringerVerlag}.
It was discussed that this collective deformation is originated from the strong attractive interaction between protons and neutrons~\cite{
  Federman1977Phys.Lett.B69_385,
  Federman1979Phys.Lett.B82_9,
  Federman1979Phys.Rev.C20_820,
  Dobaczewski1988Phys.Rev.Lett.60_2254}.
Here, atomic nuclei are deformed as a whole, 
resulting in a strong enhancement in the transition probability from the first excited state to the ground state.  
This is in marked contrast to trivial deformation seen in nuclei
with a valence nucleon outside a doubly-magic nucleus~\cite{
  Arima1954Prog.Theor.Phys.11_509,
  Minamisono1993Nucl.Phys.A559_239,
  Minamisono2001HyperfineInteract.136_225,
  Ohtsubo2012PhysRevLett109_032504}.
Since doubly-magic nuclei are, in general, spherical,
such deformation is entirely due to the valence nucleon,
thus of single-particle nature,
unless the valence nucleon induces a strong polarization effect of the core nucleus~\cite{
  Sagawa1984Nucl.Phys.A430_84}.
Although some of such atomic nuclei are known to be deformed beyond single-particle level,
the collective deformation is known to be much more significant than such a singe-particle deformation.
We note that atomic nuclei may also be deformed due to the $ \alpha $-cluster formation~\cite{
  Morinaga1956Phys.Rev.101_254,
  Morinaga1966Phys.Lett.21_78,
  Ikeda1968Prog.Theor.Phys.Suppl.E68_464,
  Kanada-Enyo1999Phys.Rev.C60_064304,
  Itagaki2001Phys.Rev.C64_014301,
  Oertzen2006Phys.Rep.432_43,
  Itoh2011Phys.Rev.C84_054308,
  Freer2014Prog.Part.Nucl.Phys.78_1,
  Baba2016Phys.Rev.C94_044303,
  Baba2017Phys.Rev.C95_064318,
  Itagaki2021Phys.Rev.C103_044303}, 
even though it is out of the scope of this paper. 
See
\cite{
  Bohr1975NuclearDeformations_W.A.Benjamin,
  Ring1980TheNuclearManyBodyProblem_SpringerVerlag} for more details.
\par
A natural question then arises:
Can atoms be deformed collectively as in atomic nuclei?
Notice that calculations of the atomic structure have been usually carried out by assuming spherical symmetry~\cite{
  Johnson1965J.Chem.Educ.42_145,
  Cohen1965J.Chem.Educ.42_397,
  Friedrich2006TheoreticalAtomicPhysics_SpringerVerlag},
and the non-sphericity has been taken care of only in a few works~\cite{
  Fertig2000Phys.Rev.A62_052511,
  Borgoo2010Comput.Phys.Commun.181_426}.
Likewise, spherical symmetry has been usually assumed to construct the pseudopotential for valence electrons in calculations of molecules and solids~\cite{
  Austin1962Phys.Rev.127_276,
  Louie1982Phys.Rev.B26_1738,
  Martin2004_CambridgeUniversityPress,
  Schwerdtfeger2011ChemPhysChem12_3143}.
It is thus widely believed that atoms are rather spherical,
in stark contrast to deformed atomic nuclei.
What is the physical origin of this difference? 
To what extent can electron density distributions in atoms be deformed collectively?
\par
To answer these questions, in this paper we calculate electron density distributions of atoms and their deformation parameters using various quantum chemical methods without assuming spherical symmetry.
Namely, we use and compare the unrestricted Hartree-Fock method~\cite{
  Hartree1928Math.Proc.Camb.Philos.Soc.24_89,
  Hartree1928Math.Proc.Camb.Philos.Soc.24_111,
  Hartree1928Math.Proc.Camb.Philos.Soc.24_426,
  Fock1930Z.Phys.61_126,
  Roothaan1951Rev.Mod.Phys.23_69,
  Pople1954J.Chem.Phys.22_571,
  Berthier1954C.R.Acad.Sci.238_91,
  McWeeny1968J.Chem.Phys.49_4852,
  Jensen2017IntroductiontoComputationalChemistry_JohnWiley&Sons},
several post-Hartree-Fock methods~\cite{
  Jensen2017IntroductiontoComputationalChemistry_JohnWiley&Sons,
  Moller1934Phys.Rev.46_618,
  Coester1958Nucl.Phys.7_421,
  Cizek1966J.Chem.Phys.45_4256,
  Cizek1971Int.J.QuantumChem.5_359,
  Pople1976Int.J.QuantumChem.10_1,
  Pople1977Int.J.QuantumChem.12_149,
  Krishnan1978Int.J.QuantumChem.14_91,
  Krishnan1980J.Chem.Phys.72_4654,
  Trucks1988Chem.Phys.Lett.147_359,
  Head-Gordon1988Chem.Phys.Lett.153_503,
  Raghavachari1990J.Phys.Chem.94_5579,
  Pople1999Rev.Mod.Phys.71_1267},
and 
the density functional theory (DFT)~\cite{
  Hohenberg1964Phys.Rev.136_B864,
  Kohn1965Phys.Rev.140_A1133,
  Kohn1999Rev.Mod.Phys.71_1253}.
By estimating the change in the energy induced by the deformation,
in nuclear systems,
we find that the spin-up and spin-down particles tend to be deformed in the same manner
due to the attractive inter-particle nuclear interaction.
In atomic systems, on the other hand, 
we find that the spin-up and spin-down particles prefer to be deformed in the opposite manner,
cancelling each other and resulting in a small deformation as a whole
due to the repulsive inter-particle Coulomb interaction.
We, thus, show that the difference between atoms and atomic nuclei in their deformability primarily originates from the different nature of the inter-particle interactions.
\par
The paper is organized as follows. 
In
section~\ref{sec:deformation},
we first give 
a brief introduction of the deformation of many-body systems,
and introduce a parameter $ \beta $ to characterize the deformation.
We then carry out quantum chemical calculations for isolated atoms
and discuss their deformations in
section~\ref{sec:numerical}.
In section~\ref{sec:model},
we qualitatively compare between atoms and nuclei using
a model to estimate a change in the energy, and discuss the origin for collective deformation.
We then summarize the paper in
section~\ref{sec:conc}. 
In 
\ref{sec:n_drop},
we discuss the deformation of neutron drops, that is, many-neutron systems trapped in a harmonic oscillator potential, and exemplify that nuclear systems can be deformed collectively.
In 
\ref{sec:gen},
a generalization of the calculations in
section~\ref{sec:model}
is presented.
\ref{sec:data} summarizes the numerical data for atomic deformation calculated in the main part of the paper.

% 
% Deformation
% -*- coding: utf-8 -*-
%%%%%%%%%%%%%%%%%%%%%%%%%%%%%%%%%%%%%%%%%%%%%%%%%% 
% 
% Deformation of atoms paper manuscript for J. Phys. B
% Deformation Part
% 
% Begin to write: 2020-07-25
% 
% Tomoya Naito, Shimpei Endo,
% Kouichi Hagino, and Yusuke Tanimura
% 
%%%%%%%%%%%%%%%%%%%%%%%%%%%%%%%%%%%%%%%%%%%%%%%%%% 
% 
\section{Collective deformation of many-fermion systems}
\label{sec:deformation}
\subsection{Deformation parameters}
\label{sec:deformation_param}
\par
Before we present the results of our calculations, let us briefly 
summarize the basic concepts of collective deformation. 
Atomic nuclei in the rare earth region,
such as $ \nuc{Sm}{154}{} $ and $ \nuc{Er}{168}{} $,
as well as those in the actinide region, such as $ \nuc{U}{234}{} $,
often exhibit characteristic spectra, in which the energy of the state with 
the total angular momentum $ I $ ($ I = 0 $, $ 2 $, $ 4 $, \ldots) is proportional to
$ I \left( I + 1 \right) $. 
These spectra are interpreted as that those nuclei are deformed in the 
ground state and show the rotational spectra given by 
\begin{equation}
  \label{eq:Erot}
  E_{\urm{rot}}
  =
  \frac{I \left( I + 1 \right) \hbar^2}{2 \ca{I}},
\end{equation}
where $ \ca{I} $ is the moment of inertia for the rotational motion. 
An important fact is that a large number of nucleons in those nuclei are 
involved in the deformation, and thus such deformation is referred to as 
collective deformation. 
Indeed, the electromagnetic transition probability from e.g.,
the $ I = 2 $ state to the $ I = 0 $ state (i.e.~the ground state) is significantly enhanced due to the 
collectivity of deformation~\cite{
  Ring1980TheNuclearManyBodyProblem_SpringerVerlag}. 
\par
The collective deformation is characterized by a non-zero value of 
the quadrupole moment $ Q_{ij} $ defined by
\begin{equation}
  \label{eq:quadre}
  Q_{ij}
  =
  \int
  \left( 3 r_i r_j - \delta_{ij} r^2 \right) \, 
  \rho \left( \ve{r} \right)
  \, d \ve{r}
  \qquad
  \text{($ i $, $ j = x $, $ y $, $ z $)}, 
\end{equation}
where $ \rho \left( \ve{r} \right) $ is the density distribution.  
Note that there exist several normalizations for the quadrupole moment tensor $ Q $.
Here, we take the normalization often used in nuclear physics~\cite{
  Ring1980TheNuclearManyBodyProblem_SpringerVerlag},
which is sometimes referred to as ``traceless quadrupole moment'' in 
the context of quantum chemistry.
\par
The symmetric matrix $ Q_{ij} $ can be diagonalized to define the cartesian axes,
$ x' $, $ y' $, and $ z' $ in the intrinsic frame. 
With the eigenvalues of $ Q_{ij} $,
the deformation parameters $ \beta_k $ can be defined as
\begin{equation}
  \label{eq:beta}
  \beta_k
  =
  \sqrt{\frac{\pi}{5}}
  \frac{Q_k}{N_{\urm{tot}} \avr{r^2}}
  \qquad
  \text{($ k = x' $, $ y' $, $ z' $)},
\end{equation}
where $ Q_k $ is an eigenvalue of $ Q_{ij} $ and
$ \avr{r^2} $ is the mean-square radius given by 
\begin{equation}
  \label{eq:RMS} 
  \avr{r^2}
  =
  \frac{
    \int
    r^2
    \rho \left( \ve{r} \right)
    \, d \ve{r}}{
    \int
    \rho \left( \ve{r} \right)
    \, d \ve{r}}
  = 
  \frac{1}{N_{\urm{tot}}}
  \int
  r^2 
  \rho \left( \ve{r} \right)
  \, d \ve{r},
\end{equation}
with $ N_{\urm{tot}} $ being the total number of particles in the system.
For atoms, $ N_{\urm{tot}} $ is the total number of electrons (the atomic number) $ Z $,
whereas for atomic nuclei it is the total number of nucleons (the mass number) $ A $. 
\par
\begin{figure}[!htb]
  \centering
  \includegraphics[width=0.75\linewidth]{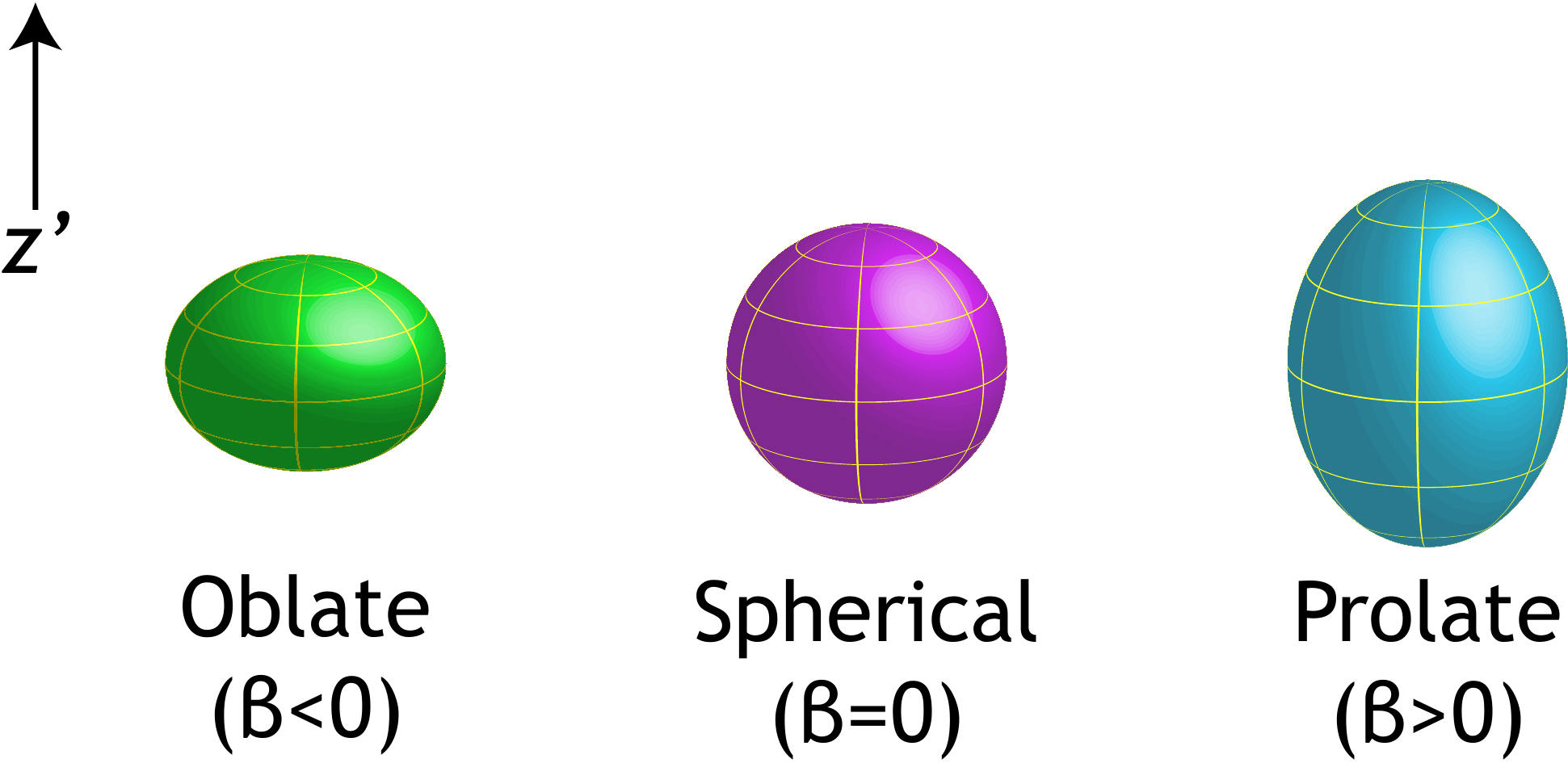}
  \caption{
    A schematic picture of the collective deformation.}
  \label{fig:schematic}
\end{figure}
\begin{figure}[!htb]
  \centering
  \includegraphics[width=0.5\linewidth]{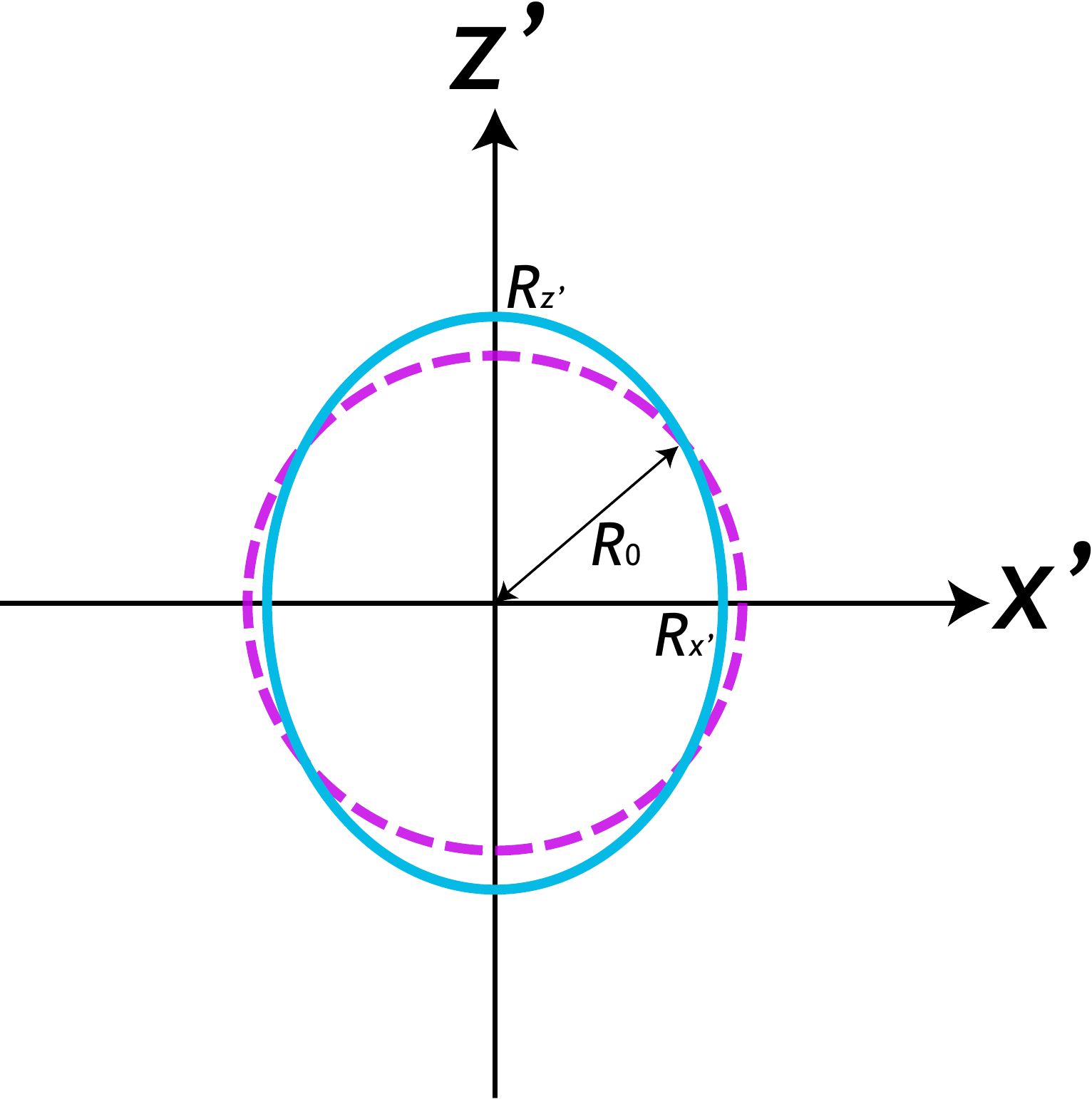}
  \caption{
    A schematic picture of a collective deformation in the $ x' z' $ plane defined in the intrinsic frame. 
    The axially symmetric shape along the $ z' $ axis is assumed. 
    The dashed and the solid curves correspond to the deformation parameter of 
    $ \beta = 0 $ and $ 0.25 $, respectively.}
  \label{fig:beta}
\end{figure}

The deformation parameter $ \beta_j $ is closely related to 
the angle-dependent radius of the system, $ R \left( \theta, \phi \right) $.  
The radius can be expanded in multipoles with 
spherical harmonics $ Y_{lm} $ as~\cite{
  Ring1980TheNuclearManyBodyProblem_SpringerVerlag}
\begin{equation}
  \label{eq:R1}
  R \left( \theta, \phi \right)
  =
  R_0
  \left[
    1
    +
    \sum_{l = 2}^{\infty}
    \sum_{m = -l}^l
    a_{lm}
    Y^*_{lm} \left( \theta, \phi \right) 
  \right].
\end{equation}
The monopole part ($ l = 0 $) in the expansion is absorbed in the definition of 
$ R_0 $, while the dipole part ($ l = 1 $) need not be considered
since, to the first order, 
it merely shifts the center of mass of the whole system without changing 
its shape. 
If one considers only the quadrupole deformation with $ l = 2 $, 
after transforming to the intrinsic frame, 
\eqref{eq:R1} is reduced to
\begin{align}
  R \left( \theta', \phi' \right)
  & =
    R_0
    \left[
    1
    +
    \beta \cos \gamma \,
    Y_{20} \left( \theta', \phi' \right)
    +
    \frac{1}{\sqrt{2}} \beta \sin \gamma \,
    Y_{22} \left( \theta', \phi' \right)
    +
    \frac{1}{\sqrt{2}} \beta \sin \gamma \,
    Y_{2-2} \left( \theta', \phi' \right)
    \right]
    \notag \\
  & = 
    R_0
    \left[
    1
    +
    \sqrt{\frac{5}{16 \pi}}
    \beta
    \left\{
    \cos \gamma
    \left( 3 \cos^2 \theta' - 1 \right)
    +
    \sqrt{3}
    \sin \gamma \,
    \sin^2 \theta' \,
    \cos 2 \phi'
    \right\}
    \right], 
    \label{eq:R2}
\end{align}
with $ a'_{20} = \beta \cos \gamma $, 
$ a'_{22} = a'_{2-2} = \frac{1}{\sqrt{2}} \beta \sin \gamma $,
and $ a'_{21} = a'_{2-1} = 0 $,
where the primes denote the quantities in the intrinsic frame.
Here, $ \beta > 0 $ and $ \gamma $ represent the degrees of elongation and triaxiality~\cite{
  Ring1980TheNuclearManyBodyProblem_SpringerVerlag}.
Due to the symmetry,
$ \gamma $ can be restricted to $ \gamma \in \left[ 0, \pi / 3 \right] $.
In addition, $ \gamma = \pi / 3 $ with $ \beta > 0 $ is identical to
$ \gamma = 0 $ with $ \beta < 0 $.
Hence, if one only considers the axially symmetric case with both positive and negative $ \beta $,
\eqref{eq:R2} can be simplified further as
\begin{equation}
  \label{eq:R3}
  R \left( \theta', \phi' \right)
  =
  R_0
  \left[
    1
    +
    \sqrt{\frac{5}{16 \pi}}
    \beta
    \left( 3 \cos^2 \theta' - 1 \right)
  \right], 
\end{equation}
with axially symmetric shape along the $ z' $ axis. 
In this case, the radii in the $ x' $, $ y' $, and $ z' $ directions are 
given by
\begin{align}
  R_{x'} = R_{y'}
  & =
    R \left( \pi/2, \phi' \right)
    =
    R_0
    \left(
    1
    -
    \sqrt{\frac{5}{16 \pi}}
    \beta
    \right), \\
  R_{z'}
  & = 
    R \left( 0, \phi' \right)
    = 
    R_0
    \left(
    1
    +
    2 \sqrt{\frac{5}{16 \pi}}
    \beta
    \right),
\end{align}
respectively.
Thus, the density distribution is spherical for $ \beta = 0 $, 
while for $ \beta > 0 $ and $ \beta < 0 $ it becomes prolate and oblate shapes, respectively,
as shown in
figure~\ref{fig:schematic}.
For a positive value of $ \beta $, the $ z' $ axis makes the longest axis 
as shown schematically in
figure~\ref{fig:beta}.
\par
The quadrupole deformation parameter $ \beta $ here corresponds to $ \beta_z $ defined by
\eqref{eq:beta}
(to the leading order, at least for a sharp-cut density distribution with the radius 
$ R \left( \theta, \phi \right) $~\cite{
  Ring1980TheNuclearManyBodyProblem_SpringerVerlag}).
Note that the condition of $ Q_x = Q_y = -Q_z / 2 $, 
i.e.~$ \beta_x = \beta_y = - \beta_z / 2 = - \beta / 2 $, 
holds for the axially symmetric case.
\par
If the particles in different spin states undergo different deformations,
as in the case for the electrons in atoms to be discussed in
section~\ref{sec:model}, 
spin-dependent deformation parameters $ \beta_{\uparrow} $ and $ \beta_{\downarrow} $ can be also defined.
Since $ \rho \left( \ve{r} \right) = \rho_{\uparrow} \left( \ve{r} \right) + \rho_{\downarrow} \left( \ve{r} \right) $ holds,
$ Q = Q_{\uparrow} + Q_{\downarrow} $ also does,
where $ \rho_{\uparrow} $ and $ \rho_{\downarrow} $ denote the densities of up-spin and down-spin fermions, respectively.
\par
It should also be noted that the discussion here is based on 
the intrinsic frame. In the laboratory frame, there is no criterion to 
choose the symmetry axis (that is, the $ z $ axis) and the system would show spherical distribution,
even when it is deformed in the intrinsic frame. 
\par
Microscopically, the collective deformation discussed in this 
section is intimately related to the mean-field theory. 
In this theory, particles move independently in a mean field potential, 
which is formed self-consistently by the interaction among the particles. 
The collective deformation occurs as a consequence of spontaneous breaking 
of rotational symmetry. 
That is, a system may be \textit{intrinsically} deformed and breaks the 
rotational symmetry even 
if the total Hamiltonian has the symmetry. 
This is actually a good advantage of the self-consistent mean-field theory used in
section~\ref{sec:numerical},
since a large part of the many-body correlations can be taken into account 
while keeping the simple picture of independent particles~\cite{
  Ring1980TheNuclearManyBodyProblem_SpringerVerlag,
  Nilsson1955Dan.Mat.Fys.Medd.29_1,
  Mottelson1955Phys.Rev.99_1615,
  Ragnarsson2005ShapesandShellsinNuclearStructure_CambridgeUniversityPress}.
\subsection{Single-particle contribution to deformation}
\label{sec:deformation_sp}
\par
In connection to the deformation in atoms discussed in the next section, 
it is instructive to compute the deformation parameters discussed in the 
previous subsection for a system with a spherical core plus a valence particle.
In this case, the quadrupole moment of the system entirely comes from the valence particle,
since
\eqref{eq:quadre} vanishes for the spherical density distribution. 
If one assumes that the wave function for the valence particle is 
given by 
\begin{equation}
  \label{eq:spwf}
  \psi_{nlms} \left( \ve{r} \right)
  =
  \frac{u_{nl} \left( r \right)}{r}
  Y_{lm} \left( \theta, \phi \right)
  \chi_s ,
\end{equation}
as a single-particle orbital 
in a spherical (effective) potential, 
where $ \chi_s $ is the spinor of the valence particle, 
the quadrupole moments of the whole system read 
$ Q_{ij} = q^{\urm{(sp)}}_{ij} $ with 
\begin{align}
  q_{ij}^{\urm{(sp)}}
  & = 
    \int
    \left( 3 r_i r_j - \delta_{ij} r^2 \right) 
    \left|
    \psi_{nlms} \left( \ve{r} \right)
    \right|^2
    \, d \ve{r}
    \notag \\
  & = 
    \int
    \left( 3 \fr{s}_i \fr{s}_j - \delta_{ij} \right)r^2
    \left|
    \psi_{nlms} \left( \ve{r} \right)
    \right|^2
    \, d \ve{r}. 
\end{align}
Here, we have introduced the notation 
$ r_i = r \fr{s}_i $ with 
\begin{equation}
  \fr{s}_i
  =
  \begin{cases}
    \sin \theta \cos \phi
    & \text{($ i = x $)}, \\
    \sin \theta \sin \phi
    & \text{($ i = y $)}, \\
    \cos \theta
    & \text{($ i = z $)}.
  \end{cases}
\end{equation}
Substituting
\eqref{eq:spwf}, one finds 
\begin{equation}
  \label{eq:sp-q2}
  q_{ij}^{\urm{(sp)}}
  =
  \int_0^\infty
  r^2\left|
    u_{nl} \left( r \right)
  \right|^2
  \, dr \,
  \int
  \left( 3 \fr{s}_i \fr{s}_j - \delta_{ij} \right)
  \left|
    Y_{lm} \left( \theta, \phi \right)
  \right|^2
  \, d \Omega, 
\end{equation}
where $ d \Omega $ is the angular part of $ d \ve{r} $.
Note that the off-diagonal components of $ q_{ij}^{\urm{(sp)}} $ vanish.
The deformation parameter for the valence particle, that is, the 
single-particle deformation parameter, then reads, 
\begin{equation}
  \label{eq:beta_sp}
  \beta_i^{\urm{(sp)}}
  =
  \sqrt{\frac{\pi}{5}}
  \frac{q_{ii}^{\urm{(sp)}}}{\avr{r^2}_{nl}}
  =
  \sqrt{\frac{\pi}{5}}
  \int
  \left( 3 \fr{s}_i^2 - 1 \right)
  \left|
    Y_{lm} \left( \theta, \phi \right)
  \right|^2
  \, d \Omega.
\end{equation}
Notice that the radial integral in
\eqref{eq:sp-q2} is cancelled with the mean-square radius 
\begin{equation}
  \avr{r^2}_{nl}
  =
  \int_0^{\infty}
  r^2
  \left| 
    u_{nl} \left( r \right)
  \right|^2
  \, dr. 
\end{equation}
The deformation parameter of the whole system
(i.e.~the core plus the valence particle)
is then given by 
\begin{equation}
  \beta_i
  = 
  \sqrt{\frac{\pi}{5}}
  \frac{q_i^{\urm{(sp)}}}{N_{\urm{tot}} \avr{r^2}},
\end{equation}
where $ \avr{r^2} =
\left[
  \left( N_{\urm{tot}} - 1 \right) \avr{r^2}_{\urm{core}} + \avr{r^2}_{nl}
\right] / N_{\urm{tot}} $ 
is the mean-square radius of the whole system 
with $ \avr{r^2}_{\urm{core}} $ being that of the core. 
If one assumes the radius of the whole system is similar to the radius of the 
valence orbital, $ \avr{r^2} \simeq \avr{r^2}_{nl} $, 
which is reasonable in the case of atomic nuclei,
the deformation parameter is given by 
\begin{equation}
  \label{eq:beta_Ntot}
  \beta_i
  \simeq
  \frac{\beta_i^{\urm{(sp)}}}{N_{\urm{tot}}}. 
\end{equation}
In contrast, in the case of atoms,
valence electrons have a larger spatial distribution
and contribute more significantly than inner core electrons to the mean-square radius $ \avr{r^2} $.
If one assumes $ \avr{r^2} \simeq \avr{r^2}_{nl} / N_{\urm{tot}} $,
i.e.~only one electron mainly contributes to $ \avr{r^2} $,
one obtains
\begin{equation}
  \label{eq:beta_Nval}
  \beta_i
  \simeq
  \beta_i^{\urm{(sp)}}.
\end{equation}
Notice that when the number of valence electrons is $ N_{\urm{val}} $,
the deformation parameter is, instead, approximately given by
\begin{equation}
  \beta_i
  \simeq
  \frac{\sum_{\urm{val}} \beta_i^{\urm{(sp)}}}{N_{\urm{val}}}.
\end{equation}
\par
Let us evaluate explicitly the single-particle deformation 
parameter for 
the pure $ p $ and $ d $ orbitals. 
To this end, we define the indices of the real spherical harmonics as follows~\cite{
  Blanco1997J.Mol.Struct.419_19}:
\begin{subequations}
  \begin{align}
    Y_p^x \left( \theta, \phi \right)
    & =
      \sqrt{\frac{3}{4 \pi}}
      \sin \theta \, \cos \phi, \\
    Y_p^y \left( \theta, \phi \right)
    & =
      \sqrt{\frac{3}{4 \pi}}
      \sin \theta \, \sin \phi, \\
    Y_p^z \left( \theta, \phi \right)
    & =
      \sqrt{\frac{3}{4 \pi}}
      \cos \theta, 
  \end{align}
\end{subequations}
and
\begin{subequations}
  \begin{align}
    Y_d^{xy} \left( \theta, \phi \right) 
    & =
      \sqrt{\frac{15}{16 \pi}}
      \sin^2 \theta \, \sin 2 \phi, \\
    Y_d^{yz} \left( \theta, \phi \right)
    & =
      \sqrt{\frac{15}{16 \pi}}
      \sin 2 \theta \, \sin \phi, \\
    Y_d^{zx} \left( \theta, \phi \right)
    & =
      \sqrt{\frac{15}{16 \pi}}
      \sin 2 \theta \, \cos \phi, \\
    Y_d^{x^2 - y^2} \left( \theta, \phi \right)
    & =
      \sqrt{\frac{15}{16 \pi}}
      \sin^2 \theta \, \cos 2 \phi, \\
    Y_d^{z^2} \left( \theta, \phi \right)
    & =
      \sqrt{\frac{5}{16 \pi}}
      \left( 3 \cos^2 \theta - 1 \right).
  \end{align}
\end{subequations}
Using these notations, 
the deformation parameter $ \beta^{\urm{(sp)}}_i $ for $ Y_p^j $ is found to be 
\begin{equation}
  \label{eq:beta_single_p}
  \beta_i^{\urm{(sp)}}
  =
  \begin{cases}
    \frac{4}{5} \sqrt{\frac{\pi}{5}}
    = 0.6341
    & \text{($ i = j $)}, \\
    - \frac{2}{5} \sqrt{\frac{\pi}{5}}
    = -0.3171
    & \text{(Otherwise)}. 
  \end{cases}
\end{equation}
For the $ d $-orbitals, the deformation parameter 
for $ Y_d^{jk} $
($ \left( j, k \right) = \left( x, y \right) $, $ \left( y, z \right) $, $ \left( z, x \right) $)
is given by 
\begin{subequations}
  \label{eq:beta_single_d}
  \begin{equation}
    \label{eq:beta_single_dxy}
    \beta_i^{\urm{(sp)}}
    =
    \begin{cases}
      - \frac{4}{7} \sqrt{\frac{\pi}{5}}
      = -0.4530
      & \text{($ i \ne j $ and $ i \ne k $)}, \\
      \frac{2}{7} \sqrt{\frac{\pi}{5}}
      = 0.2265
      & \text{(Otherwise)}, 
    \end{cases}
  \end{equation}
  while the deformation parameters for 
  $ Y_d^{x^2 - y^2} $ and 
  $ Y_d^{z^2} $ read 
  \begin{equation}
    \label{eq:beta_single_dxsqysq}
    \beta_i^{\urm{(sp)}}
    =
    \begin{cases}
      - \frac{4}{7} \sqrt{\frac{\pi}{5}}
      = -0.4530
      & \text{($ i = z $)}, \\
      \frac{2}{7} \sqrt{\frac{\pi}{5}}
      = 0.2265
      & \text{(Otherwise)}, 
    \end{cases} 
  \end{equation}
  and 
  \begin{equation}
    \label{eq:beta_single_dzsq}
    \beta_i^{\urm{(sp)}}
    =
    \begin{cases}
      \frac{4}{7} \sqrt{\frac{\pi}{5}}
      = 0.4530
      & \text{($ i = z $)}, \\
      - \frac{2}{7} \sqrt{\frac{\pi}{5}}
      = -0.2265
      & \text{(Otherwise)}, 
    \end{cases} 
  \end{equation}
\end{subequations}
respectively.
Note that due to higher-order deformations,  
the $ d $ orbitals are not axially symmetric except the $ d_{z^2} $ orbital,
even though triaxiality vanishes ($ \gamma = 0 $).

% 
% Numerical Calculation
% -*- coding: utf-8 -*-
%%%%%%%%%%%%%%%%%%%%%%%%%%%%%%%%%%%%%%%%%%%%%%%%%% 
% 
% Deformation of atoms paper manuscript for J. Phys. B
% Numerical Calculation Part
% 
% Begin to write: 2020-07-25
% 
% Tomoya Naito, Shimpei Endo,
% Kouichi Hagino, and Yusuke Tanimura
% 
%%%%%%%%%%%%%%%%%%%%%%%%%%%%%%%%%%%%%%%%%%%%%%%%%% 
% 
\section{Quantum Chemical Calculation}
\label{sec:numerical}
\par
In this section, a possibility of the deformation of isolated atoms
is studied numerically with the computational code \textsc{Gaussian}~\cite{
  Gaussian}.
In the \textsc{Gaussian} code, the wave function and the energy of the ground state of atomic and molecular systems are numerically calculated by using the Gaussian-type basis expansion.
With this code, all-electron calculation can be performed for a wide range of atoms ($ Z \le 54 $).
\par
First, our calculation setup is explained in
section~\ref{sec:numerical_setup}.
Then, the results and their basic explanations are presented in
sections~\ref{sec:numerical_systematic} and \ref{sec:numerical_method},
where we shall see that open-shell atoms have non-zero deformation parameters.
To understand clearly the mechanism of such deformations,
we discuss the results in more details in
section~\ref{sec:numerical_sp}
by analyzing the deformations of each single-particle orbital.
\par
For convenience, the Hartree atomic unit
$ m_e = \hbar = e^2 = a_{\urm{B}} = 4 \pi \epsilon_0 = 1 $ is used,
where $ m_e $ is the mass of electrons,
$ \epsilon_0 $ is the permittivity of vacuum, and
$ a_{\urm{B}} = 5.291 8 \ldots \times 10^{-11} \, \mathrm{m} $
is the Bohr radius~\cite{
  Mohr2016Rev.Mod.Phys.88_035009,
  CODATA}.
The unity of the electric quadrupole moment in the Hartree atomic unit corresponds to
$ 4.486 6 \ldots \times 10^{-40} \, \mathrm{C} \, \mathrm{m}^2
= 1.345 0 \ldots \, \mathrm{Debye} \, \mathrm{\AA} $.
Here, we use quadrupole moment of \textit{density} itself, instead of \textit{charge} density,
and thus the electric quadrupole moment corresponds to
$ -e $ ($ < 0 $) times of $ Q_k $ in this paper.
\subsection{Calculation setup and deformation parameters}
\label{sec:numerical_setup}
\par
An isolated neutral atom with the atomic number $ Z $ in the non-relativistic scheme is described by the Hamiltonian 
\begin{equation}
  \label{eq:Hamil_general}
  H
  =
  -
  \frac{1}{2}
  \sum_{j = 1}^Z
  \laplace_j
  +
  \sum_{j = 1}^Z
  V_{\urm{ext}} \left( \ve{r}_j \right) 
  +
  \sum_{1 \le j < k \le Z}
  V_{\urm{int}} \left( \ve{r}_j, \ve{r}_k \right),
\end{equation}
where $ \ve{r}_j $ is the coordinate of the particle $ j $.
The external potential $ V_{\urm{ext}} $ and
the electron-electron interaction $ V_{\urm{int}} $ read 
\begin{align}
  V_{\urm{ext}} \left( \ve{r} \right)
  & =
    - \frac{Z}{r},
    \label{eq:ext} \\
  V_{\urm{int}} \left( \ve{r}_j, \ve{r}_k \right)
  & =
    \frac{1}{r_{jk}},
    \label{eq:int}
\end{align}
respectively, 
where $ r = \left| \ve{r} \right| $ and $ r_{jk} = \left| \ve{r}_j - \ve{r}_k \right| $.
Here, the atomic nucleus is assumed to be a point charge with infinitely heavy mass, and the spin-orbit interaction is neglected.
\par
In this paper, the many-body Schr\"{o}dinger equation is approximately solved
using mainly the unrestricted Hartree-Fock (UHF) method~\cite{
  Roothaan1951Rev.Mod.Phys.23_69,
  Pople1954J.Chem.Phys.22_571,
  Berthier1954C.R.Acad.Sci.238_91,
  McWeeny1968J.Chem.Phys.49_4852,
  Jensen2017IntroductiontoComputationalChemistry_JohnWiley&Sons}.
The quadrupole deformation parameters of atoms have not yet been directly observed with sufficient accuracy to compare with our theoretical results.
Moreover, it is generally known to be difficult to systematically estimate theoretical errors in quantum chemical calculations.
For this reason, we regard the difference among the results of various methods and bases
as crude estimates of the error in the quadrupole moment $ Q_i $
and the correspoinding deformation parameter $ \beta $.
\par
To assess the dependence of the many-body methods,
several post-Hartree-Fock methods
and 
the DFT~\cite{
  Hohenberg1964Phys.Rev.136_B864,
  Kohn1965Phys.Rev.140_A1133,
  Kohn1999Rev.Mod.Phys.71_1253}
with the PZ81 local density approximation
(LDA) exchange-correlation functional~\cite{
  Perdew1981Phys.Rev.B23_5048}
are also employed.
As for the post-Hartree-Fock methods, we choose 
the configuration interaction~\cite{
  Pople1976Int.J.QuantumChem.10_1,
  Pople1977Int.J.QuantumChem.12_149,
  Krishnan1980J.Chem.Phys.72_4654,
  Pople1999Rev.Mod.Phys.71_1267}
with double excitations (CID) and single-double excitations (CISD),
the coupled-cluster method~\cite{
  Coester1958Nucl.Phys.7_421,
  Cizek1966J.Chem.Phys.45_4256,
  Cizek1971Int.J.QuantumChem.5_359}
with double excitations (CCD) and single-double excitations (CCSD),
and
M{\o}ller-Plesset many-body perturbation theory~\cite{
  Moller1934Phys.Rev.46_618}
with second (MP2), third (MP3), and fourth (MP4) orders~\cite{
  Krishnan1978Int.J.QuantumChem.14_91,
  Head-Gordon1988Chem.Phys.Lett.153_503,
  Trucks1988Chem.Phys.Lett.147_359,
  Raghavachari1990J.Phys.Chem.94_5579}.
For the MP4 calculation, the excitations are restricted to single, double, and quadruple excitations (MP4SDQ),
which is simply referred to as MP4.
The calculations are performed with these post Hartree-Fock methods,
i.e.~CID, CISD, CCD, CCSD, MP2, MP3, and MP4,
based on the UHF calculation.
In these post Hartree-Fock calculations, the core is not treated as frozen.
\par
To perform the many-body calculations,
the basis expansions with the 
6-31+G~\cite{
  Hehre1972J.Chem.Phys.56_2257,
  Clark1983J.Comput.Chem.4_294},
dAug-CC-pV5Z~\cite{
  Woon1993J.Chem.Phys.98_1358,
  Peterson1994J.Chem.Phys.100_7410},
and
STO-3G~\cite{
  Hehre1969J.Chem.Phys.51_2657}
bases are used.
It should be noted that the different bases are optimized to different ranges of atoms~\cite{
  Gaussian}:
The STO-3G basis can be applied from $ \mathrm{H} $ atom to $ \mathrm{Xe} $ atom.
The applicable ranges of the 6-31+G and the dAug-CC-pV5Z bases are limited to atoms from $ \mathrm{H} $ to $ \mathrm{Kr} $.
In addition, the dAug-CC-pV5Z basis cannot be applied to $ \mathrm{Mg} $, $ \mathrm{K} $, or $ \mathrm{Ca} $ atoms.
In this paper, we mainly employ the 6-31+G basis,
while the results with the dAug-CC-pV5Z and STO-3G bases are also shown in
\ref{sec:data}.
\par
To perform the calculation with the \textsc{Gaussian} code, the multiplicity of the ground states has to be assumed,
which are taken from the database provided by the National Institute of Standards and Technology (NIST)~\cite{
  NIST_ASD}.
We do not assume any symmetry for the initial condition, nor use the filling approximation,
in which valence orbitals in open shells are filled with equal occupation probabilities~\footnote{
  For example, in the case of $ \mathrm{B} $,
  $ 2p_x $, $ 2p_y $, and $ 2p_z $ orbitals are occupied with the probability of $ 1/3 $.
  Notice that the wave function in the filling approximation is different from
  the coherent superposition of those orbitals,
  $ \ket{\text{val}} = \left( \ket{2p_x} + \ket{2p_y} + \ket{2p_z} \right) / \sqrt{3} $,
  even though both of them lead to a spherical density distribution.}.
With this procedure,
non-spherical electron density can be described properly.
After the calculation is converged, the valence-electron configurations are evaluated by using the natural orbital analysis~\cite{
  Foster1980J.Am.Chem.Soc.102_7211,
  Carpenter1988J.Mol.Struct.169_41,
  Reed1988Chem.Rev.88_899,
  NBO}.
\par
After we obtain the electron density $ \rho \left( \ve{r} \right) $ with these quantum chemical methods,
the mean-square radius $ \avr{r^2} $ and the quadrupole moments $ Q_i $ are calculated
according to
\eqref{eq:quadre} and \eqref{eq:RMS}, respectively.
Here, we define the coordinate frame so that the quadrupole moment tensor $ Q $ is diagonalized
(see
\eqref{eq:beta}).
\subsection{Systematic behaviors of atomic deformations and their basis dependence}
\label{sec:numerical_systematic}
\par
\begin{figure}[!htb]
  \centering
  \includegraphics[width=1.0\linewidth]{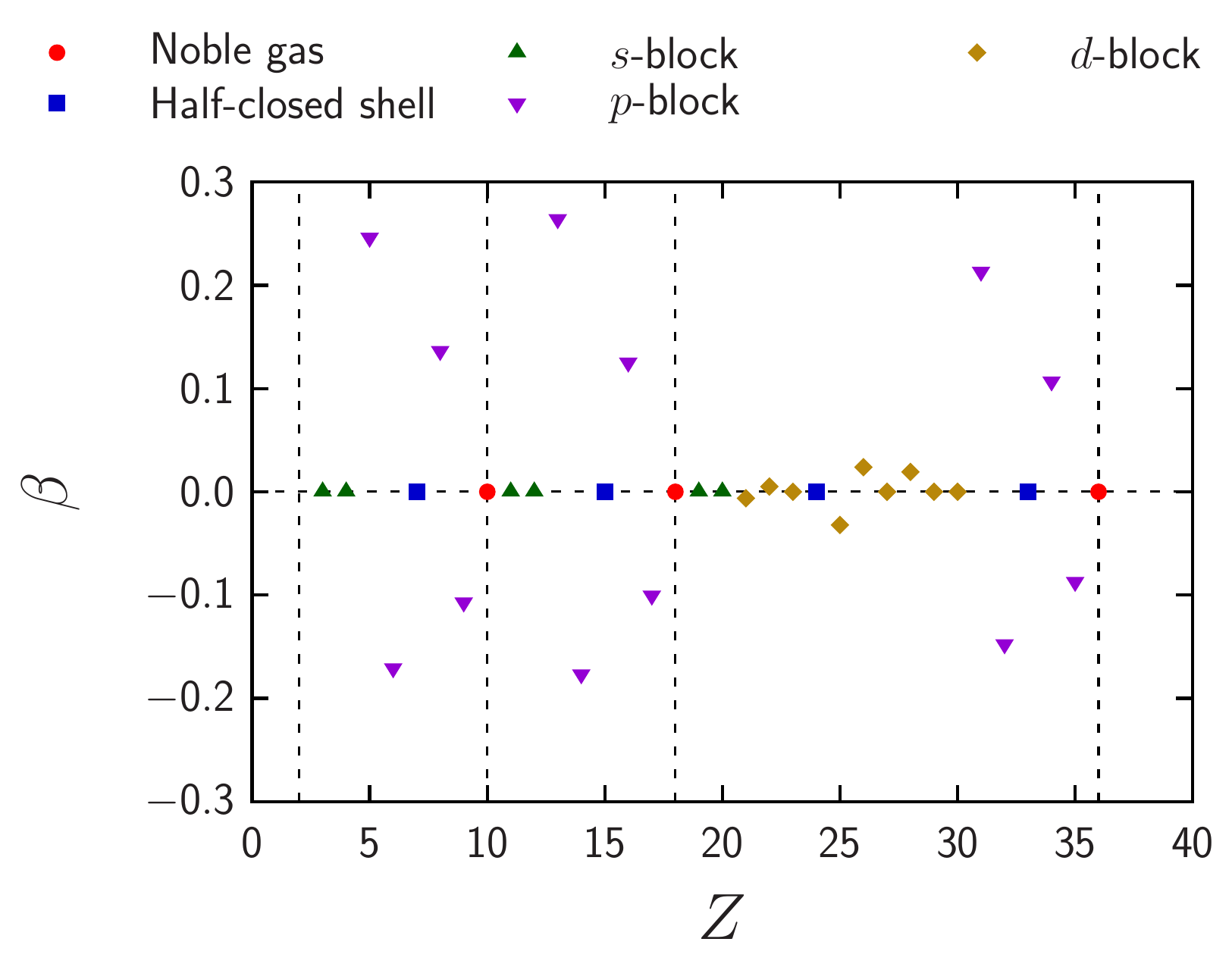}
  \caption{
    Deformation parameter $ \beta $ as a function of the atomic number $ Z $ from $ \mathrm{Li} $ ($ Z = 3 $) to $ \mathrm{Kr}$ ($ Z = 36 $)
    calculated with the unrestricted Hartree-Fock method with the 6-31+G basis.
    The filled circles and the filled squares show noble-gas and half-closed-shell atoms, respectively.
    The triangles, the inverse-triangles, and the diamonds show the others;
    atoms with the outer-most open-shell $ s $ orbitals ($ s $ block),
    $ p $ orbitals ($ p $ block), and $ d $ orbitals ($ d $ block), respectively.
    The vertical dashed lines denote the closed-shell atoms.}
  \label{fig:Gaussian_UHF_631pG_all}
\end{figure}

Let us now numerically investigate deformation in atoms.
We first show in
figure~\ref{fig:Gaussian_UHF_631pG_all}
the deformation parameter $ \beta $
for atoms from $ \mathrm{Li} $ ($ Z = 3 $) to $ \mathrm{Kr}$ ($ Z = 36 $)
calculated with the unrestricted Hartree-Fock method with the 6-31+G basis.
See Table~\ref{tab:Gaussian_UHF_631pG_all} in
\ref{sec:data} for the actual values.
In
figure~\ref{fig:Gaussian_UHF_631pG_all},
we use different symbols to classify the nature of the atoms:
The filled circles show noble-gas (i.e.~closed-shell) atoms.
The filled squares show half-closed-shell atoms;
atoms with three valence electrons occupying the outer-most $ p $ orbitals,
or with five valence electrons occupying the outer-most $ d $ orbitals.
The triangles, the inverse triangles, and the diamonds show the others;
atoms with $ s $, $ p $, and $ d $ orbitals for the outer-most open shell, respectively~\cite{
  Cotton1995_JohnWiley&Sons}.
Note that the octupole moments for all the calculated atoms are found to be zero.
\par
\textbf{Spherical atoms}: According to
figure~\ref{fig:Gaussian_UHF_631pG_all},
all the noble-gas atoms are spherical ($ \beta = 0 $), as is expected.
Such trivial results are guaranteed by the fact that all the orbitals for any given principal and azimuthal quantum numbers, $ n $ and $ l $, are either fully occupied or unoccupied,
and thus their distributions are, in total, spherical.
In other words,
$ \sum_{m = -l}^l \left| Y_{lm} \left( \theta, \phi \right) \right|^2 = \left( 2l + 1 \right) / 4 \pi $
is a constant, independent of the angles $ \theta $ and $ \phi $.
\par
For the other atoms, the electronic configuration of their cores is the same as that of the noble-gas atoms.
Hence, as long as the core density is not deformed due to non-trivial many-body effects,
i.e.~the polarization due to the interaction between the core and valence electrons,
the deformation of the atom is expected to come only from the valence electrons.
Possibility of such non-trivial deformation will be studied in details in
section~\ref{sec:numerical_sp},
but we can easily conclude before going to such details that the $ s $-block and half-closed-shell atoms should be spherical ($ \beta = 0 $),
as shown in
figure~\ref{fig:Gaussian_UHF_631pG_all}.
The valence electrons of the $ s $-block atoms by definition only occupy an $ s $ orbital,
and the many-body effects between the valence electrons and the spherically symmetric core necessarily results in a spherical electron density distribution.
This is the case also for the half-closed-shell atoms:
Atom with three electrons occupying valence $ p $ orbitals, or those with five electrons occupying valence $ d $ orbitals, are spherical in both cases,
since
$ \alpha $-spin electrons occupy all the $ \left( 2l + 1 \right) $ states with $ m = -l $, $ -l+1 $, \ldots, $ l $
due to the Hund rule~\cite{
  Hund1925Z.Phys.34_296,
  Hund1925Z.Phys.33_345,
  Slater1929Phys.Rev.34_1293,
  Hongo2004J.Chem.Phys.121_7144,
  Oyamada2010J.Chem.Phys.133_164113}. 
\par
\textbf{Open-shell atoms}: For the other open-shell atoms ($ p $-block and $ d $-block atoms),
the deformation parameters $ \beta $ are generally non-zero.
The deformation parameters for the atoms with the same group show similar tendencies,
i.e.~$ \left| \beta \right| \simeq 0.1 $--$ 0.3 $ for the $ p $-block atoms and
$ \left| \beta \right| \lesssim 0.01 $ for the $ d $-block atoms.
This shows that the nature of the few valence open-shell electrons plays a major role in determining $ \beta $ of the whole atom.
In the context of nuclear physics, an atomic nucleus with the deformation parameter of $ \simeq \pm 0.1 $ or larger is regarded as a deformed nucleus (see
\ref{sec:n_drop}).
It is thus tempting to regard the atoms with $ \left| \beta \right| \gtrsim 0.1 $ as significantly deformed as nuclei.
However, this deformation originates from single-particle valence orbitals and thus it is misleading to regard it as collective deformation.
For example, the deformation of the $ p $-block atoms can be understood as the valence $ p $ orbitals mainly contributing to the deformation.
Indeed, $ \left| \beta \right| \simeq 0.1 $--$ 0.3 $ of the $ p $-block atoms in
figure~\ref{fig:Gaussian_UHF_631pG_all}
are similar values in order of magnitude as that of a single $ p $ orbital given by
\eqref{eq:beta_Nval} and \eqref{eq:beta_single_p}.
On the other hand, the deformations of the $ d $-block atoms are much smaller than those of the $ p $-block atoms
(see and compare with
\eqref{eq:beta_single_d}).
This implies that one cannot regard the deformation of $ d $-block atoms as originating only from the single-orbital deformation.
The mechanism of the deformation and the physical origin of this difference will be discussed in
section~\ref{sec:numerical_sp}:
We will conclude there that the $ p $-block atoms can be regarded as an almost inert core plus valence $ p $-electrons,
while many-body effects between the core and the valence electrons are relevant in the $ d $-block atoms.
In the $ d $-block atoms, the core electrons tend to collectively cancel the deformation of the valence $ d $ orbitals.
\par
Exceptionally $ \beta = 0 $ open-shell atoms are
$ \mathrm{V} $ ($ Z = 23 $),
$ \mathrm{Co} $ ($ Z = 27 $),
$ \mathrm{Cu} $ ($ Z = 29 $),
and $ \mathrm{Zn} $ ($ Z = 30 $).
In the case of $ \mathrm{Cu} $ and $ \mathrm{Zn} $ atoms,
ten of the valence electrons completely occupy the valence $ d $ orbital and the remaining one or two valence electrons occupy the $ s $ orbital.
Thus, they are trivially spherical, in the same manner as the $ s $-block atoms.
In the case of $ \mathrm{V} $ and $ \mathrm{Co} $ atoms,
on the other hand, three or eight valence electrons occupy the $ d $ orbitals.
At first sight, such open-shell $ d $-block atoms do not seem to be trivially $ \beta = 0 $.
However, we note that they can be $ \beta = 0 $ if $ d_{x^2-y^2} $ and $ d_{z^2} $ orbitals are occupied (unoccupied) at the same time~\footnote{
  Higher-order deformations, such as hexadecapole deformation, exist
  even though they are considerably small.}.
For example, $ d^8 $ occupation may lead to $ \beta = 0 $ if the unoccupied orbitals are $ d_{x^2-y^2} $ and $ d_{z^2} $ orbitals.
This is the case for $ \mathrm{Co} $ atom.
For $\mathrm{V}$ atom, $ d^3 $ occupation leads to $ \beta = 0 $ because $ d_{x^2-y^2} $ and $ d_{z^2} $ orbitals are occupied simultaneously.
\par
\textbf{Basis dependence}: In order to guarantee that our results presented in this section and in
figure~\ref{fig:Gaussian_UHF_631pG_all}
do not depend on a specific choice of the basis in the calculation,
we repeat calculations with different basis sets:
In addition to the 6-31+G basis used in
figure~\ref{fig:Gaussian_UHF_631pG_all}
and Table~\ref{tab:Gaussian_UHF_631pG_all} in
\ref{sec:data},
we also use the dAug-CC-pV5Z and STO-3G bases.
The results are shown in Table~\ref{tab:Gaussian_UHF_dCCpV_all} and Table~\ref{tab:Gaussian_UHF_STO_all} in
\ref{sec:data}, respectively.
We find that as long as the configurations of the valence electrons are the same, the deformation parameters $ \beta $ are essentially the same, hence basis independent.
We note, however, that there are a few results with the STO-3G basis which are different from those with the dAug-CC-pV5Z and 6-31+G bases, even when the configurations are the same.
This may be because the STO-3G basis set spans a smaller space than the other basis sets,
and thus the tail region of the electron density distribution $ \rho \left( \ve{r} \right) $ is not accurately described compared with the others.
\subsection{Many-body method dependence}
\label{sec:numerical_method}
\par
\begin{figure}[!htb]
  \centering
  \includegraphics[width=1.0\linewidth]{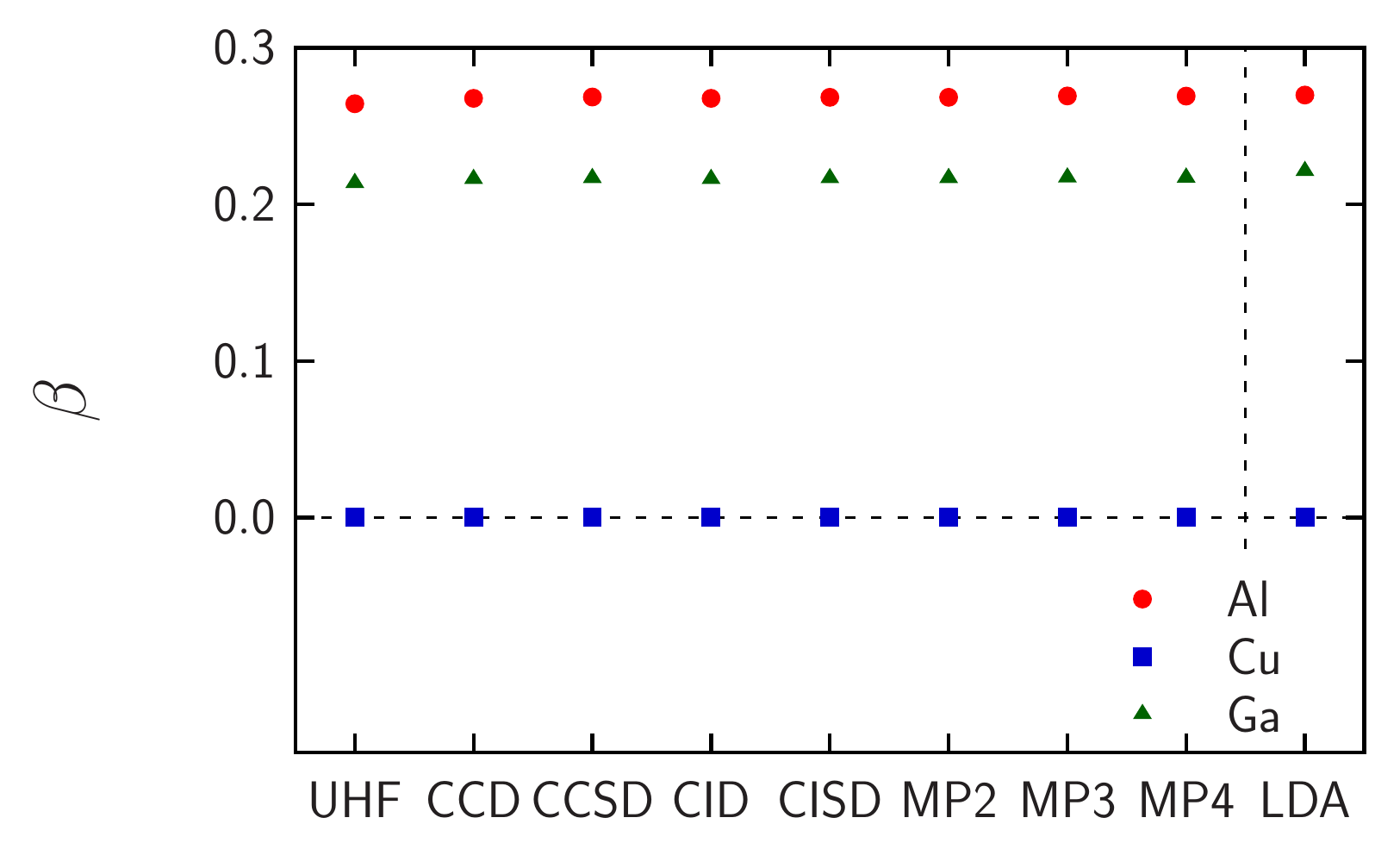}
  \caption{
    Dependence of the deformation parameter $ \beta $
    on many-body methods with the 6-31+G basis.
    Data for $ \mathrm{Al} $ ($ Z = 13 $),
    $ \mathrm{Cu} $ ($ Z = 29 $), and $ \mathrm{Ga} $ ($ Z = 31 $)
    are shown with circles, squares, and triangles, respectively.
    For more details, see the text.}
  \label{fig:Gaussian_631pG_Al_Cu_Ga}
\end{figure}

Our results presented in the previous section essentially remain the same, 
even if we employ other many-body methods than the unrestricted Hartree-Fock (UHF) method.
As benchmark examples,
$ \mathrm{Al} $ ($ Z = 13 $),
$ \mathrm{Cu} $ ($ Z = 29 $), and
$ \mathrm{Ga} $ ($ Z = 31 $) are selected.
In
figure~\ref{fig:Gaussian_631pG_Al_Cu_Ga},
we show the deformation parameters $ \beta $ for those atoms calculated with a wide variety of different many-body methods with the 6-31+G basis.
The data are shown in Table~\ref{tab:Gaussian_631pG_Al_Cu_Ga} in
\ref{sec:data}.
We find that the deformation parameters $ \beta $ calculated with the post-Hartree-Fock methods
(CCD, CCSD, CID, CISD, MP2, MP3, and MP4) are almost the same as those with the UHF.
The excellent agreement of the UHF and the post-Hartree-Fock methods in
figure~\ref{fig:Gaussian_631pG_Al_Cu_Ga}
implies that the correlations beyond Hartree-Fock may be irrelevant to the quadrupole deformation.
\par
We have also performed the DFT calculation
(indicated by LDA in
figure~\ref{fig:Gaussian_631pG_Al_Cu_Ga})
and found that the result also agrees excellently with the UHF and post-Hartree-Fock methods.
Since the DFT method is not directly linked to the UHF calculation,
there is no \textit{a priori} reason why $ \beta $ should be the same.
In principle, owing to the Hohenberg-Kohn theorem, the DFT calculation provides the exact, i.e.~the full CI, density, if the exact exchange-correlation functional were known.
In practice, the DFT results should depend on the choice of the employed exchange-correlation functional,
since only approximated functional have been known.
The excellent agreement between the DFT (LDA) and the other methods in
figure~\ref{fig:Gaussian_631pG_Al_Cu_Ga},
therefore, guarantees that finite deformation parameters presented in the previous section and following sections are not at all artifacts of a specific many-body approximation method.
\par
We note that all the above results and discussions in this section hold true for the other bases:
We have performed the same calculation also with dAug-CC-pV5Z basis as shown in Table~\ref{tab:Gaussian_dCCpV_Al_Cu_Ga} in
\ref{sec:data},
and obtained essentially the same results as those with the 6-31+G basis in
figure~\ref{fig:Gaussian_631pG_Al_Cu_Ga}
and Table~\ref{tab:Gaussian_631pG_Al_Cu_Ga}.
\subsection{Analyses with single-particle orbitals}
\label{sec:numerical_sp}
\par
To understand further the physical mechanism of the atomic deformations,
let us define the deformation parameter of each single-particle orbital
$ \beta_{\urm{max}}^{\urm{(sp)}} $ by $ \beta_i^{\urm{(sp)}} $ for the direction of the symmetry axis of the orbital.
Here, the symmetry axis of the orbital may be different from that of the whole atoms,
i.e.~the $ i $ axis for $ p_i $ orbital,
the $ k $ axis for $ d_{ij} $ orbital ($ i \ne j \ne k $),
and the $ z $ axis for the other $ d $ orbitals.
Angular-momentum mixing can occur due to the deformation,
so that single-particle orbitals in these methods can, in general, show $ \beta $ different from those of the pure spherical harmonics.
Specifically, as we will see, an $ s $ orbital can show a non-zero $ \beta $
due to the mixture with $ d $ orbitals.
Note that $ \beta_{\urm{max}}^{\urm{(sp)}} $ is the deformation parameter for each single-particle orbital (see
\eqref{eq:beta_sp}),
which is divided by the radius of each orbital.
Therefore, $ \beta_i $ of the whole atom is not a simple sum of $ \beta_{\urm{max}}^{\urm{(sp)}} $.
Thus, a care must be taken when using them, but they are still useful in understanding the mechanism of the deformation microscopically.
\par
\textbf{Deformation of single-particle orbitals}: Table~\ref{tab:Q_valence} shows $ \beta_{\urm{max}}^{\urm{(sp)}} $ for selected valence orbitals of some open-shell atoms.
All the data are calculated by the unrestricted Hartree-Fock calculation with the 6-31+G basis. First, we find that the deformation parameters of the valence $ p $ and $ d $ orbitals 
are almost the same as those of the pure $ p $ and $ d $ orbitals given by
\eqref{eq:beta_single_p}--\eqref{eq:beta_single_dzsq}.
This suggests that even when the atom is deformed,
the single-particle orbitals of the valence $ p $ and $ d $ electrons are described almost as a product of the radial function and the pure spherical harmonics.
In other words,
the valence $ p $ and $ d $ electrons feel an almost perfectly spherical effective (Hartree-Fock or Kohn-Sham) potential.
We can, therefore, conclude that the density of the electrons is almost perfectly spherical in the central region due to the strong Coulomb interaction of the nucleus,
resulting in the spherical effective potential for the valence electrons.
The electron density is deformed only in the surface and the tail regions of the atoms described by the valence electrons.
\par
\textbf{Deformations in the surface and degeneracy lifting}: Such deformations lead to a non-spherical effective potential in the surface and the tail regions.
There are several evidences for this.
Firstly, the outer-most $ s $ orbital can have non-zero $ \beta $ and become non-spherical.
This is shown in Table~\ref{tab:Q_valence} as a non-zero $ \beta $ of the $ 4s $ orbitals,
which tend to cancel the deformation of the valence $ 3d $ orbitals.
This is due to the screening effect.
That is, the outer-most $ 4s $ electrons try to cancel the deformed electron density of the open-shell valence $ 3d $ electrons.
We will discuss this $ s $-orbital deformation in more details later in this section taking $ \mathrm{Sc} $ atom as an example.
\par
Secondly, the orbitals with the same $ n $ and $ l $ become non-degenerate in energy in some cases.
This can be seen in Table~\ref{tab:Q_Sc_MO_main},
where the quadrupole moment of $ q_{\urm{max}}^{\urm{(sp)}} $,
the square radius $ \avr{r^2} $, and the eigenenergies of the single-particle orbitals $ \epsilon_j $,
for $ \mathrm{Sc} $ atoms are shown.
As one can see, the eigenenergies of the $ p_z $ states become different from those of $ p_x $ and $ p_y $ states.
Since the deformation is axially symmetric and thus there is only one specific direction,
i.e.~the orbitals along $ x $ and $ y $ axes
($ p_x $ and $ p_y $ orbitals or $ d_{yz} $ and $ d_{zx} $ orbitals) remain degenerate.
This is similar to the Stark effect~\cite{
  Schiff1968QuantumMechanics_McGrawHill}
if the difference of spherical and deformed effective potentials is treated as an external electric field.
We note, however, that the energy difference for the core orbitals is the order of $ 0.05 \, \mathrm{Hartree} \simeq 1 \, \mathrm{eV} $ or less, and is rather small.
This is because they have small overlap with the valence wave functions creating the non-spherical effective potential.
\par
On the other hand, for the open-shell outer-most valence electrons,
such liftings of degeneracy can be relevant.
For example, the outer-most electron of $ \mathrm{Al} $ atom occupies the $ 3p_z $ orbital.
If the system is completely spherical, the $ 3p_z $ orbital is degenerate with the $ 3p_x $ and $ 3p_y $ orbitals,
while due to the deformation the $ 3p_x $ and $ 3p_y $ orbitals are non-degenerate to the $ 3p_z $ orbital.
The eigenenergy of the $ 3p_z $ orbital is $ -0.2099 \, \mathrm{Hartree} $,
while those of the $ 3p_x $ and $ 3p_y $ orbitals are $ 0.0121 \, \mathrm{Hartree} $.
Due to this degeneracy lifting, the last electron prefers to occupy the $ 3p_z $ orbital instead of the equal filling of the $ 3p_x $, $ 3p_y $, and $ 3p_z $ orbitals, leading to the deformation.
It is noted that this degeneracy lifting of $ p $ states results from the second-order perturbation:
It is caused by the deformation of the core and the $ 3s $ orbital,
which is induced by the existence of the $ 3p_z $ electron.
\par
These results suggest that the dominant part of the effective potential is created by the spherical central region.
The non-spherical contribution to the effective potential of the valence deformed electrons is sub-dominant and can be treated as small correction.
\begin{table}[!htb]
  \centering
  \caption{
    Deformation parameters for the selected single-particle orbitals.}
  \label{tab:Q_valence}
  \begin{indented}
  \item[]
    \begin{tabular}{llD{.}{.}{4}}
      \br
      Atom & Orbital & \multicolumn{1}{c}{$ \beta_{\urm{max}}^{\urm{(sp)}} $} \\ \hline
      $ \mathrm{Al} $ & $ \alpha 3p_z $         &  0.6341 \\
      \mr
      $ \mathrm{Sc} $ & $ \alpha 3p_z $         &  0.6341 \\
      $ \mathrm{Sc} $ & $ \alpha 3p_x $         &  0.6341 \\
      $ \mathrm{Sc} $ & $ \alpha 3d_{yz} $      & -0.4530 \\
      $ \mathrm{Sc} $ & $ \alpha 4s $           & -0.0098 \\
      $ \mathrm{Sc} $ & $ \beta  4s $           &  0.0830 \\
      \mr
      $ \mathrm{Ti} $ & $ \alpha 3p_x $         &  0.6341 \\
      $ \mathrm{Ti} $ & $ \alpha 3p_z $         &  0.6341 \\
      $ \mathrm{Ti} $ & $ \alpha 3d_{zx} $      & -0.4530 \\
      $ \mathrm{Ti} $ & $ \alpha 4s $           & -0.0063 \\
      $ \mathrm{Ti} $ & $ \beta  4s $           & -0.0530 \\
      \mr
      $ \mathrm{Ni} $ & $ \alpha 3p_z $         &  0.6341 \\
      $ \mathrm{Ni} $ & $ \alpha 3d_{x^2-y^2} $ & -0.4530 \\
      $ \mathrm{Ni} $ & $ \alpha 3d_{xy} $      & -0.4530 \\
      $ \mathrm{Ni} $ & $ \alpha 3d_{yz} $      & -0.4530 \\
      $ \mathrm{Ni} $ & $ \alpha 3d_{z^2} $     &  0.4388 \\
      $ \mathrm{Ni} $ & $ \alpha 4s $           & -0.0163 \\
      \mr
      $ \mathrm{Ga} $ & $ \alpha 3p_x $         &  0.6341 \\
      $ \mathrm{Ga} $ & $ \alpha 3p_z $         &  0.6341 \\
      $ \mathrm{Ga} $ & $ \alpha 3d_{x^2-y^2} $ & -0.4530 \\
      $ \mathrm{Ga} $ & $ \alpha 3d_{yz} $      & -0.4530 \\
      $ \mathrm{Ga} $ & $ \alpha 3d_{z^2} $     &  0.4576 \\
      $ \mathrm{Ga} $ & $ \alpha 4s $           & -0.0008 \\
      $ \mathrm{Ga} $ & $ \alpha 4p_z $         &  0.6341 \\
      \br
    \end{tabular}
  \end{indented}
\end{table}
\par
\textbf{Screening effect}: Let us scrutinize deformations of each orbital with $ \mathrm{Sc} $ ($ Z = 21 $) atom in Table~\ref{tab:Q_Sc_MO_main} (results for other atoms are also shown in Table~\ref{tab:Q_Al_MO}--\ref{tab:Q_Ga_MO} in
\ref{sec:data}).
Here, $ \mathrm{Sc} $ atom has filled $ 1s $--$ 3p $ orbitals,
in addition to one $ 3d_{xy} $ electron and two $ 4s $ electrons.
The quadrupole moments $ q^{\urm{(sp)}}_z $ of the filled $ 1s $--$ 3p $ orbitals completely cancel each other, resulting in undeformed spherical core.
The square radii of the $ 1s $--$ 3p $ orbitals are much smaller than those of $ 3d_{xy} $ and $ 4s $ orbitals,
so that we can essentially regard $ \mathrm{Sc} $ atom as a system of one $ 3d_{xy} $ electron and two $ 4s $ electrons orbiting around an almost perfectly inert small spherical core.
The electron in the $ 3d_{xy} $ in the $ \alpha $-spin state is oblate $ q^{\urm{(sp)}}_z = -2.1642 < 0 $.
This results in an oblate shape of $ \mathrm{Sc} $ atom as a whole,
but the value $ q^{\urm{(sp)}}_z = -0.4189 $ is much smaller than the quadrupole moment of the $ 3d_{xy} $ electron.
This can be explained by the screening effect:
We can see in Table~\ref{tab:Q_Sc_MO_main} that the electron in the $ 4s $ orbital with $ \beta $-spin state deforms so significantly that the quadrupole moment of the $ \beta 4s $ electron cancels that of the $ \alpha 3d_{xy} $ orbital.
To be more precise, $ \left| q^{\urm{(sp)}}_z \right| $ of $ \beta 4s $ is
almost the same but slightly smaller than $ \left| q^{\urm{(sp)}}_z \right| $ of $ \alpha 3d_{xy} $,
which suggests that the quadrupole moment of the $ \alpha 3d_{xy} $ electron is almost but not perfectly cancelled, resulting in a small but finite deformation of $ \mathrm{Sc} $ atom.
On the other hand, the $ 4s $ electron in the other spin state (i.e.~$ \alpha 4s $) is deformed much less than $ \beta 4s $ electron, and does not contribute so much to the screening.
This is because the Coulomb repulsion, which is the key for the screening effect (see
section~\ref{sec:model}),
is relevant between different spin states, while the Pauli principle makes the effects of the Coulomb repulsion between the same spins weaker.
\par
Similar situations also occur in the other atoms presented in Table~\ref{tab:Q_valence}.
For $ \mathrm{Ti} $ atom (see also Table~\ref{tab:Q_Ti_MO} in
\ref{sec:data}),
the large deformation is due to the $ \alpha 3d_{yz} $ and the $ \alpha 3d_{zx} $ orbitals, and in total the deformation is prolate ($ \beta > 0 $).
The $ \beta 4s $ orbital is deformed to cancel the deformation of the $ \alpha 3d $ orbitals.
The $ \alpha 4s $ orbital is also deformed but because the effects of the Coulomb repulsion is weakened by the Pauli principle, the deformation is much smaller.
For $ \mathrm{Ni} $ atom  (Table~\ref{tab:Q_Ni_MO} in
\ref{sec:data}),
the large prolate deformation created by the $ \beta 3d_{z^2} $ orbital is partly, but not fully, cancelled by the $ \alpha 4s $ electron~\footnote{
  Since the eigenenergies of $ d_{x^2 - y^2} $ and $ d_{z^2} $ are identical if the system is completely spherical,
  the state with the last outer-most electron occupying $ d_{x^2 - y^2} $ is almost degenerate with the state with $ d_{z^2} $.
  The deformation parameters $ \beta $ of these two states may have opposite sign,
  since $ \beta^{\urm{(sp)}} $ of $ d_{x^2 - y^2} $ and $ d_{z^2} $ orbitals have the same absolute value but with the opposite sign
  (see
  \eqref{eq:beta_single_dxsqysq} and \eqref{eq:beta_single_dzsq}).
  Therefore, there exist two states close in energy to each other which have opposite sign of $ \beta $ with almost the same absolute value.}.
\par
In the case of $ \mathrm{Ga} $ atom (see also Table~\ref{tab:Q_Ga_MO} in
\ref{sec:data}),
the large prolate deformation is due to the $ \alpha 4p_z $ orbital $ q^{\urm{(sp)}}_z = 11.1659 > 0 $.
With the same argument, we might expect that the $ \beta 4s $ orbital should be largely deformed and cancel that of the $ \alpha 4p_z $ orbital.
However, the $ \beta 4s $ orbital has rather small quadrupole moment $ q^{\urm{(sp)}}_z = 0.0139 $.
It, therefore, does not cancel the deformation of the $ \alpha 4p_z $ orbital, resulting in $ Q_z = 11.1576 $ of $ \mathrm{Ga} $ atom as a whole.
This is because $ \avr{r^2} $ of the $ 4s $ orbital is much smaller than that of the $ \alpha 4p_z $ orbital,
and thus the wave function of the $ 4s $ orbital does not have enough overlap region with the $ \alpha 4p_z $ orbital to cancel its deformation. 
\par
We can also see the importance of the radii by comparing $ \mathrm{Sc} $, $ \mathrm{Ti} $, and $ \mathrm{Ni} $ data in Tables~\ref{tab:Q_Sc_MO_main}, \ref{tab:Q_Ti_MO}, and \ref{tab:Q_Ni_MO}, respectively.
For $ \mathrm{Sc} $ atom, the quadrupole moment of the valence $ d $ electron is almost perfectly cancelled.
On the other hand, it is marginally cancelled for $ \mathrm{Ti} $ atom,
while it is slightly cancelled for $ \mathrm{Ni} $ atom.
This behavior can be easily understood by noting that radius of the valence $ d $ orbital gets smaller as the atomic number increases:
$ \avr{r^2} = 3.8 $, $ 2.9 $, and $ 1.5 \, \mathrm{Bohr}^2 $ for $ \mathrm{Sc} $, $ \mathrm{Ti} $, and $ \mathrm{Ni} $ atoms, respectively.
The overlap between the valence $ d $ orbital and the $ 4s $ orbital gets smaller,
which explains why the screening effects gets weaker as the atomic number increases in these atoms.
\par
Moreover, the deformation of the $ d $-block atoms becomes further smaller due to the radius of the valence single-particle orbitals,
compared to that of the $ p $-block atoms.
The order of $ \sum_m \avr{r^2}_{n1} = 3 \avr{r^2}_{n1} $ for the valence $ p $ orbital is the same as that of $ Z \avr{r^2} $ for the whole atom
and thus $ \left| \beta \right| $ of the $ p $-block open-shell atoms are large,
whereas the order of $ \sum_m \avr{r^2}_{n2} = 5 \avr{r^2}_{n2} $ for the valence $ d $ orbital is smaller than that of $ Z \avr{r^2} $,
since the $ s $ orbital with the principal quantum number $ n + 1 $ is, in general, also occupied.
\par
\textbf{Systematic behavior of $ \beta $}: With all these physical arguments, we can now fully understand the results in
figure~\ref{fig:Gaussian_UHF_631pG_all}:
We have found there that the $ p $-block and $ d $-block atoms can be deformed,
but the deformation is much smaller for the $ d $-block atoms compared with the $ p $-block atoms showing single-particle-like deformation.
This result can be explained by noting that the screening effects of the outer-most $ s $ orbital are rather significant for the $ d $-block atoms,
as discussed above for $ \mathrm{Sc} $, $ \mathrm{Ti} $, and $ \mathrm{Ni} $ atoms.
On the other hand, the outer-most valence $ p $ orbital of the $ p $-block atom has much smaller overlap with the outer-most $ s $ orbital,
so that the deformation caused by the valence $ p $ orbital remains unscreened.
\par
This difference can also be understood by using deformation of the effective potential.
As discussed in the beginning of this subsection,
the effective potential remains almost spherical, while in the tail region it becomes anisotropic
mainly due to the outer-most $ p $ or $ d $ orbitals.
Consequently, such deformation can affect the orbitals which have larger spatial distribution.
On the one hand, the outer-most $ s $ orbital of the $ d $-block atoms,
which has comparable $ \avr{r^2} $ to the valence orbital,
is thus significantly affected by the anisotropic effective potential created by the valence $ d $ electrons.
Accordingly, the $ s $ orbital can be deformed largely.
On the other hand, the outer-most $ s $ orbital of the $ p $-block atoms has smaller $ \avr{r^2} $ than the outer-most $ p $ orbital.
Thus, the $ s $ orbital cannot be affected by such deformation of the effective potential, and, as a result, deformed largely.
\par
Since deformation of the $ p $-block atoms originates from single-particle-like deformation,
atoms whose configurations of valence electrons are $ p^1 $ and $ p^4 $ always show prolate ($ \beta > 0 $) deformation and the last electron occupies $ p_z $ orbital.
In contrast,
atoms whose configurations of valence electrons are $ p^2 $ and $ p^5 $ always show oblate ($ \beta < 0 $) deformation and the last two electrons occupy $ p_x $ and $ p_y $ orbital.
Since deformation of the $ d $-block atoms induces the screening as discussed above,
the deformation is small and there is no obvious tendency.
\begin{table}[!htb]
  \centering
  \caption{
    Mean-square radius $ \avr{r^2}_{nl} $, 
    single-particle quadrupole moment $ q_z^{\urm{(sp)}} $,
    and deformation parameter $ \beta_z^{\urm{(sp)}} $
    for single-particle orbitals of $ \mathrm{Sc} $ atom
    calculated by the unrestricted Hartree-Fock method with 6-31+G basis.
    The single-particle energy of each orbital $ \epsilon_j $ and corresponding orbital are also shown.
    The Hartree atomic unit is used for $ \epsilon_j $, $ \avr{r^2}_{nl} $, and $ q_z^{\urm{(sp)}} $.}
  \label{tab:Q_Sc_MO_main}
  \begin{indented}
  \item[]
    \begin{tabular}{rlD{.}{.}{4}D{.}{.}{4}D{.}{.}{4}D{.}{.}{4}}
      \br
      Spin & Orbital & \multicolumn{1}{c}{$ \epsilon_j $} & \multicolumn{1}{c}{$ \avr{r^2}_{nl} $} & \multicolumn{1}{c}{$ q^{\urm{(sp)}}_z $} & \multicolumn{1}{c}{$ \beta_z^{\urm{(sp)}} $} \\
      \mr
      $ \alpha $ & $ 1s $      & -165.8952 &  0.0073 &  0.0000 &  0.0000 \\
      $ \alpha $ & $ 2s $      &  -19.0956 &  0.1393 &  0.0000 &  0.0000 \\
      $ \alpha $ & $ 2p_z $    &  -15.6861 &  0.1162 &  0.0930 &  0.6341 \\
      $ \alpha $ & $ 2p_x $    &  -15.6830 &  0.1164 & -0.0466 & -0.3171 \\
      $ \alpha $ & $ 2p_y $    &  -15.6830 &  0.1164 & -0.0466 & -0.3171 \\
      $ \alpha $ & $ 3s $      &   -2.5988 &  1.3396 & -0.0031 & -0.0018 \\
      $ \alpha $ & $ 3p_x $    &   -1.6232 &  1.6265 & -0.6506 & -0.3171 \\
      $ \alpha $ & $ 3p_y $    &   -1.6232 &  1.6265 & -0.6506 & -0.3171 \\
      $ \alpha $ & $ 3p_z $    &   -1.5893 &  1.6107 &  1.2886 &  0.6341 \\
      $ \alpha $ & $ 3d_{xy} $ &   -0.3375 &  3.7873 & -2.1642 & -0.4530 \\
      $ \alpha $ & $ 4s $      &   -0.2171 & 17.4198 & -0.2152 & -0.0098 \\
      \mr
      $ \beta $  & $ 1s $      & -165.8951 &  0.0073 &  0.0000 &  0.0000 \\
      $ \beta $  & $ 2s $      &  -19.0832 &  0.1391 &  0.0001 &  0.0007 \\
      $ \beta $  & $ 2p_z $    &  -15.6837 &  0.1162 &  0.0929 &  0.6341 \\
      $ \beta $  & $ 2p_x $    &  -15.6657 &  0.1161 & -0.0464 & -0.3171 \\
      $ \beta $  & $ 2p_y $    &  -15.6657 &  0.1161 & -0.0464 & -0.3171 \\
      $ \beta $  & $ 3s $      &   -2.5451 &  1.3370 &  0.0387 &  0.0230 \\
      $ \beta $  & $ 3p_z $    &   -1.5798 &  1.6156 &  1.2925 &  0.6341 \\
      $ \beta $  & $ 3p_x $    &   -1.5273 &  1.6196 & -0.6478 & -0.3171 \\
      $ \beta $  & $ 3p_y $    &   -1.5273 &  1.6196 & -0.6478 & -0.3171 \\
      $ \beta $  & $ 4s $      &   -0.2051 & 18.5388 &  1.9406 &  0.0830 \\
      \mr
      \multicolumn{3}{l}{Total}            & 53.1312 & -0.4189 \\
      \br
    \end{tabular}    
  \end{indented}
\end{table}
\par
\textbf{Single-particle versus many-body effects}: As shown above, the deformations of the electron density in atoms at most originate from the single-particle orbitals of a few valence electrons, and thus there is no collective deformation.
Rather, many-body effects in atoms disfavor deformations.
With such tiny deformations, the effective potential is slightly deformed only in the surface and the tail (i.e.~the valence electron) regions,
but the dominant part remains spherical.
This justifies the use of spherically symmetric effective potentials or density functionals in the conventional atomic structure calculations~\cite{
  Parr1989Density-FunctionalTheoryofAtomsandMolecules_OxfordUniversityPress,
  Martin2004_CambridgeUniversityPress,
  Friedrich2006TheoreticalAtomicPhysics_SpringerVerlag}.
\par
This is in stark contrast with nuclear systems.
It is well known that nuclei can be deformed significantly via collective many-body effects,
and they often have $ \left| \beta \right| \gtrsim 0.3 $~\cite{
  Ring1980TheNuclearManyBodyProblem_SpringerVerlag,
  Serot1986AdvancesinNuclearPhysics16_}.
In
\ref{sec:n_drop}, we exemplify this with model calculations for nuclear many-body systems.
In the following section, we attempt to explain with a simple qualitative model why there is no collective deformation in the atoms whereas it exists in atomic nuclei.

% 
% Model Discussion
% -*- coding: utf-8 -*-
%%%%%%%%%%%%%%%%%%%%%%%%%%%%%%%%%%%%%%%%%%%%%%%%%% 
% 
% Deformation of atoms paper manuscript for J. Phys. B
% Model Part
% 
% Begin to write: 2020-07-26
% 
% Tomoya Naito, Shimpei Endo,
% Kouichi Hagino, and Yusuke Tanimura
% 
%%%%%%%%%%%%%%%%%%%%%%%%%%%%%%%%%%%%%%%%%%%%%%%%%% 
% 
\section{Qualitative discussion}
\label{sec:model}
\par
In the previous section and in
\ref{sec:n_drop},
we have shown that electrons in atoms are much less likely to deform than nucleons in atomic nuclei.
We argue in this section that this difference physically originates from the nature of the inter-particle interactions:
the repulsive Coulomb interaction between the electrons and the attractive nuclear interaction between the nucleons.
It should be noticed that the isoscalar proton-neutron interaction, which is strong attractive, has been argued to play an important role in nuclear deformation~\cite{
  Federman1977Phys.Lett.B69_385,
  Federman1979Phys.Lett.B82_9,
  Federman1979Phys.Rev.C20_820,
  Dobaczewski1988Phys.Rev.Lett.60_2254}.
While this is true in general,
we show in
\ref{sec:n_drop} that neutron drops can also be deformed
even though there is only one type of particles, 
and thus, there is no neutron-proton interaction.
The deformations in the neutron drops can be qualitatively understood by our analytical argument given in this section.
\par
To illustrate this point, we closely follow
\cite{
  Bohigas1979Phys.Rep.51_267,
  Bertsch1997Phys.Rev.C56_839}.
We consider a system with $ N $ particles where the total-spin component, denoted as $ S_z $, is conserved; 
$ \left[ \hat{H}, S_z \right] = 0 $.
By taking the $ N $-body wave function as a simultaneous eigenstate of $ S_z $ and $ \hat{H} $,
the number of particles in each spin state is conserved,
and we can regard the system as composed of
$ N_{\uparrow} $ spin-up and
$ N_{\downarrow} $  spin-down particles
($ N = N_{\uparrow} + N_{\downarrow} $).
We consider deforming wave functions
$ \Psi_{\urm{sph}} \left( \ve{r}_1, \ldots, \ve{r}_N \right) $
associated with a spherical density
with the following coordinate transformation:
\begin{align}
  \label{eq:wf_def}
  & \Psi_{\urm{def}}
    \left(
    x_1, y_1, z_1,
    \ldots,
    x_N, y_N, z_N
    \right)
    \notag \\
  = & \, 
      e^{N_{\uparrow} \left( 2 \beta_{\uparrow} - \alpha_{\uparrow} \right)/2} 
      e^{N_{\downarrow} \left( 2 \beta_{\downarrow} - \alpha_{\downarrow} \right)/2} 
      \Psi_{\urm{sph}}
      \left(
      X_1, Y_1, Z_1,
      \ldots,
      X_N, Y_N, Z_N
      \right),
\end{align}
where
$ X_j = e^{\beta_{\uparrow}} x_j $,
$ Y_j = e^{\beta_{\uparrow}} y_j $, and
$ Z_j = e^{- \alpha_{\uparrow}} z_j $
for $ j = 1 $, $ 2 $, \ldots, $ N_{\uparrow} $
and 
$ X_j = e^{\beta_{\downarrow}} x_j $,
$ Y_j = e^{\beta_{\downarrow}} y_j $, and
$ Z_j = e^{- \alpha_{\downarrow}} z_j $
for $ j = N_{\uparrow} + 1 $, $ N_{\uparrow} + 2 $, \ldots, $ N $.
The coefficient $ e^{N_a \left( 2 \beta_a - \alpha_a \right)/2} $
($ a = {\uparrow} $, $ {\downarrow} $)
is due to the normalization,
which appears since, in the $ \alpha_a \ne 2 \beta_a $ case, the volume of the system is not conserved.
Correspondingly, the single-particle densities of up- and down-spin states are also deformed as follows:
\begin{subequations}
  \begin{align}
    \rho_{\urm{def}, \, {\uparrow}}
    \left( x, y, z \right)
    & = \, 
      e^{2 \beta_{\uparrow} - \alpha_{\uparrow}}
      \rho_{\urm{sph}, \, {\uparrow}}
      \left(
      e^{\beta_{\uparrow}} x, e^{\beta_{\uparrow}} y, e^{- \alpha_{\uparrow}} z
      \right), \\
    \rho_{\urm{def}, \, {\downarrow}}
    \left( x, y, z \right)
    & = 
      e^{2 \beta_{\downarrow} - \alpha_{\downarrow}}
      \rho_{\urm{sph}, \, {\downarrow}}
      \left(
      e^{\beta_{\downarrow}} x, e^{\beta_{\downarrow}} y, e^{- \alpha_{\downarrow}} z
      \right).
  \end{align}
\end{subequations}
The coefficient $ e^{2 \beta_a - \alpha_a} $ also originates from the normalization coefficient in
\eqref{eq:wf_def},
or equivalently, the particle-number conservation;
\begin{equation}
  \int
  \rho_{\urm{def}, \, a} \left( \ve{r} \right)
  \, d \ve{r}
  =
  \int
  \rho_{\urm{sph}, \, a} \left( \ve{r} \right)
  \, d \ve{r}
  =
  N_a.
\end{equation}
\par
Due to the saturation property,
the atomic nuclei are expected to be deformed with the constant volume.
For such deformation, $ \alpha_a = 2 \beta_a $ holds.
Furthermore, as seen in
section~\ref{sec:numerical},
atoms are also deformed with $ \alpha = 2 \beta $.
Hence, in the following discussion, we will show the calculation with $ \alpha_a = 2 \beta_a $,
and results with $ \alpha_a \ne 2 \beta_a $ will be shown in
\ref{sec:gen},
where we will reach essentially the same conclusion as those shown in this section.
Note that $ \alpha_a = 2 \beta_a $ corresponds to the quadrupole deformation:
It makes the wave function shrunken in the $ x $- and $ y $-axes,
and elongated in the $ z $-axis directions when $ \beta_a $ is positive.
The quadrupole deformation of a wave function
$ \ket{\Psi_{\urm{sph}}} $
associated with a spherical density for 
the Hamiltonian $ \hat{H} = \hat{T} + V_{\urm{ext}} + V_{\urm{int}} $,
consisting of the kinetic energy operator $ \hat{T} $,
central external potential $ V_{\urm{ext}} $,
and the interaction $ V_{\urm{int}} $,
can be equivalently performed with the following canonical transformation:
\begin{equation}
  \ket{\Psi_{\urm{def}}}
  =
  e^{\beta_{\uparrow}   \left[ \hat{H}, \hat{Q}_{\uparrow}   \right]}
  e^{\beta_{\downarrow} \left[ \hat{H}, \hat{Q}_{\downarrow} \right]}
  \ket{\Psi_{\urm{sph}}}.
\end{equation}
Here, $ \hat{Q}_a $ is the quadrupole operator for each spin state defined as
\begin{align}
  \hat{Q}_{\uparrow}
  & =
    m
    \sum_{j = 1}^{N_{\uparrow}}
    \left[
    z_j^2
    -
    \frac{1}{2}
    \left(
    x_j^2  + y_j^2
    \right)
    \right], \\
  \hat{Q}_{\downarrow}
  & =
    m
    \sum_{j = N_{\uparrow} + 1}^N
    \left[
    z_j^2
    -
    \frac{1}{2}
    \left(
    x_j^2  + y_j^2
    \right)
    \right],
\end{align}
where $ m $ and $ x_j $, $ y_j $, $ z_j $ are the mass and coordinates of the particles.
\par
When the absolute value of the deformation parameter $ \left| \beta_a \right| $ is small,
the one-body density $ \rho_{\urm{def}, \, a} $ of the system is written as follows:
\begin{align}
  \rho_{\urm{def}, \, a}
  \left( x, y, z \right)
  = & \, 
      \rho_{\urm{sph}, \, a}
      \left( e^{\beta_a} x, e^{\beta_a} y, e^{- 2 \beta_a} z \right)
      \notag \\
  = & \,
      \rho_{\urm{sph}, \, a} \left( r \right)
      +
      f^{\left( 1 \right)}_a \left( \ve{r} \right)
      \frac{d \rho_{\urm{sph}, \, a} \left( r \right)}{dr}
      +
      f^{\left( 2 \right)}_a \left( \ve{r} \right)
      \frac{d^2 \rho_{\urm{sph}, \, a} \left( r \right)}{dr^2}
      +
      O \left( \beta^3 \right), 
      \label{eq:1density_partialwave} \\
  f^{\left( 1 \right)}_a \left( \ve{r} \right)
  = & \,
      \beta_a
      \frac{x^2 + y^2 - 2 z^2}{r}
      +
      \beta_a^2
      \frac{x^2 + y^2 + 4 z^2}{r}
      \notag \\
    & \, 
      -
      \frac{\beta_a^2}{2}
      \frac{x^4 + y^4 + 4 z^4}{r^3}
      +
      \beta_a^2
      \frac{2 \left( x^2 + y^2 \right) z^2 - x^2 y^2}{r^3}
      \notag \\
  = & \,
      -
      \sqrt{\frac{16 \pi}{5}}
      \beta_a
      r Y_{20} \left( \theta, \phi \right)
      +
      \beta_a^2
      \left[
      2 r + \sqrt{\frac{16 \pi}{5}} r Y_{20} \left( \theta, \phi \right)
      \right]
      -
      \frac{8 \pi}{5}
      \beta_a^2
      r \left[ Y_{20} \left( \theta, \phi \right) \right]^2
      \label{eq:f1} \\
  f^{\left( 2 \right)}_a \left( \ve{r} \right)
  = & \,
      \frac{\beta_a^2}{2}
      \frac{x^4 + y^4 + 4 z^4}{r^2}
      -
      \beta_a^2
      \frac{2 \left( x^2 + y^2 \right) z^2 - x^2 y^2}{r^2}
      \notag \\
  = & \,
      \frac{8 \pi}{5}
      \beta_a^2
      r^2 \left[ Y_{20} \left( \theta, \phi \right) \right]^2.
      \label{eq:f2} 
\end{align}
Let us consider how the energy of the system changes when we induce an infinitesimal quadrupole deformation
$ \left| \beta_a \right| \ll 1 $
with this transformation to a spherically symmetric system
$ \rho_{\urm{sph}} \left( \ve{r} \right) = \rho_{\urm{sph}} \left( r \right) $.
\par
\textbf{Kinetic energy}: The kinetic energy reads
\begin{align}
  T_{\urm{def}, \, a}
  & =
    \brakket{\Psi_{\urm{def}, \, a}}{\hat{T}}{\Psi_{\urm{def}, \, a}}
    \notag \\
  & = 
    \left(
    \frac{2}{3}
    e^{2 \beta_a}
    +
    \frac{1}{3}
    e^{- 4 \beta_a}
    \right)
    \brakket{\Psi_{\urm{sph}, \, a}}{\hat{T}}{\Psi_{\urm{sph}, \, a}}
    \notag \\
  & =
    \left(
    \frac{2}{3}
    e^{2 \beta_a}
    +
    \frac{1}{3}
    e^{- 4 \beta_a}
    \right)
    T_{\urm{sph}, \, a}
    \label{eq:kin}
\end{align}
One can therefore see that the kinetic energy increases with the deformation 
\begin{equation}
  \label{eq:kin_ene_inc}
  \Delta T_a
  \defeq 
  T_{\urm{def}, \, a}
  -
  T_{\urm{sph}, \, a}
  \simeq
  4 
  \beta_a^2
  T_{\urm{sph}, \, a}
  >
  0.
\end{equation}
We also note that the deformation of a spin component does not affect the kinetic energy of the other spin component.
\par
\textbf{Central external potential energy}: The central external potential energy of a system with density $ \rho $ is, in general, written as
\begin{equation}
  E_{\urm{ext}}
  =
  \int
  \rho \left( \ve{r} \right)
  V_{\urm{ext}} \left( r \right)
  \, d \ve{r}.
\end{equation}
Therefore, by using
\eqref{eq:1density_partialwave},
the central external potential changes as
\begin{align}
  & \Delta E_{\urm{ext}, \, a}
    \notag \\
  \defeq & \,
           E_{\urm{ext}, \, a}^{\urm{def}}
           -
           E_{\urm{ext}, \, a}^{\urm{sph}}
           \notag \\
  = & \,
      \int
      \left[
      \rho_{\urm{def}, \, a} \left( \ve{r} \right)
      -
      \rho_{\urm{sph}, \, a} \left( r \right)
      \right]
      V_{\urm{ext}} \left( r \right)
      \, d \ve{r}
      \notag \\
  \simeq & \,
           \int
           f^{\left( 1 \right)}_a \left( \ve{r} \right)
           \frac{d \rho_{\urm{sph}, \, a} \left( r \right)}{dr}
           V_{\urm{ext}} \left( r \right)
           \, d \ve{r}
           +
           \int
           f^{\left( 2 \right)}_a \left( \ve{r} \right)
           \frac{d^2 \rho_{\urm{sph}, \, a} \left( r \right)}{dr^2}
           V_{\urm{ext}} \left( r \right)
           \, d \ve{r}
           \notag \\
  = & \,
      \frac{32 \pi}{5}
      \beta_a^2 
      \int_0^{\infty}
      \frac{d \rho_{\urm{sph}, \, a} \left( r \right)}{dr}
      V_{\urm{ext}} \left( r \right)
      r^3 \, dr
      +
      \frac{8 \pi}{5} 
      \beta_a^2
      \int_0^{\infty}
      \frac{d^2 \rho_{\urm{sph}, \, a} \left( r \right)}{dr^2}
      V_{\urm{ext}} \left( r \right)
      r^4 \, dr
      \notag \\
  = & \,
      -
      \frac{32 \pi}{5} 
      \beta_a^2
      \int_0^{\infty}
      \rho_{\urm{sph}, \, a} \left( r \right)
      \frac{d}{dr}
      \left[
      V_{\urm{ext}} \left( r \right)
      r^3
      \right]
      \, dr
      +
      \frac{8 \pi}{5} 
      \beta_a^2 
      \int_0^{\infty}
      \rho_{\urm{sph}, \, a} \left( r \right)
      \frac{d^2}{dr^2}
      \left[
      V_{\urm{ext}} \left( r \right)
      r^4
      \right]
      \, dr,
      \label{eq:ext_full}
\end{align}
where we have assumed
$ V_{\urm{ext}} \left( r \right) \sim r^d $ ($ d > -3 $) at $ r \to 0 $
in performing the partial integral in the final line.
\par
Let us consider in particular the power-law type of external potential
$ V_{\urm{ext}} \left( r \right) = c r^d $
with given constants $ c $ and $ d $.
Substituting this form into
\eqref{eq:ext_full},
one obtains
\begin{equation}
  \label{eq:ext_subs}
  \Delta E_{\urm{ext}, \, a}
  \simeq 
  \left[
    -
    \frac{8}{5} \left( d + 3 \right)
    +
    \frac{2}{5} \left( d + 4 \right) \left( d + 3 \right)
  \right]
  \beta_a^2 
  E_{\urm{ext}, \, a}^{\urm{sph}}
\end{equation}
In the case of the Coulomb potential ($ d = -1 $),
\eqref{eq:ext_subs} reads
\begin{equation}
  \label{eq:ext_Coul}
  \Delta E_{\urm{ext}, \, a}
  \simeq 
  -
  \frac{4}{5}
  \beta_a^2 
  E_{\urm{ext}, \, a}^{\urm{sph}}
  > 0 .
\end{equation}
In the case of the harmonic oscillator potential ($ d = 2 $),
\eqref{eq:ext_subs} reads
\begin{equation}
  \label{eq:ext_HO}
  \Delta E_{\urm{ext}, \, a}
  \simeq
  4
  \beta_a^2 
  E_{\urm{ext}, \, a}^{\urm{sph}}
  > 0 .
\end{equation}
One can therefore see that
the external potential energy increases with the deformation for both the Coulomb and harmonic oscillator potentials.
As will be discussed later, they represent the Coulomb potential created by the atomic nucleus in atomic systems,
and an external harmonic trap for a neutron drop discussed in
\ref{sec:n_drop}, respectively.
We also note that the deformation of a spin component does not affect the potential energy of the other spin component.
\par
\textbf{Interaction energy}: The interaction part of the energy, on the other hand, can either decrease or increase, and hence, favor or disfavor deformation depending on the nature of the interaction.
Also, as we shall show, the spins play an important role.
To see this point, let us consider the direct (Hartree) part of the two-body interaction energy
between the $ a $ and $ b $ spin states characterized by one-body density
$ \rho_{\urm{def}, \, a} \left( \ve{r} \right) $:
\begin{equation}
  \label{eq:int_def}
  E_{\urm{int}, \, ab}^{\urm{def}}
  =
  \frac{1}{2}
  \iint
  V_{\urm{int}}^{ab} \left( \left| \ve{r} - \ve{r}' \right| \right)
  \rho_{\urm{def}, \, a} \left( \ve{r} \right)
  \rho_{\urm{def}, \, b} \left( \ve{r}' \right)
  \, d \ve{r}
  \, d \ve{r}',
\end{equation}
where $ V_{\urm{int}}^{ab} $ is the interaction between a particle with spin $ a $ and that with spin $ b $.
The total interaction energy,
as long as we only consider the direct part,
can then be written as
$ E_{\urm{int}, \, \urm{tot}}^{\urm{def}}
=
E_{\urm{int}, \, {\uparrow} {\uparrow}}^{\urm{def}}
+
2 E_{\urm{int}, \, {\uparrow} {\downarrow}}^{\urm{def}}
+
E_{\urm{int}, \, {\downarrow} {\downarrow}}^{\urm{def}} $.
Notice that the multipole expansion of the interaction reads
\begin{equation}
  \label{eq:int_partial}
  V_{\urm{int}}^{ab} \left( \left| \ve{r} - \ve{r}' \right| \right)
  =
  \sum_{l, \, m}
  V_l^{ab} \left( r, r' \right)
  Y_{lm} \left( \theta, \phi \right)
  Y_{lm}^* \left( \theta', \phi' \right).
\end{equation}
Substituting
\eqref{eq:1density_partialwave} and \eqref{eq:int_partial} into
\eqref{eq:int_def},
one obtains the change of the interaction energy induced by the deformation
$ \Delta E_{\urm{int}, \, ab} \defeq E_{\urm{int}, \, ab}^{\urm{def}} - E_{\urm{int}, \, ab}^{\urm{sph}} $ as
\begin{equation}
  \label{eq:ab_int}
  \Delta E_{\urm{int}, \, ab} 
  =
  \beta_a^2
  \,
  \Delta \epsilon_{\urm{int}, \, ab}^{\left( 1 \right)}
  +
  \beta_b^2
  \,
  \Delta \epsilon_{\urm{int}, \, ab}^{\left( 2 \right)}
  +
  \beta_a \beta_b
  \,
  \Delta \epsilon_{\urm{int}, \, ab}^{\left( 3 \right)}
  +
  O \left( \beta^3 \right),
\end{equation}
with
\begin{subequations}
  \label{eq:uab_int}
  \begin{align}
    \Delta \epsilon_{\urm{int}, \, ab}^{\left( 1 \right)}
    = & \,
        \frac{16 \pi}{5} 
        \int_0^{\infty} 
        \int_0^{\infty} 
        \frac{d \rho_{\urm{sph}, \, a} \left( r \right)}{dr}
        \rho_{\urm{sph}, \, b} \left( r' \right)
        V_{0}^{ab} \left( r, r' \right)
        r^3 r'^2 \, dr \, dr'
        \notag \\
      & \, 
        +         
        \frac{4 \pi}{5}  
        \int_0^{\infty} 
        \int_0^{\infty} 
        \frac{d^2 \rho_{\urm{sph}, \, a} \left( r \right)}{dr^2}
        \rho_{\urm{sph}, \, b} \left( r' \right)
        V_0^{ab} \left( r, r' \right)
        r^4 r'^2 \, dr \, dr', \\
    \Delta \epsilon_{\urm{int}, \, ab}^{\left( 2 \right)}
    = & \,
        \frac{16 \pi}{5} 
        \int_0^{\infty} 
        \int_0^{\infty} 
        \rho_{\urm{sph}, \, a} \left( r \right)
        \frac{d \rho_{\urm{sph}, \, b} \left( r' \right)}{dr'}
        V_0^{ab} \left( r, r' \right)
        r^2 r'^3 \, dr \, dr'
        \notag \\
      & \, 
        +         
        \frac{4 \pi}{5}  
        \int_0^{\infty} 
        \int_0^{\infty} 
        \rho_{\urm{sph}, \, a} \left( r \right)
        \frac{d^2 \rho_{\urm{sph}, \, b} \left( r' \right)}{dr'^2}
        V_0^{ab} \left( r, r' \right)
        r^2 r'^4 \, dr \, dr'
        =
        \Delta \epsilon_{\urm{int}, \, ba}^{\left( 1 \right)}, \\
    \Delta \epsilon_{\urm{int}, \, ab}^{\left( 3 \right)}
    = & \,
        \frac{8 \pi}{5}
        \int_0^{\infty} 
        \int_0^{\infty} 
        \frac{d \rho_{\urm{sph}, \, a} \left( r \right)}{dr}
        \frac{d \rho_{\urm{sph}, \, b} \left( r' \right)}{dr'}
        V_2^{ab} \left( r, r' \right)
        r^3 r'^3 \, dr \, dr'.
  \end{align}
\end{subequations}
To be more explicit, for the parallel spin states, one finds
\begin{equation}
  \label{eq:ab_int_aa}
  \Delta E_{\urm{int}, \, aa} 
  =
  \beta_a^2
  \left( 
    \Delta \epsilon_{\urm{int}, \, aa}^{\left( 1 \right)}
    +
    \Delta \epsilon_{\urm{int}, \, aa}^{\left( 2 \right)}
    +
    \Delta \epsilon_{\urm{int}, \, aa}^{\left( 3 \right)}
  \right)
  +
  O \left( \beta^3 \right)
\end{equation}
and for the anti-parallel spin states ($ a \ne b $) 
\begin{equation}
  \label{eq:ab_int_ab}
  \Delta E_{\urm{int}, \, ab}
  =
  \beta_{\uparrow}^2
  \,
  \Delta \epsilon_{\urm{int}, \, ab}^{\left( 1 \right)}
  +
  \beta_{\downarrow}^2
  \,
  \Delta \epsilon_{\urm{int}, \, ab}^{\left( 2 \right)}
  +
  \beta_{\uparrow} \beta_{\downarrow}
  \,
  \Delta \epsilon_{\urm{int}, \, ab}^{\left( 3 \right)}
  +
  O \left( \beta^3 \right).
\end{equation}
\par
First, let us consider the case of atomic nuclei.
We note that the interaction between nucleons in vacuum are not yet rigorously known,
but it basically comprises of relatively long-range attractive interactions
originating from exchanges of several mesons and a short-range repulsive interaction~\cite{
  Yukawa1935Proc.Phys.Math.Soc.Jpn.Third17_48,
  Stoks1994Phys.Rev.C49_2950,
  Wiringa1995Phys.Rev.C51_38,
  Machleidt2001Phys.Rev.C63_024001,
  Ishii2007Phys.Rev.Lett.99_022001}.
These combined together have a net attractive effect between nucleons,
which results in a bound state of 
a proton and a neutron (i.e.~deuteron)~\cite{
  Krane1988IntroductoryNuclearPhysics_JohnWiley&Sons,
  Povh2008ParticlesandNuclei_SpringerVerlag},
and in a large negative scattering length between nucleons~\cite{
  Wiringa1995Phys.Rev.C51_38,
  Machleidt2001Phys.Rev.C63_024001}.
We also note that an effective in-medium nucleon-nucleon interaction used in nuclear many-body calculations in general has a net attractive effect~\cite{
  Bender2003Rev.Mod.Phys.75_121}.
For our qualitative discussion in this section, a simple delta-function interaction
$ V_{\urm{int}}^{ab} \left( \ve{r} \right) = - g_{ab} \delta \left( \ve{r} \right) $
suffices to model this net nuclear interaction between the nucleons.
The parameter $ g_{ab} $ is positive when the interaction is attractive, such as for nuclear systems,
whereas it is negative when the interaction is repulsive.
The multipole expansion of this interaction reads
\begin{equation}
  \label{eq:delta}
  V_0^{ab} \left( r, r' \right)
  =
  V_2^{ab} \left( r, r' \right)
  =
  - g_{ab} \frac{\delta \left( r - r' \right)}{rr'}.
\end{equation}
Substituting
\eqref{eq:delta} into
\eqref{eq:ab_int} and \eqref{eq:uab_int},
one obtains
\begin{align}
  & \Delta E_{\urm{int}, \, ab}
    \notag \\
  \simeq & \,
           -
           \frac{16 \pi g_{ab}}{5}
           \beta_a^2
           \int_0^{\infty}
           \frac{d \rho_{\urm{sph}, \, a} \left( r \right)}{dr}
           \rho_{\urm{sph}, \, b} \left( r \right)
           r^3 \, dr
           -
           \frac{4 \pi g_{ab}}{5}
           \beta_a^2
           \int_0^{\infty}
           \frac{d^2 \rho_{\urm{sph}, \, a} \left( r \right)}{dr^2}
           \rho_{\urm{sph}, \, b} \left( r \right)
           r^4 \, dr
           \notag \\
  & \,
    - 
    \frac{16 \pi g_{ab}}{5}
    \beta_b^2
    \int_0^{\infty}
    \rho_{\urm{sph}, \, a} \left( r \right)
    \frac{d \rho_{\urm{sph}, \, b} \left( r \right)}{dr}
    r^3 \, dr
    -
    \frac{4 \pi g_{ab}}{5}
    \beta_b^2
    \int_0^{\infty}
    \rho_{\urm{sph}, \, a} \left( r \right)
    \frac{d^2 \rho_{\urm{sph}, \, b} \left( r \right)}{dr^2}
    r^4 \, dr
    \notag \\
  & \,
    -
    \frac{8 \pi g_{ab}}{5}
    \beta_a \beta_b
    \int_0^{\infty}
    \frac{d \rho_{\urm{sph}, \, a} \left( r \right)}{dr}
    \frac{d \rho_{\urm{sph}, \, b} \left( r \right)}{dr}
    r^4 \, dr
    \notag \\
  = & \, 
      \frac{4 \pi g_{ab}}{5}
      \left( \beta_a - \beta_b \right)^2
      \int_0^{\infty}
      \frac{d \rho_{\urm{sph}, \, a} \left( r \right)}{dr}
      \frac{d \rho_{\urm{sph}, \, b} \left( r \right)}{dr}
      r^4 \, dr.
      \label{eq:int_delta}
\end{align}
Thus, on the one hand, the interaction energy for the same spin component does not change with the deformation:
$ \Delta E_{\urm{int}, \, {\uparrow} {\uparrow}}
=
\Delta E_{\urm{int}, \, {\downarrow} {\downarrow}}
= 0 $.
This can be also shown using
\eqref{eq:int_def}
without using the multipole expansion.
This is due to a scaling symmetry of the delta function interaction.
On the other hand, the interaction energy of the anti-parallel spin states
$ E_{\urm{int}, \, {\uparrow} {\downarrow}}^{\urm{def}} $ 
can be non-zero.
Indeed, in the nuclear system, the one-body density is more or less constant in the interior region due to its saturation property,
$ d \rho_{\urm{sph}} /dr \simeq 0 $,
and it suddenly decreases at the surface region $ d \rho_{\urm{sph}} / dr < 0 $.
Thus, it is reasonable to assume $ d \rho_{\urm{sph}} / dr \le 0 $ in the most region.
Therefore, for the anti-parallel spin states $ a \ne b $,
the integral of
\eqref{eq:int_delta} is positive.
For the attractive interaction $ g_{{\uparrow} {\downarrow}} > 0 $,
it is energetically favorable to have $ \beta_a = \beta_b $,
so that there is no energy increase with the deformation,
i.e.~$ \Delta E_{\urm{int}, \, {\uparrow} {\downarrow}} \simeq 0 $.
Hence, the spin-up and -down particles deform in the same manner, i.e.~$ \rho_{\urm{def}, \, {\uparrow}} \left( \ve{r} \right)
=
\rho_{\urm{def}, \, {\downarrow}} \left( \bm{r} \right) $.
We note that we have assumed in the above argument the simplest delta-function interaction.
In reality, the modern nuclear effective interaction, such as the Skyrme interaction, contains the terms simulating finite-range effects
(e.g., $ t_1 $ and $ t_2 $ terms in
\eqref{eq:skyrme}).
This finite-range effects can make the energy of the system decrease with the nuclear deformation,
in contrast to what have been found above with a delta interaction.
The above argument can thus only explain why the condition $ \beta_a = \beta_b $ is favored:
Once the nuclear deformation occurs with the finite-range effects or other reasons,
the nuclei deform in a spin-independent manner due to the attractive nature of interaction.
\par
In contrast, if the interaction was repulsive $ g_{{\uparrow} {\downarrow}} < 0$,
the energy can decrease with the deformation
$ \Delta E_{\urm{int}, \, {\uparrow} {\downarrow}} < 0 $.
In particular, the condition $ \beta_a = - \beta_b $ is favored,
so that the spin-up and spin-down particles undergo the opposite quadrupole deformation.
They cancel with each other when one considers the total density
$
\rho_{\urm{def}} \left( \ve{r} \right)
=
\rho_{\urm{def}, \, {\uparrow}} \left( \ve{r} \right)
+
\rho_{\urm{def}, \, {\downarrow}} \left( \ve{r} \right) $,
so that the net deformation of the whole system is suppressed.
\par
Next, let us consider the case of atoms.
The multipole expansion of the Coulomb interaction reads
\begin{equation}
  \label{eq:Coul}
  V_l \left( r, r' \right)
  =
  \frac{4 \pi}{2l + 1}
  \frac{r_{<}^l}{r_{>}^{l+1}}, 
\end{equation}
where $ r_{<} $ and $ r_{>} $ are smaller and greater ones of $ r $ and $ r' $, respectively.
Substituting
\eqref{eq:Coul} into
\eqref{eq:uab_int},
one obtains
\begin{subequations}
  \begin{align}
    \Delta \epsilon_{\urm{int}, \, ab}^{\left( 1 \right)}
    = & \,
        \frac{64 \pi^2}{5}
        \int_0^{\infty}
        \int_0^{\infty}
        \frac{d \rho_{\urm{sph}, \, a} \left( r \right)}{dr}
        \rho_{\urm{sph}, \, b} \left( r' \right)
        \frac{1}{r_{>}}
        r^3 r'^2 \, dr \, dr'
        \notag \\
      & \, 
        +
        \frac{16 \pi^2}{5}
        \int_0^{\infty}
        \int_0^{\infty}
        \frac{d^2 \rho_{\urm{sph}, \, a} \left( r \right)}{dr^2}
        \rho_{\urm{sph}, \, b} \left( r' \right)
        \frac{1}{r_{>}}
        r^4 r'^2 \, dr \, dr'
        \notag \\
    = & \,
        \frac{16 \pi^2}{5}
        \int_0^{\infty}
        \int_0^{\infty}
        \theta \left( r - r' \right)
        \frac{d \rho_{\urm{sph}, \, a} \left( r \right)}{dr}
        \rho_{\urm{sph}, \, b} \left( r' \right)
        r^2 r'^2 \, dr \, dr', \\
    \Delta \epsilon_{\urm{int}, \, ab}^{\left( 3 \right)}
    = & \,
        \frac{32 \pi^2}{25}
        \int_0^{\infty}
        \int_0^{\infty}
        \frac{d \rho_{\urm{sph}, \, a} \left( r \right)}{dr}
        \frac{d \rho_{\urm{sph}, \, b} \left( r' \right)}{dr'}
        \frac{r_{<}^2}{r_{>}^3}
        r^3 r'^3 \, dr \, dr'
        \notag \\
    = & \,
        \frac{32 \pi^2}{25}
        \int_0^{\infty}
        \int_0^{\infty}
        \frac{d \rho_{\urm{sph}, \, a} \left( r \right)}{dr}
        \frac{d \rho_{\urm{sph}, \, b} \left( r' \right)}{dr'}
        \left[
        \theta \left( r - r' \right) r'^5
        +
        \theta \left( r' - r \right) r^5
        \right]
        \, dr \, dr',
        \label{eq:int_Coul}
  \end{align}
\end{subequations}
where $ \theta \left( r \right) $ is the Heaviside step function defined by
\begin{equation}
  \theta \left( r - r' \right)
  =
  \begin{cases}
    0 & \text{($ r < r' $)}, \\
    1 & \text{($ r > r' $)}.
  \end{cases}
\end{equation}
In the atomic system, the one-body density $ \rho_{\urm{sph}} $ is generally a decreasing function of $ r $,
except for the surface region
(which will be discussed in this section),
so that it is reasonable to assume $ d \rho_{\urm{sph}} /dr \le 0 $ in the most region.
Hence,
one can find
$ \Delta \epsilon_{\urm{int}, \, ab}^{\left( 1 \right)}
= \Delta \epsilon_{\urm{int}, \, ba}^{\left( 2 \right)} < 0 $
and 
$ \Delta \epsilon_{\urm{int}, \, ab}^{\left( 3 \right)} > 0 $.
One can also see for the parallel spin states 
\begin{align}
  & \Delta \epsilon_{\urm{int}, \, aa}^{\left( 1 \right)}
    +
    \Delta \epsilon_{\urm{int}, \, aa}^{\left( 2 \right)}
    +
    \Delta \epsilon_{\urm{int}, \, aa}^{\left( 3 \right)}
    \notag \\
  & =
    \frac{32 \pi^2}{5}
    \int_0^{\infty}
    \int_0^{\infty}
    \theta \left( r - r' \right)
    \frac{d \rho_{\urm{sph}, \, a} \left( r \right)}{dr}
    \rho_{\urm{sph}, \, a} \left( r' \right) 
    \left(
    r^2 - r'^2
    \right)
    r'^2
    \, dr \, dr'
    < 
    0.
    \label{eq:usumeval}
\end{align}
Therefore, when one considers the interaction energy between the parallel spin
[equation~\eqref{eq:ab_int_aa}],
one finds that the right-hand side of
is negative, and thus the Coulomb interaction favors deformation.
This is a well-known fact in a nuclear fission problem,
when the deformed Coulomb energy is evaluated with the liquid-drop model~\cite{
  Bohr1939Phys.Rev.56_426,
  Ring1980TheNuclearManyBodyProblem_SpringerVerlag}.
In addition, if one considers the interaction between the different spin
[equation~\eqref{eq:ab_int_ab}],
deformation with the different sign
$ \beta_{\uparrow} = - \beta_{\downarrow} $ 
is energetically favored
because of the positive
$ \Delta \epsilon_{\urm{int}, \, {\uparrow} {\downarrow}}^{\left( 3 \right)} $.
The effect of
$ \Delta \epsilon_{\urm{int}, \, {\uparrow} {\downarrow}}^{\left( 1 \right)} $ and
$ \Delta \epsilon_{\urm{int}, \, {\uparrow} {\downarrow}}^{\left( 2 \right)} $ 
also favors deformation
because
$ \Delta \epsilon_{\urm{int}, \, {\uparrow} {\downarrow}}^{\left( 1 \right)}
= \Delta \epsilon_{\urm{int}, \, {\downarrow} {\uparrow}}^{\left( 2 \right)} < 0 $.
Therefore, the Coulomb repulsive interaction favors spin-dependent deformation:
$ \rho_{\uparrow} $ and $ \rho_{\downarrow} $ makes the opposite quadrupole deformations.
When one considers the deformation of the total density
$ \rho_{\urm{def}} \left( \ve{r} \right)
=
\rho_{\urm{def}, \, {\uparrow}} \left( \ve{r} \right)
+
\rho_{\urm{def}, \, {\downarrow}} \left( \ve{r} \right) $,
they cancel with each other and the total deformation would be small.
These conclusions are in line with those for the repulsive delta function interaction.
However, in contrast to the delta function interaction,
there exists non-vanishing contribution from the parallel spin states
$ \Delta E_{\urm{int}, \, aa} $.
\par 
\textbf{Overall discussion}: From these qualitative discussions, we can see that difference in the nature of the interaction is the key in understanding why atoms do not tend to deform collectively while nuclei do.
The kinetic energy and the central external potential energy increase with the deformation,
hence disfavor the deformation
(see
\eqref{eq:kin_ene_inc}, \eqref{eq:ext_Coul}, and \eqref{eq:ext_HO}).
The interaction energy, on the other hand, may decrease with the deformation depending on the nature of the interaction.
In this connection, we mention that  
the well-known Hund rule~\cite{
  Hund1925Z.Phys.33_345,
  Hund1925Z.Phys.34_296,
  Slater1929Phys.Rev.34_1293,
  Hongo2004J.Chem.Phys.121_7144,
  Oyamada2010J.Chem.Phys.133_164113}
is also intimately related to the repulsive nature of electron-electron interaction, 
as has been argued in
\cite{
  Povh2015ParticlesandNuclei_Springer-Verlag}.
If this decrease is greater than the increase of the kinetic and external potential energies, the system would deform.
One finds above that an attractive short-range interaction,
modeled by the delta function interaction,
favors deformation in a spin-independent manner
$ \beta_{\uparrow} = \beta_{\downarrow} $
(see
\eqref{eq:int_delta}).
The repulsive Coulomb interaction energy, on the other hand,
favors spin-dependent deformation $ \beta_{\uparrow} = - \beta_{\downarrow} $,
in a manner to cancel with each other the deformation of the total density
$ \rho_{\urm{def}} \left( \ve{r} \right)
=
\rho_{\urm{def}, \, {\uparrow}} \left( \ve{r} \right)
+
\rho_{\urm{def}, \, {\downarrow}} \left( \ve{r} \right) $
(see
\eqref{eq:usumeval} and its discussion below).
The former represents nuclear systems, and the latter represents atomic systems.
The above results thus qualitatively explain why nuclear systems can deform significantly while atomic systems cannot.
Moreover, according to the discussion with short-range repulsive interaction
($ g_{{\uparrow} {\downarrow}} < 0 $),
this behavior depends on whether the interaction is repulsive or attractive,
while the range of the interaction does not play an important role.
The above argument based on two internal states, spin up and down states, can be readily extended to nuclear systems by considering four internal states, proton-up, proton-down, neutron-up, neutron-down states.
While this provides a qualitative way to understand the nuclear deformations,
it is important to notice that there are several spin-isospin channels in a two-nucleon system, and the relative strength of a nucleon-nucleon interaction for different channels may become rather important in nuclear structure.
In particular, it is known that the proton-neutron isoscalar interaction plays an important role in nuclear deformation, as has been discussed in
\cite{
  Federman1977Phys.Lett.B69_385,
  Federman1979Phys.Lett.B82_9,
  Federman1979Phys.Rev.C20_820,
  Dobaczewski1988Phys.Rev.Lett.60_2254}.
While this is true in general,
we notice that the neutron drops can also be deformed even without the proton-neutron isoscalar interaction,
as shown in
\ref{sec:n_drop}.
\par
This model discussion is closely related to the screening effects of electrons numerically found and discussed in
section~\ref{sec:numerical}.
The screening effect occurs due to the repulsive Coulomb interaction, which aims to decrease any charge imbalance.
If, for example, the spin-up electrons make a positive quadrupole deformation, the down spins attempt to make a negative quadrupole deformation to attain the charge neutrality.
Therefore, the above argument based on the multipole component of the interaction is consistent with the screening effect.
Indeed, we have found in
section~\ref{sec:numerical}
that
the deformation of the valence $ d $-orbital is cancelled by the deformation of the opposite spin in the $ s $-orbital for the $ d $-block atoms.
This is in line with the above model calculation.
We have also found in
section~\ref{sec:numerical}
that the deformations of atoms occur only in the valence electron region for $ p $- and $ d $-block atoms.
We can also explain this by using the above model:
The central external potential energy, favoring spherical shape, is great in the central region.
Therefore, even if the interaction energy is negative and favors deformation, it is difficult to overcome the positive potential energy.
It is thus rather unlikely for the central region of the atoms to undergo significant deformations.
In the valence electron region, on the other hand, the potential energy becomes small and it is easier for the interaction energy to overcome it.
This qualitatively explains why atoms can be deformed only in the valence electron region. 
\par 
As the above discussions for the quadrupole deformation may also apply to the higher order multipole deformations at any order, we can conjecture the following statement for general infinitesimal deformations:
The multipole component of the inter-particle interaction must be attractive for the system to undergo significant collective deformations.
The Coulomb interaction between the electrons has repulsive multipole components for all orders, and therefore the deformations of up-spin and down-spin states cancel with each other for all multipole orders, resulting in a small total deformation as numerically found in the previous section.
The nuclear interaction, on the other hand, may undergo spin-independent deformations when a multipole component of the interaction is attractive and strong enough.
\par
We note that we have assumed in the above model argument that $ d \rho_{\urm{sph}} / dr \le 0 $.
While this assumption is valid for closed shell atoms and also for open-shell atoms in their central region,
the electron density usually oscillates for open-shell atoms in the surface region:
The electron density in the surface region is mostly determined by the wave function of the outer-most valence electrons $ \psi_{nlms} \left( r, \theta, \phi \right) $,
which show oscillations with $ r $ when $ n > 1 $.
Thus, the above model argument in the valence electron region should be taken with some cautions.
\par
We also note that we have neglected in the above discussion the exchange-correlation part in evaluating the interaction energy
[equation~\eqref{eq:int_def}].
In the Hartree-Fock theory and the DFT,
the exchange or exchange-correlation part is generally found to be sub-dominant than the direct part contributions for electrons in atoms~\cite{
  Parr1989Density-FunctionalTheoryofAtomsandMolecules_OxfordUniversityPress,
  Kohn1999Rev.Mod.Phys.71_1253,
  Martin2004_CambridgeUniversityPress,
  Engel2011_Springer-Verlag}.
We have, therefore, only considered the direct part in our basic qualitative discussions above,
while one needs to include the exchange-correlation part for more elaborated descriptions.
In particular, for electrons in the surface region of atoms (i.e.~the valence orbital),
the exchange-correlation part becomes more important compared with the central region, because the density is rather small and thus the electrons are strongly correlated~\cite{
  Wigner1934Phys.Rev.46_1002,
  Parr1989Density-FunctionalTheoryofAtomsandMolecules_OxfordUniversityPress,
  Kohn1999Rev.Mod.Phys.71_1253}.
Our model discussion on the surface region of atoms, therefore, should be at best qualitative one.
In the nuclear DFT, on the other hand, the direct and exchange parts are 
usually not considered separately.
Rather, a short-range attractive effective interaction
$ V_{\urm{int}} \left( \left| \ve{r} - \ve{r}' \right| \right) $
in
\eqref{eq:int_def} is often used whose shape and parameters are chosen
to reproduce well nuclear properties measured in experiments~\cite{
  Vautherin1972Phys.Rev.C5_626,
  Vautherin1973Phys.Rev.C7_296,
  Bender2003Rev.Mod.Phys.75_121}.
Thus, the above argument based on
\eqref{eq:int_def},
ostensibly considering only the direct part of interaction energy, naturally contains many-body
correlation effects,
including those due to the three-body and tensor interaction, 
and therefore should be valid even though atomic nuclei
are rather strongly correlated. 
\par
While we have shown that atoms in their ground states are unlikely to be deformed significantly, we may expect larger deformations in more exotic atomic systems.
As can be seen in our qualitative discussion above,
atoms may be deformed if the condition $ d \rho_{\urm{sph}} / dr \le 0 $ does not hold or if the exchange-correlation term dominates the direct term.
These are often the cases for highly excited atoms.
In particular, we can expect that Rydberg atoms~\cite{
  Gallagher2005Rydbergatoms_CambridgeUniversityPress,
  Saffman2010Rev.Mod.Phys.82_2313,
  Luukko2017Phys.Rev.Lett.119_203001},
which have electrons in excited states with extremely large principal quantum numbers,
may show significant deformations.
While the Rydberg atoms with one electron excited to such orbital will trivially show a deformed shape representing the single-particle orbital, they may show non-trivial collective deformations when more than one electrons are excited in the Rydberg orbits~\cite{
  Luukko2017Phys.Rev.Lett.119_203001}.

% 
% Conclusion
% -*- coding: utf-8 -*-
%%%%%%%%%%%%%%%%%%%%%%%%%%%%%%%%%%%%%%%%%%%%%%%%%% 
% 
% Deformation of atoms paper manuscript for J. Phys. B
% Conclusion Part
%
% Begin to write: 2020-07-26
%
% Tomoya Naito, Shimpei Endo,
% Kouichi Hagino, and Yusuke Tanimura
% 
%%%%%%%%%%%%%%%%%%%%%%%%%%%%%%%%%%%%%%%%%%%%%%%%%%
%
\section{Conclusion}
\label{sec:conc}
\par
We have investigated numerically whether the electron density distribution of isolated atoms can be deformed.
To this end, we have calculated the deformation parameters for various atoms with a wide variety of different many-body methods,
and have found that the noble-gas, half-shell-closed, and $ s $-block atoms are spherical,
while the $ p $- and $ d $-block atoms are deformed.
The deformations of the $ p $- and $ d $-block atoms are neither collective nor significantly deformed.
We have shown that the core part remains spherical even for these atoms, and their deformations originate from a few valence electrons.
Therefore, the deformations of the atoms are at most of a single-particle nature of a few valance electrons:
Atoms do not deform collectively, in contrast to nuclei.
We have also shown that the many-body effects of electrons in atoms tend to make the deformations smaller due to the screening effect. 
\par
Owing to the deformation, the self-consistent effective mean-field potential for $ p $- and $ d $-block atoms is slightly deformed in the surface and the tail region.
Therefore, the core electrons are still eigenstates of angular momentum 
and are described by single spherical harmonics,
whereas the valence electrons are not.
We have found numerically that the degeneracy of the valence electrons are indeed lifted due to this effect, but the effect is $ 0.05 \, \mathrm{Hartree} $ or much less.
Hence, calculations of the atomic structure with spherical symmetry are still justified.
\par
We have compared the atomic deformation to the nuclear deformation by using a qualitative model. 
We have argued that the difference between them originates from the properties of the inter-particle interactions.
On the one hand, in the case of atomic nuclei, the interaction is of attractive in net, and the deformation of the spin-up component favors to be the same as that of the spin-down component.
Consequently, the collective deformation can occur. 
On the other hand, in the case of atoms, the interaction is purely repulsive,
and the deformation of the spin-up component favors to be the opposite of that of the spin-down component,
cancelling each other.
Accordingly, the collective deformation does not occur.
The apparent difference originates from the sign of the inter-particle interaction, 
rather than its range.
\par
While the collective deformations of nuclei have clearly been observed by the gamma-ray spectroscopy of their rotational spectra of
\eqref{eq:Erot},
the atomic deformations studied in this work are neither significant nor collective,
and therefore should not show similar rotational spectra.
One rather needs to directly probe the density distribution of the electrons to experimentally test the deformations found by our quantum chemical calculations.
It would be rather challenging but we expect that they may be directly observed with recent atomic and molecular physics technology,
such as photo-ionization microscopy~\cite{
  Stodolna2013Phys.Rev.Lett.110_213001,
  Cohen2016Phys.Rev.A94_013414},
or tomographic imaging~\cite{
  Itatani2004Nature432_867}.

% 
% Acknowledgement
% -*- coding: utf-8 -*-
%%%%%%%%%%%%%%%%%%%%%%%%%%%%%%%%%%%%%%%%%%%%%%%%%% 
% 
% Deformation of atoms paper manuscript for J. Phys. B
% Acknowledge Part
%
% Begin to write: 2020-07-24
%
% Tomoya Naito, Shimpei Endo,
% Kouichi Hagino, and Yusuke Tanimura
% 
%%%%%%%%%%%%%%%%%%%%%%%%%%%%%%%%%%%%%%%%%%%%%%%%%%
%
\section*{Acknowledgment}
\par
We thank
Gianluca Col\`{o},
Yoshiko Kanada-En'yo,
Haozhao Liang,
Takashi Nakatsukasa,
and
Kenichi Yoshida 
for fruitful discussion.
The RIKEN iTHEMS program,
the JSPS Grant-in-Aid for JSPS Fellows under Grant No.~JP19J20543,
and JSPS KAKENHI Grant Nos.~JP19K03861, JP19K21028, and JP21H00116 are acknowledged.
The numerical calculations were performed on cluster computers at the RIKEN iTHEMS program.

\appendix
%
% Neutron Drop
% -*- coding: utf-8 -*-
%%%%%%%%%%%%%%%%%%%%%%%%%%%%%%%%%%%%%%%%%%%%%%%%%% 
% 
% Deformation of atoms paper manuscript for J. Phys. B
% Neutron Drop Part
% 
% Begin to write: 2020-07-26
% 
% Tomoya Naito, Shimpei Endo,
% Kouichi Hagino, and Yusuke Tanimura
% 
%%%%%%%%%%%%%%%%%%%%%%%%%%%%%%%%%%%%%%%%%%%%%%%%%% 
% 
\section{Deformation of neutron drop in a harmonic trap}
\label{sec:n_drop}
\par
In this Appendix, we discuss an illustrative case of deformed 
systems: that is, a neutron drop trapped in a spherical harmonic potential. 
Even though neutron drops are fictitious, they provide useful 
systems to benchmark many-body theories~\cite{
  Pudliner1996Phys.Rev.Lett.76_2416,
  Smerzi1997Phys.Rev.C56_2549,
  Gandolfi2011Phys.Rev.Lett.106_012501,
  Bogner2011Phys.Rev.C84_044306,
  Maris2013Phys.Rev.C87_054318,
  Shen2018Phys.Rev.C97_054312,
  Shen2019Phys.Rev.C99_034322,
  Wang2021Chin.Phys.C45_064103},
As the density can be controlled by changing the strength of the trapping potential.
They have also been utilized in connection with neutron-rich nuclei and neutron stars~\cite{
  Zhao2016Phys.Rev.C94_041302}.
By using this model, we show in this Appendix that, within HF theory, 
the nuclear systems are indeed collectively deformed due to the attractive nuclear
force:
The systems are collectively deformed when the inter-particle attraction
is strong enough, while they do not if the interaction is repulsive.
\par
Although the interaction between neutrons is attractive, in contrast to the 
interaction between electrons, 
typical nuclear interactions are not strong enough so that 
systems consisting solely of neutrons do not bound by themselves. 
To study properties of many-neutron systems, one thus needs to introduce 
an external confining potential, analogously to the electron-nucleus external potential in atoms. 
In this Appendix, we use an isotropic harmonic oscillator potential, 
\begin{equation}
  \label{eq:ho}
  V_{\urm{ext}}
  \left( r \right)
  =
  \frac{1}{2}
  m_n
  \Omega^2
  r^2, 
\end{equation}
where $ m_n = 939.6 \, \mathrm{MeV} / c^2 $ is the neutron mass~\cite{
  Zyla2020Prog.Theor.Exp.Phys.2020_083C01},
to localize the neutrons in neutron drops. 
In this study, we set $ \hbar \Omega = 5 \, \mathrm{MeV} $.
\par
To take into account the interaction between neutrons, 
we employ the HF theory with the Skyrme effective interaction~\cite{
  Vautherin1972Phys.Rev.C5_626,
  Vautherin1973Phys.Rev.C7_296}.
The Skyrme interaction is a zero-range force with momentum and density dependences, 
given by
\begin{align}
  v_{\urm{Sk}}
  = & \, 
      t_0
      \left( 1 + x_0 P_{\sigma} \right)
      \delta \left( \ve{r}_1 - \ve{r}_2 \right)
      -
      \frac{t_1}{8}
      \left( 1 + x_1 P_{\sigma} \right)
      \left[
      \overleftarrow{\Nabla}^2
      \delta \left( \bm{r}_1 - \bm{r}_2 \right)
      +
      \delta \left( \bm{r}_1 - \bm{r}_2 \right)
      \overrightarrow{\Nabla}^2
      \right]
      \notag \\
    & \, 
      +
      \frac{t_2}{4}
      \left( 1 + x_2 P_{\sigma} \right)
      \left[
      \overleftarrow{\Nabla}
      \delta \left( \bm{r}_1 - \bm{r}_2 \right)
      \cdot
      \overrightarrow{\Nabla}
      \right]
      +
      \frac{t_3}{6}
      \left( 1 + x_3 P_{\sigma} \right)
      \delta \left( \bm{r}_1 - \bm{r}_2 \right)
      \left\{
      \rho \left( \frac{\bm{r}_1 + \bm{r}_2}{2} \right)
      \right\}^{\alpha}
      \notag \\
    & \,
      +
      i \frac{W_0}{4}
      \left[
      \overleftarrow{\Nabla}
      \times
      \delta \left( \bm{r}_1 - \bm{r}_2 \right)
      \overrightarrow{\Nabla}
      \right]
      \cdot
      \left( \bm{\sigma}_1 + \bm{\sigma}_2 \right),
      \label{eq:skyrme}
\end{align}
where
$ \overrightarrow{\Nabla} = \overrightarrow{\Nabla}_1 - \overrightarrow{\Nabla}_2 $
acting on the right and
$ \overleftarrow{\Nabla} = \overleftarrow{\Nabla}_1 - \overleftarrow{\Nabla}_2 $ 
acting on the left,
and
$ P_{\sigma} = \left( 1 + \bm{\sigma}_1 \cdot \bm{\sigma}_2 \right) / 2 $ 
is the spin-exchange operator. 
The coefficients $ t_i $, $ x_i $ ($ i = 0 $--$ 3 $),
$ \alpha $, and $ W_0 $ are the parameters to be
adjusted to
properties of the homogeneous nucleon gas, called nuclear matter,
and experimentally observed nuclear properties. 
Among these, the $ t_0 $ and $ t_3 $ terms determine the bulk property of nuclei. 
The former is attractive,
and the latter with density dependence ($ \alpha > 0 $) is repulsive. 
The repulsive $ t_3 $ term, which becomes dominant over the other terms at high densities, 
is responsible for the saturation of nuclear matter density. 
\par
In our study, we adopt the SkM* parameter set~\cite{
  Bartel1982Nucl.Phys.A386_79}
given in Table~\ref{tb:skmstar}.
In order to see how the deformation
property depends on the strength and the sign of the interaction, 
we scale the strength parameters of the Skyrme interaction 
as $ t_i \to f t_i $ for $ i = 0 $, $ 1 $, $ 2 $ and
$ t_3 \to \left| f \right| t_3 $ with $ f $ being varied within
the range $ \left[ -1, 1 \right] $,
where $ f = 1 $ corresponds to the original.
The $ t_3 $ term is kept repulsive to avoid collapse of the system.
In our numerical calculations, reflection and axial symmetries are imposed.  
\begin{table}[!htb]
  \centering
  \caption{
    Values of
    $ t_0 $ ($ \mathrm{MeV} \, \mathrm{fm}^3 $),
    $ t_1 $ ($ \mathrm{MeV} \, \mathrm{fm}^5 $),
    $ t_2 $ ($ \mathrm{MeV} \, \mathrm{fm}^5 $), 
    $ t_3 $ ($ \mathrm{MeV} \, \mathrm{fm}^{3 \left( 1 + \alpha \right)} $),
    $ x_0 $, $ x_1 $, $ x_2 $, $ x_3 $, $ \alpha $, and
    $ W_0 $ ($ \mathrm{MeV} \, \mathrm{fm}^5 $) of
    the SkM* parameter set~\cite{
      Bartel1982Nucl.Phys.A386_79}.}
  \label{tb:skmstar}
  \begin{indented}
  \item[]
    \begin{tabular}{cccccccccc}
      \br
      \multicolumn{1}{c}{$ t_0 $} & \multicolumn{1}{c}{$ t_1 $} & \multicolumn{1}{c}{$ t_2 $} & \multicolumn{1}{c}{$ t_3 $} & \multicolumn{1}{c}{$ x_0 $} & \multicolumn{1}{c}{$ x_1 $} & \multicolumn{1}{c}{$ x_2 $} & \multicolumn{1}{c}{$ x_3 $} & \multicolumn{1}{c}{$ \alpha $} & \multicolumn{1}{c}{$ W_0 $} \\
      \mr
      $ -2645 $ & $ 410 $ & $ -135 $ & $ 15595 $ & $ 0.09 $ & $ 0 $ & $ 0 $ & $ 0 $ & $ 1/6 $ & $ 130 $ \\
      \br
    \end{tabular}
  \end{indented}
\end{table}
\par
To find HF solutions for a neutron drop, we perform several HF iterations 
for each system starting from prolate and oblate configurations. 
Most often, a prolate (oblate) initial condition ends up with a prolate (oblate) 
solution for open-shell systems. 
For closed-shell systems, on the other hand, a spherical 
solution is obtained regardless of the initial deformation
if it is energetically favored. 
\par
\begin{figure}[!htb]
  \centering
  \includegraphics[width=1.0\linewidth]{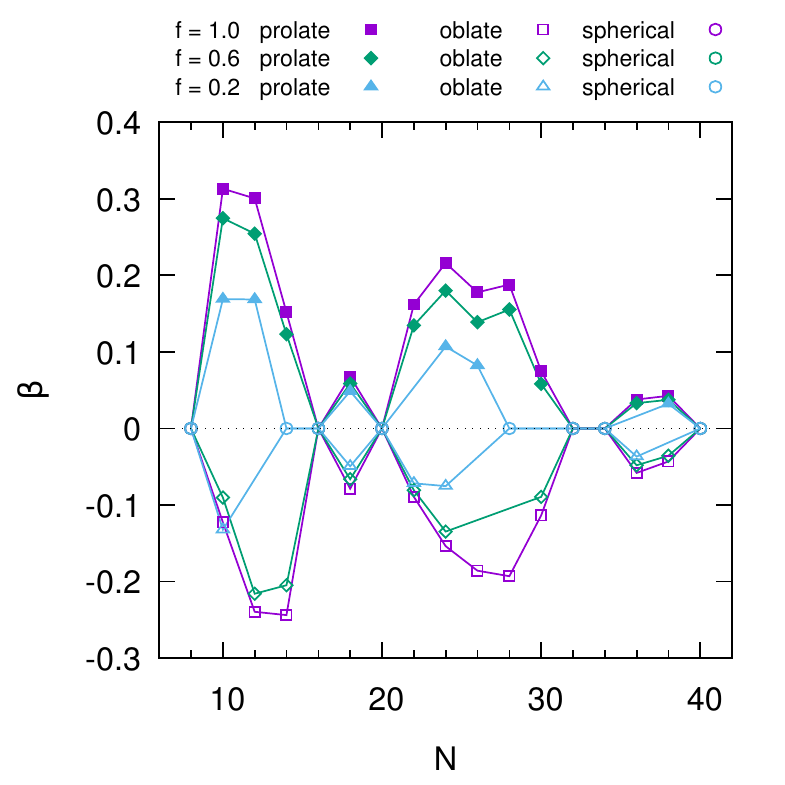}
  \caption{
    Deformation parameters of HF solutions for the neutron drops
    $ {}^{\text{$ 8 $--$ 40 $}} n $ with $ \hbar \Omega = 5 \, \mathrm{MeV} $.
    Here, $ f $ is the scaling factor of the strengths of Skyrme interaction
    (see the main text for details).
    When there are two solutions both on prolate and oblate sides, both are plotted in the figure.}
  \label{fig:def_ndrops}
\end{figure}

Figure~\ref{fig:def_ndrops}
shows the deformation parameters of the neutron drops 
$ {}^{\text{$ 8 $--$ 40 $}} n $ with attractive ($ f > 0 $) interactions.
Here, $ {}^{N} n $ denotes the system with $ N $ neutrons. 
The spherical solutions are shown by open circles. 
The prolate and oblate solutions are respectively plotted by solid and open symbols. 
Squares, diamonds, and triangles correspond to the solutions for
$ f = 1 $, $ 0.6 $, and $ 0.2 $, respectively. 
As can be seen from the figure, many of the neutron drops examined here have both prolate 
and oblate solutions, often at similar energies to each other. 
The quadrupole deformation tends to be smaller as the strengths 
of the attractive interaction decrease. 
This behavior of the deformation can be understood as follows: 
as the interaction is weakened, the effect of the spherical external field becomes 
more significant relative to the interaction,
which is consistent with the discussion shown in
section~\ref{sec:model}.
As a result, the system tends to 
adjust its density distribution closer to the spherical shape to reduce the potential 
energy of the external field.
The systems with $ N = 8 $, $ 16 $, $ 20 $, $ 32 $, $ 34 $, and $ 40 $ are spherical for all the values 
of the scaling parameter $ f $.
They correspond to closed-shell configurations, \textit{e.g.},
$ \left( 1s_{1/2} \right)^2 \left( 1p_{3/2} \right)^4 \left( 1p_{1/2} \right)^2 $ for $ {}^{8} n $,
$ \left( 1s_{1/2} \right)^2 \left( 1p_{3/2} \right)^4 \left( 1p_{1/2} \right)^2 \left( 1d_{5/2} \right)^6 \left( 2s_{1/2} \right)^2$ for $ {}^{16} n $, 
and
$ \left( 1s_{1/2} \right)^2 \left( 1p_{3/2} \right)^4 \left( 1p_{1/2} \right)^2 \left( 1d_{5/2} \right)^6 \left( 2s_{1/2} \right)^2 \left( 1d_{3/2} \right)^4 $ for $ {}^{20} n $. 
\par
We have also carried out HF calculations with repulsive ($ f < 0 $) interactions, 
but we obtained spherical solutions for only a few systems with 
closed-shell configurations, and no deformed solution was found. 
Although the pairing correlation would stabilize the convergence 
and allows us to study repulsive systems as well as the attractive ones, 
in this appendix we only consider systems with variable attractive 
interaction within the HF theory. 
\par
\begin{figure}[!htb]
  \centering
  \includegraphics[width=1.0\linewidth]{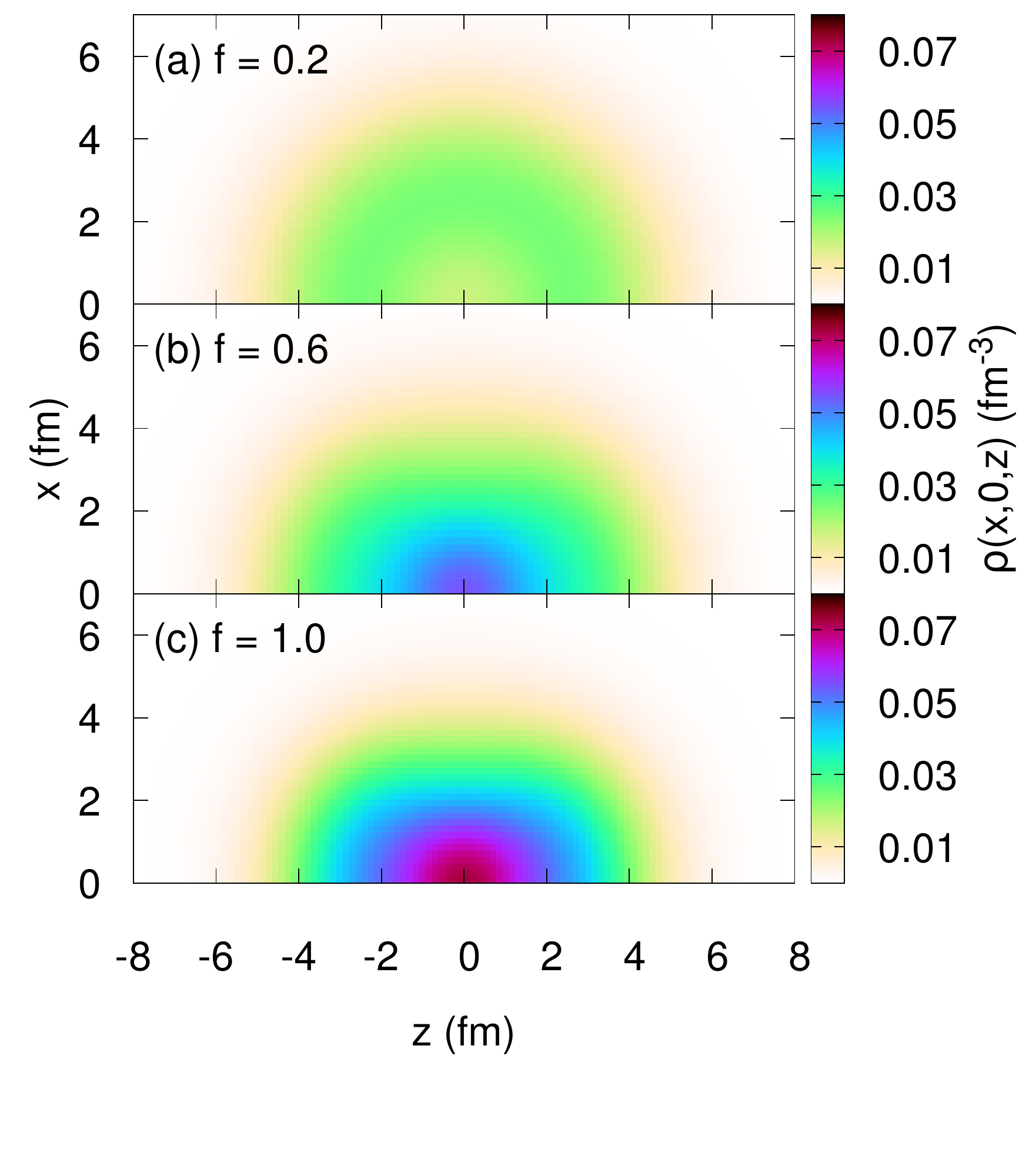}
  \caption{
    Densities of the prolate HF solutions on $ xz $ plane of $ {}^{14} n $ for
    (a) $ f = 0.2 $, 
    (b) $ f = 0.6 $, and
    (c) $ f = 1.0 $.
    The horizontal ($z$) axis is the axis of symmetry.}
  \label{fig:dens_n14_pro}
\end{figure}
\begin{figure}[!htb]
  \centering
  \includegraphics[width=1.0\linewidth]{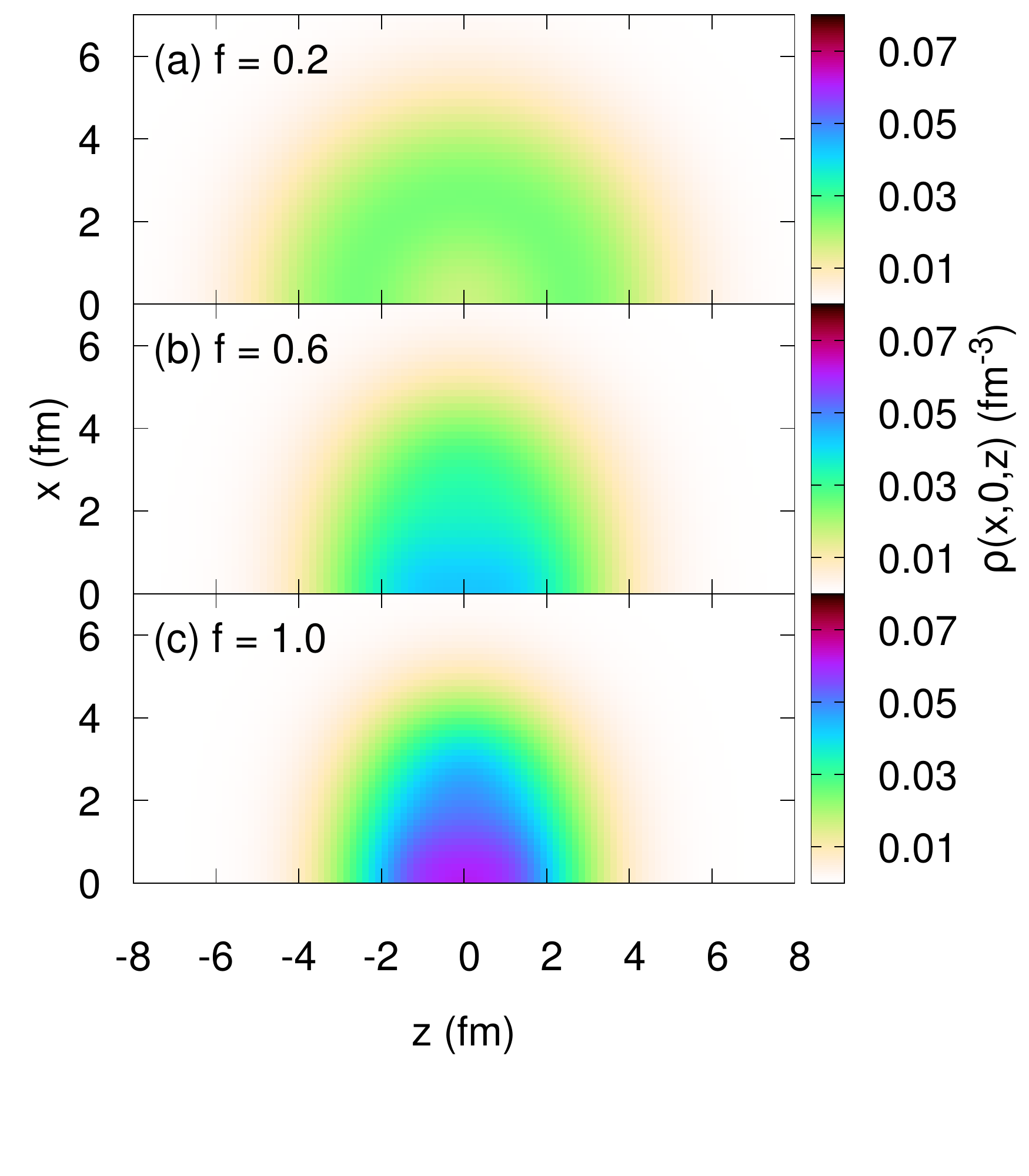}
  \caption{
    Same as
    figure~\ref{fig:dens_n14_pro}
    but for the oblate HF solutions.}
  \label{fig:dens_n14_obl}
\end{figure}

In
figures~\ref{fig:dens_n14_pro} and \ref{fig:dens_n14_obl},
we show the neutron density distributions of prolate and oblate solutions, respectively,
of $ {}^{14} n $ for different values of $ f $.  
In both cases, we see clearly that the density becomes tighter and more deformed as 
$ f $ changes from $ 0.2 $ to $ 1 $. 
\par
Evidently, neutrons trapped in an isotropic harmonic-oscillator potential can be collectively 
deformed for open-shell systems,
in contrast to electrons in atoms, as is argued in
section~\ref{sec:model}. 
As expected, the spherically symmetric external field tends to suppress the 
deformation parameter as well as the radius of the system.
This conclusion is consistent with the qualitative discussion in
section~\ref{sec:model}.
%

% 
% General
% -*- coding: utf-8 -*-
%%%%%%%%%%%%%%%%%%%%%%%%%%%%%%%%%%%%%%%%%%%%%%%%%% 
% 
% Deformation of atoms paper manuscript for J. Phys. B
% General calculation
% 
% Begin to write: 2021-01-03
% 
% Tomoya Naito, Shimpei Endo,
% Kouichi Hagino, and Yusuke Tanimura
% 
%%%%%%%%%%%%%%%%%%%%%%%%%%%%%%%%%%%%%%%%%%%%%%%%%% 
% 
\section{Generalization of Section~\ref{sec:model}}
\label{sec:gen}
\par
In this Appendix, we will show the general case (i.e.~$ \alpha_a \ne 2 \beta_a $) of the calculations shown in
section~\ref{sec:model}.
Firstly, the single-particle densities of up- and down-spin states are defined as follows:
\begin{equation}
  \rho_{\urm{def}, \, a} \left( \ve{r} \right)
  =
  \frac{1}{2}
  \sum_{j = 1}^N
  \int
  \Psi_{\urm{def}}^* \left( \ve{r}_1, \ldots, \ve{r}_N \right)
  \left( 1 \pm \sigma_{zj} \right)
  \delta \left( \ve{r} - \ve{r}_j \right)
  \Psi_{\urm{def}} \left( \ve{r}_1, \ldots, \ve{r}_N \right)
  \, d \ve{r}_1 \, \ldots \, d \ve{r}_N,
\end{equation}
where $ {+} $ in $ 1 \pm \sigma_{zj} $ for $ a = {\uparrow} $ and $ {-} $ for $ a = {\downarrow} $.
For simplicity, the spin coordinates of particles
($ {\uparrow} $ for particle $ j = 1 $, $ 2 $, \ldots, $ N_{\uparrow} $
and
$ {\uparrow} $ for particle $ j = N_{\uparrow} + 1 $, $ N_{\uparrow} + 2 $, \ldots, $ N $)
are omitted.
Using the notation of $ \ve{R}_a $, $ \ve{r} $, $ \ve{R}_j $, and $ \ve{r}_j $
which denote
$ \ve{R}_a = \left( e^{\beta_a} x, e^{\beta_a} y, e^{- \alpha_a} z \right) $,
$ \ve{r} = \left( x, y, z \right) $,
$ \ve{R}_j = \left( e^{\beta_j} x, e^{\beta_j} y, e^{- \alpha_j} z \right) $,
$ \ve{r}_j = \left( x_j, y_j, z_j \right) $, 
and the relation 
$ \delta \left( \ve{r} - \ve{r}_j \right) = e^{- \left( \alpha_a - 2 \beta_a \right)} \delta \left( \ve{R}_a - \ve{R}_j \right) $,
one obtains 
\begin{align}
  & \rho_{\urm{def}, \, a} \left( \ve{r} \right)
    \notag \\
  = & \,
      \frac{1}{2}
      \sum_{j = 1}^N
      \int
      \Psi_{\urm{sph}}^* \left( \ve{R}_1, \ldots, \ve{R}_N \right)
      \left( 1 \pm \sigma_{zj} \right)
      \delta \left( \ve{r} - \ve{r}_j \right)
      \Psi_{\urm{sph}} \left( \ve{R}_1, \ldots, \ve{R}_N \right)
      \, d \ve{R}_1 \, \ldots \, d \ve{R}_N
      \notag \\
  = & \,
      \frac{1}{2}
      \sum_{j = 1}^N
      e^{2 \beta_a - \alpha_a}
      \int
      \Psi_{\urm{sph}}^* \left( \ve{R}_1, \ldots, \ve{R}_N \right)
      \left( 1 \pm \sigma_{zj} \right)
      \delta \left( \ve{R}_a - \ve{R}_j \right)
      \Psi_{\urm{sph}} \left( \ve{R}_1, \ldots, \ve{R}_N \right)
      \, d \ve{R}_1 \, \ldots \, d \ve{R}_N
      \notag \\
  = & \,
      e^{2 \beta_a - \alpha_a}
      \rho_{\urm{sph}, \, a} \left( \ve{R}_a \right).
\end{align}
\par
First of all,
when the absolute value of the deformation parameters $ \left| \alpha_a \right| $ and  $ \left| \beta_a \right| $ are small,
the one-body density $ \rho_{\urm{def}, \, a} $ of the system reads
\begin{align}
  \rho_{\urm{def}, \, a}
  \left( x, y, z \right)
  = & \, 
      \rho_{\urm{sph}, \, a}
      \left( e^{\beta_a} x, e^{\beta_a} y, e^{- \alpha_a} z \right)
      \notag \\
  = & \,
      \left[
      1 + \left( 2 \beta_a - \alpha_a \right) + \frac{1}{2} \left( 2 \beta_a - \alpha_a \right)^2
      \right]
      \rho_{\urm{sph}, \, a} \left( r \right)
      \notag \\
    & \, 
      +
      f^{\left( 1 \right)}_a \left( \ve{r} \right)
      \frac{d \rho_{\urm{sph}, \, a} \left( r \right)}{dr}
      +
      f^{\left( 2 \right)}_a \left( \ve{r} \right)
      \frac{d^2 \rho_{\urm{sph}, \, a} \left( r \right)}{dr^2}
      +
      O \left( \beta^3 \right),
      \label{eq:gen_1density_partialwave} \\
  f^{\left( 1 \right)}_a \left( \ve{r} \right)
  = & \,
      \left[ 1 + \left( 2 \beta_a - \alpha_a \right) \right]
      \left( \beta_a \frac{x^2 + y^2}{r} - \alpha_a \frac{z^2}{r} \right)
      \notag \\
    & \,
      +
      \beta_a^2 \frac{x^2 + y^2}{r}
      +
      \alpha_a^2 \frac{z^2}{r}
      -
      \frac{\beta_a^2}{2}
      \frac{x^4 + y^4}{r^3}
      -
      \frac{\alpha_a^2}{2}
      \frac{z^4}{r}
      -
      \beta_a^2
      \frac{x^2 y^2}{r^3}
      +
      \alpha_a \beta_a
      \frac{\left( x^2 + y^2 \right) z^2}{r^3},
      \label{eq:gen_f1} \\
  f^{\left( 2 \right)}_a \left( \ve{r} \right)
  = & \,
      \frac{\beta_a^2}{2}
      \frac{x^4 + y^4}{r^2}
      +
      \frac{\alpha_a^2}{2}
      \frac{z^4}{r^2}
      +
      \beta_a^2
      \frac{x^2 y^2}{r^2}
      -
      \alpha_a \beta_a
      \frac{\left( x^2 + y^2 \right) z^2}{r^2}.
      \label{eq:gen_f2} 
\end{align}
The angular integration of
$ f^{\left( 1 \right)}_a $ and $ f^{\left( 2 \right)}_a $
can be performed analytically as 
\begin{align}
  F^{\left( 1 \right)}_a \left( r \right)
  & \defeq
    \int f^{\left( 1 \right)}_a \left( \ve{r} \right) \, d \Omega
    =
    \frac{2 \pi}{15} r
    \left(
    17 \alpha_a^2
    -
    36 \alpha_a \beta_a
    +
    52 \beta_a^2
    -
    10 \alpha_a
    +
    20 \beta_a
    \right), 
    \label{eq:gen_F1} \\
  F^{\left( 2 \right)}_a \left( r \right)
  & \defeq
    \int f^{\left( 2 \right)}_a \left( \ve{r} \right) \, d \Omega
    =
    \frac{2 \pi}{15} r^2
    \left(
    3 \alpha_a^2
    -
    4 \alpha_a \beta_a
    +
    8 \beta_a^2
    \right),
    \label{eq:gen_F2}
\end{align}
respectively.
Let us consider using this transformation how the energy of the system changes when an infinitesimal quadrupole deformation
$ \left| \alpha_a \right| \ll 1 $ and $ \left| \beta_a \right| \ll 1 $
is induced to a spherically symmetric system
$ \rho_{\urm{sph}} \left( \ve{r} \right) = \rho_{\urm{sph}} \left( r \right) $.
\par
\textbf{Kinetic energy}: The kinetic energy reads
\begin{align}
  T_{\urm{def}, \, a}
  & =
    \brakket{\Psi_{\urm{def}, \, a}}{\hat{T}}{\Psi_{\urm{def}, \, a}}
    \notag \\
  & = 
    \left(
    \frac{2}{3}
    e^{2 \beta_a}
    +
    \frac{1}{3}
    e^{- 2 \alpha_a}
    \right)
    \brakket{\Psi_{\urm{sph}, \, a}}{\hat{T}}{\Psi_{\urm{sph}, \, a}}
    \notag \\
  & =
    \left(
    \frac{2}{3}
    e^{2 \beta_a}
    +
    \frac{1}{3}
    e^{- 2 \alpha_a}
    \right)
    T_{\urm{sph}, \, a}.
    \label{eq:gen_kin}
\end{align}
\par
\textbf{Central external potential energy}: By using
\eqref{eq:gen_1density_partialwave}
and the angular integral
\eqref{eq:gen_F1} and \eqref{eq:gen_F2},
the central external potential changes as
\begin{align}
  & \Delta E_{\urm{ext}, \, a}
    \notag \\
  \simeq & \,
           \left[
           \left( 2 \beta_a - \alpha_a \right) + \frac{1}{2} \left( 2 \beta_a - \alpha_a \right)^2
           \right]
           \int 
           \rho_{\urm{sph}, \, a} \left( r \right)
           V_{\urm{ext}} \left( r \right)
           \, d \ve{r}
           \notag \\
  & \, 
    +
    \int
    f^{\left( 1 \right)}_a \left( \ve{r} \right)
    \frac{d \rho_{\urm{sph}, \, a} \left( r \right)}{dr}
    V_{\urm{ext}} \left( r \right)
    \, d \ve{r}
    +
    \int
    f^{\left( 2 \right)}_a \left( \ve{r} \right)
    \frac{d^2 \rho_{\urm{sph}, \, a} \left( r \right)}{dr^2}
    V_{\urm{ext}} \left( r \right)
    \, d \ve{r}
    \notag \\
  = & \,
      4 \pi
      \left[
      \left( 2 \beta_a - \alpha_a \right) + \frac{1}{2} \left( 2 \beta_a - \alpha_a \right)^2
      \right]
      \int_0^{\infty}
      \rho_{\urm{sph}, \, a} \left( r \right)
      V_{\urm{ext}} \left( r \right)
      r^2 \, dr
      \notag \\
  & \, 
    +
    \int_0^{\infty}
    F^{\left( 1 \right)}_a \left( r \right)
    \frac{d \rho_{\urm{sph}, \, a} \left( r \right)}{dr}
    V_{\urm{ext}} \left( r \right)
    r^2 \, dr
    +
    \int_0^{\infty}
    F^{\left( 2 \right)}_a \left( r \right)
    \frac{d^2 \rho_{\urm{sph}, \, a} \left( r \right)}{dr^2}
    V_{\urm{ext}} \left( r \right)
    r^2 \, dr
    \notag \\
  = & \,
      4 \pi
      \left[
      \left( 2 \beta_a - \alpha_a \right) + \frac{1}{2} \left( 2 \beta_a - \alpha_a \right)^2
      \right]
      \int_0^{\infty}
      \rho_{\urm{sph}, \, a} \left( r \right)
      V_{\urm{ext}} \left( r \right)
      r^2 \, dr
      \notag \\
  & \, 
    -
    \int_0^{\infty}
    \rho_{\urm{sph}, \, a} \left( r \right)
    \frac{d}{dr}
    \left[
    F^{\left( 1 \right)}_a \left( r \right)
    V_{\urm{ext}} \left( r \right)
    r^2
    \right]
    \, dr
    +
    \int_0^{\infty}
    \rho_{\urm{sph}, \, a} \left( r \right)
    \frac{d^2}{dr^2}
    \left[
    F^{\left( 2 \right)}_a \left( r \right)
    V_{\urm{ext}} \left( r \right)
    r^2
    \right]
    \, dr,
    \label{eq:gen_ext_full}
\end{align}
where the external potential 
$ V_{\urm{ext}} \left( r \right) \sim r^d $ ($ d > -3 $)
is assumed to behave as $ r \to 0 $.
We assume the form of the external potential as
$ V_{\urm{ext}} \left( r \right) = c r^d $
with given constants $ c $ and $ d $.
Substituting this form into
\eqref{eq:gen_ext_full},
we get
\begin{align}
  \Delta E_{\urm{ext}, \, a}
  = & \,
      \left[
      -
      \left( 2 \beta_a - \alpha_a \right)
      +
      \frac{1}{2} \left( 2 \beta_a - \alpha_a \right)^2
      \right.
      \notag \\
    & \,
      \left.
      -
      \frac{1}{30} \left( d + 3 \right)
      \left( 17 \alpha_a^2 - 36 \alpha_a \beta_a + 52 \beta_a^2 - 10 \alpha_a + 20 \beta_a \right)
      \right.
      \notag \\
    & \,
      \left.
      +
      \frac{1}{30} \left( d + 4 \right) \left( d + 3 \right)
      \left( 3 \alpha_a^2 - 4 \alpha_a \beta_a + 8 \beta_a^2 \right)
      \right]
      E_{\urm{ext}, \, a}^{\urm{sph}}.
      \label{eq:gen_ext_subs}
\end{align}
In the case of the Coulomb potential ($ d = -1 $),
\eqref{eq:gen_ext_subs} reads
\begin{align}
  & \Delta E_{\urm{ext}, \, a}
    \notag \\
  = & \, 
      \left[
      -
      \left( 2 \beta_a - \alpha_a \right)
      +
      \frac{1}{2} \left( 2 \beta_a - \alpha_a \right)^2
      -
      \frac{1}{15}
      \left( 8 \alpha_a^2 - 24 \alpha_a \beta_a + 28 \beta_a^2 - 10 \alpha_a + 20 \beta_a \right)
      \right]
      E_{\urm{ext}, \, a}^{\urm{sph}},
      \label{eq:gen_ext_Coul}
\end{align}
while in the case of the harmonic oscillator potential ($ d = 2 $),
\eqref{eq:gen_ext_subs} reads
\begin{align}
  & \Delta E_{\urm{ext}, \, a}
    \notag \\
  = & \,
      \left[
      -
      \left( 2 \beta_a - \alpha_a \right)
      +
      \frac{1}{2} \left( 2 \beta_a - \alpha_a \right)^2
      +
      \frac{1}{6}
      \left( \alpha_a^2 + 12 \alpha_a \beta_a - 4 \beta_a^2 - 10 \alpha_a + 20 \beta_a \right)
      \right]
      E_{\urm{ext}, \, a}^{\urm{sph}}.
      \label{eq:gen_ext_HO}
\end{align}
\par
\textbf{Interaction energy}: Substituting
\eqref{eq:gen_1density_partialwave} and \eqref{eq:int_partial} into
\eqref{eq:int_def},
one obtains
\begin{align}
  E_{\urm{int}, \, ab}^{\urm{def}}
  \simeq & \,
           c_a c_b
           E_{\urm{int}, \, ab}^{\urm{sph}}
           \notag \\
         & \,
           +
           \frac{c_a}{2}
           \int_0^{\infty}
           \int_0^{\infty}
           F^{\left( 1 \right)}_b \left( r' \right)
           \rho_{\urm{sph}, \, a} \left( r \right)
           \frac{d \rho_{\urm{sph}, \, b} \left( r' \right)}{dr'}
           V_0^{ab} \left( r, r' \right)
           r^2 r'^2 \, dr \, dr'
           \notag \\
         & \, 
           +
           \frac{c_b}{2}
           \int_0^{\infty}
           \int_0^{\infty}
           F^{\left( 1 \right)}_a \left( r \right)
           \frac{d \rho_{\urm{sph}, \, a} \left( r \right)}{dr}
           \rho_{\urm{sph}, \, b} \left( r' \right)
           V_0^{ab} \left( r, r' \right)
           r^2 r'^2 \, dr \, dr'
           \notag \\
         & \,
           +
           \frac{c_a}{2}
           \int_0^{\infty}
           \int_0^{\infty}
           F^{\left( 2 \right)}_b \left( r' \right)
           \rho_{\urm{sph}, \, a} \left( r \right)
           \frac{d^2 \rho_{\urm{sph}, \, b} \left( r' \right)}{dr'^2}
           V_0^{ab} \left( r, r' \right)
           r^2 r'^2 \, dr \, dr'
           \notag \\
         & \, 
           +
           \frac{c_b}{2}
           \int_0^{\infty}
           \int_0^{\infty}
           F^{\left( 2 \right)}_a \left( r \right)
           \frac{d^2 \rho_{\urm{sph}, \, a} \left( r \right)}{dr^2}
           \rho_{\urm{sph}, \, b} \left( r' \right)
           V_0^{ab} \left( r, r' \right)
           r^2 r'^2 \, dr \, dr'
           \notag \\
         & \,
           +
           \frac{1}{8 \pi}
           \int_0^{\infty}
           \int_0^{\infty}
           F^{\left( 1 \right)}_a \left( r \right)
           F^{\left( 1 \right)}_b \left( r' \right)
           \frac{d \rho_{\urm{sph}, \, a} \left( r \right)}{dr}
           \frac{d \rho_{\urm{sph}, \, b} \left( r' \right)}{dr'}
           V_0^{ab} \left( r, r' \right)
           r^2 r'^2 \, dr \, dr'
           \notag \\
         & \,
           +
           \frac{8 \pi}{45}
           \left( \alpha_a + \beta_a \right)
           \left( \alpha_b + \beta_b \right)
           \int_0^{\infty}
           \int_0^{\infty}
           \frac{d \rho_{\urm{sph}, \, a} \left( r \right)}{dr}
           \frac{d \rho_{\urm{sph}, \, b} \left( r' \right)}{dr'}
           V_2^{ab} \left( r, r' \right)
           r^3 r'^3 \, dr \, dr',
           \label{eq:gen_ab_int} 
\end{align}
where
$ c_a = 1 + \left( 2 \beta_a - \alpha_a \right) + \frac{1}{2} \left( 2 \beta_a - \alpha_a \right)^2 $.
\par
In the case of atomic nuclei,
substituting
\eqref{eq:delta} into
\eqref{eq:gen_ab_int},
one obtains
\begin{align}
  E_{\urm{int}, \, ab}^{\urm{def}}
  \simeq & \,
           c_a c_b
           E_{\urm{int}, \, ab}^{\urm{sph}}
           \notag \\
         & \,
           - 
           \frac{\pi c_a g_{ab}}{15}
           \left( 17 \alpha_b^2 - 36 \alpha_b \beta_b + 52 \beta_b^2 - 10 \alpha_b + 20 \beta_b \right)
           \int_0^{\infty}
           \rho_{\urm{sph}, \, a} \left( r \right)
           \frac{d \rho_{\urm{sph}, \, b} \left( r \right)}{dr}
           r^3 \, dr
           \notag \\
         & \,
           -
           \frac{\pi c_b g_{ab}}{15}
           \left( 17 \alpha_a^2 - 36 \alpha_a \beta_a + 52 \beta_a^2 - 10 \alpha_a + 20 \beta_a \right)
           \int_0^{\infty}
           \rho_{\urm{sph}, \, b} \left( r \right)
           \frac{d \rho_{\urm{sph}, \, a} \left( r \right)}{dr}
           r^3 \, dr
           \notag \\
         & \,
           -
           \frac{\pi c_a g_{ab}}{15}
           \left( 3 \alpha_b^2 - 4 \alpha_b \beta_b + 8 \beta_b^2 \right)
           \int_0^{\infty}
           \rho_{\urm{sph}, \, a} \left( r \right)
           \frac{d^2 \rho_{\urm{sph}, \, b} \left( r \right)}{dr^2}
           r^4 \, dr
           \notag \\
         & \,
           -
           \frac{\pi c_b g_{ab}}{15}
           \left( 3 \alpha_a^2 - 4 \alpha_a \beta_a + 8 \beta_a^2 \right)
           \int_0^{\infty}
           \rho_{\urm{sph}, \, b} \left( r \right)
           \frac{d^2 \rho_{\urm{sph}, \, a} \left( r \right)}{dr^2}
           r^4 \, dr
           \notag \\
         & \,
           -
           \frac{\pi g_{ab}}{450}
           \left( 17 \alpha_a^2 - 36 \alpha_a \beta_a + 52 \beta_a^2 - 10 \alpha_a + 20 \beta_a \right)
           \notag \\
         & \,
           \qquad \times
           \left( 17 \alpha_b^2 - 36 \alpha_b \beta_b + 52 \beta_b^2 - 10 \alpha_b + 20 \beta_b \right)
           \int_0^{\infty}
           \frac{d \rho_{\urm{sph}, \, a} \left( r \right)}{dr}
           \frac{d \rho_{\urm{sph}, \, b} \left( r \right)}{dr}
           r^4 \, dr
           \notag \\
         & \,
           -
           \frac{8 \pi g_{ab}}{45}
           \left( \alpha_a + \beta_a \right)
           \left( \alpha_b + \beta_b \right)
           \int_0^{\infty}
           \frac{d \rho_{\urm{sph}, \, a} \left( r \right)}{dr}
           \frac{d \rho_{\urm{sph}, \, b} \left( r \right)}{dr}
           r^4 \, dr.
           \label{eq:gen_int_delta}
\end{align}
\par
In the case of atoms,
substituting
\eqref{eq:Coul} into
\eqref{eq:gen_ab_int},
one obtains
\begin{align}
  E_{\urm{int}, \, ab}^{\urm{def}}
  \simeq & \,
           c_a c_b
           E_{\urm{int}, \, ab}^{\urm{sph}}
           \notag \\
         & \,
           +
           \frac{4 \pi^2 c_a}{15}
           \left( 17 \alpha_b^2 - 36 \alpha_b \beta_b + 52 \beta_b^2 - 10 \alpha_b + 20 \beta_b \right)
           \notag \\
         & \,
           \qquad \times
           \int_0^{\infty} 
           \int_0^{\infty} 
           \rho_{\urm{sph}, \, a} \left( r \right)
           \frac{d \rho_{\urm{sph}, \, b} \left( r' \right)}{dr'}
           \frac{1}{r_{>}}
           r^2 r'^3 \, dr \, dr'
           \notag \\
         & \,
           +
           \frac{4 \pi^2 c_b}{15}
           \left( 17 \alpha_a^2 - 36 \alpha_a \beta_a + 52 \beta_a^2 - 10 \alpha_a + 20 \beta_a \right)
           \notag \\
         & \,
           \qquad \times 
           \int_0^{\infty} 
           \int_0^{\infty} 
           \frac{d \rho_{\urm{sph}, \, a} \left( r \right)}{dr}
           \rho_{\urm{sph}, \, b} \left( r' \right)
           \frac{1}{r_{>}}
           r^3 r'^2 \, dr \, dr'
           \notag \\
         & \,
           +
           \frac{4 \pi^2 c_a}{15}
           \left( 3 \alpha_b^2 - 4 \alpha_b \beta_b + 8 \beta_b^2 \right)
           \int_0^{\infty} 
           \int_0^{\infty} 
           \rho_{\urm{sph}, \, a} \left( r \right)
           \frac{d^2 \rho_{\urm{sph}, \, b} \left( r' \right)}{dr'^2}
           \frac{1}{r_{>}}
           r^2 r'^4 \, dr \, dr'
           \notag \\
         & \,
           +
           \frac{4 \pi^2 c_b}{15}
           \left( 3 \alpha_a^2 - 4 \alpha_a \beta_a + 8 \beta_a^2 \right)
           \int_0^{\infty} 
           \int_0^{\infty} 
           \frac{d^2 \rho_{\urm{sph}, \, a} \left( r \right)}{dr^2}
           \rho_{\urm{sph}, \, b} \left( r' \right)
           \frac{1}{r_{>}}
           r^4 r'^2 \, dr \, dr'
           \notag \\
         & \,
           +
           \frac{2 \pi^2}{225}
           \left( 17 \alpha_a^2 - 36 \alpha_a \beta_a + 52 \beta_a^2 - 10 \alpha_a + 20 \beta_a \right)
           \notag \\
         & \,
           \qquad \times
           \left( 17 \alpha_b^2 - 36 \alpha_b \beta_b + 52 \beta_b^2 - 10 \alpha_b + 20 \beta_b \right)
           \notag \\
         & \,
           \qquad \times
           \int_0^{\infty} 
           \int_0^{\infty} 
           \frac{d \rho_{\urm{sph}, \, a} \left( r \right)}{dr}
           \frac{d \rho_{\urm{sph}, \, b} \left( r' \right)}{dr'}
           \frac{1}{r_{>}}
           r^3 r'^3 \, dr \, dr'
           \notag \\
         & \,
           +
           \frac{32 \pi^2}{225}
           \left( \alpha_a + \beta_a \right)
           \left( \alpha_b + \beta_b \right)
           \int_0^{\infty} 
           \int_0^{\infty} 
           \frac{d \rho_{\urm{sph}, \, a} \left( r \right)}{dr}
           \frac{d \rho_{\urm{sph}, \, b} \left( r' \right)}{dr'}
           \frac{r_{<}^2}{r_{>}^3}
           r^3 r'^3 \, dr \, dr'.
           \label{eq:gen_int_Coul}
\end{align}
\par
Substituting $ \alpha_a = 2 \beta_a $ into equations shown here,
one can obtain equations shown in
section~\ref{sec:model}.
As long as the small deformation is considered, the conclusion shown in
section~\ref{sec:model}
remains.

% 
% Data
% -*- coding: utf-8 -*-
%%%%%%%%%%%%%%%%%%%%%%%%%%%%%%%%%%%%%%%%%%%%%%%%%% 
% 
% Deformation of atoms paper manuscript for J. Phys. B
% Data Part
% 
% Begin to write: 2020-08-03
% 
% Tomoya Naito, Shimpei Endo,
% Kouichi Hagino, and Yusuke Tanimura
% 
%%%%%%%%%%%%%%%%%%%%%%%%%%%%%%%%%%%%%%%%%%%%%%%%%% 
% 
\section{Numerical data}
\label{sec:data}
\par
All the numerical data related to the main part of the paper 
are shown in Tables \ref{tab:Gaussian_UHF_631pG_all}--\ref{tab:Q_Ga_MO}.
\begin{table}[!htb]
  \centering
  \caption{
    The total energies, 
    the root-mean-square radii, $ Z \avr{r^2} $, the quadrupole moments, $ Q $,
    and the deformation parameters, $ \beta $, 
    calculated with the unrestricted Hartree-Fock method with the 6-31+G basis.
    Configurations of the valence electrons calculated by the natural orbital analysis are also shown.
    The Hartree atomic unit is used for total energies, $ Z \avr{r^2} $, and $ Q $.}
  \label{tab:Gaussian_UHF_631pG_all}
  {\tiny
    \begin{tabular}{lrcD{.}{.}{4}D{.}{.}{4}D{.}{.}{4}D{.}{.}{4}D{.}{.}{4}D{.}{.}{4}D{.}{.}{4}D{.}{.}{4}l}
      \br
      \multicolumn{1}{c}{Atom} & \multicolumn{1}{c}{Z} & \multicolumn{1}{c}{Mult.} & \multicolumn{1}{c}{Energy} & \multicolumn{1}{c}{$ Z \avr{r^2} $} & \multicolumn{1}{c}{$ Q_x $} & \multicolumn{1}{c}{$ Q_y $} & \multicolumn{1}{c}{$ Q_z $} & \multicolumn{1}{c}{$ \beta_x $} & \multicolumn{1}{c}{$ \beta_y $} & \multicolumn{1}{c}{$ \beta_z $} & \multicolumn{1}{c}{Config.} \\
      \mr
      Li &  3 & 2 &    -7.4315  & 18.2800 &  0.0000 &  0.0000 &   0.0000 &  0.0000 &  0.0000 &  0.0000 & $ 2s^{ 1.00} $ \\
      Be &  4 & 1 &   -14.5696  & 17.3430 &  0.0000 &  0.0000 &   0.0000 &  0.0000 &  0.0000 &  0.0000 & $ 2s^{ 2.00} $ \\
      B  &  5 & 2 &   -24.5237  & 15.9020 & -2.4718 & -2.4718 &  +4.9435 & -0.1232 & -0.1232 & +0.2464 & $ 2s^{ 2.00} \, 2p^{ 1.00} $ \\
      C  &  6 & 3 &   -37.6809  & 13.8214 & +1.4877 & +1.4877 &  -2.9754 & +0.0853 & +0.0853 & -0.1706 & $ 2s^{ 2.00} \, 2p^{ 2.00} $ \\
      N  &  7 & 4 &   -54.3863  & 12.1756 &  0.0000 &  0.0000 &   0.0000 &  0.0000 &  0.0000 &  0.0000 & $ 2s^{ 2.00} \, 2p^{ 3.00} $ \\
      O  &  8 & 3 &   -74.7835  & 11.3088 & -0.9747 & -0.9747 &  +1.9496 & -0.0683 & -0.0683 & +0.1367 & $ 2s^{ 2.00} \, 2p^{ 3.99} \, 3p^{ 0.01} $ \\
      F  &  9 & 2 &   -99.3675  & 10.3172 & +0.6952 & +0.6952 &  -1.3904 & +0.0534 & +0.0534 & -0.1068 & $ 2s^{ 2.00} \, 2p^{ 5.00} $ \\
      Ne & 10 & 1 &  -128.4835  &  9.4321 &  0.0000 &  0.0000 &   0.0000 &  0.0000 &  0.0000 &  0.0000 & $ 2s^{ 2.00} \, 2p^{ 6.00} $ \\
      \mr
      Na & 11 & 2 &  -161.8414  & 27.1816 &  0.0000 &  0.0000 &   0.0000 &  0.0000 &  0.0000 &  0.0000 & $ 3s^{ 1.00} $ \\
      Mg & 12 & 1 &  -199.5953  & 29.6206 &  0.0000 &  0.0000 &   0.0000 &  0.0000 &  0.0000 &  0.0000 & $ 3s^{ 2.00} $ \\
      Al & 13 & 2 &  -241.8545  & 33.4055 & -5.5694 & -5.5694 & +11.1387 & -0.1322 & -0.1322 & +0.2643 & $ 3s^{ 2.00} \, 3p^{ 1.00} $ \\
      Si & 14 & 3 &  -288.8288  & 32.2974 & +3.5934 & +3.5934 &  -7.1869 & +0.0882 & +0.0882 & -0.1764 & $ 3s^{ 2.00} \, 3p^{ 2.00} $ \\
      P  & 15 & 4 &  -340.6894  & 30.3267 &  0.0000 &  0.0000 &   0.0000 &  0.0000 &  0.0000 &  0.0000 & $ 3s^{ 2.00} \, 3p^{ 3.00} $ \\
      S  & 16 & 3 &  -397.4722  & 29.2999 & -2.3199 & -2.3199 &  +4.6397 & -0.0628 & -0.0628 & +0.1255 & $ 3s^{ 2.00} \, 3p^{ 4.00} $ \\
      Cl & 17 & 2 &  -459.4438  & 27.7141 & +1.7509 & +1.7509 &  -3.5020 & +0.0501 & +0.0501 & -0.1002 & $ 3s^{ 2.00} \, 3p^{ 5.00} $ \\
      Ar & 18 & 1 &  -526.7731  & 26.0836 &  0.0000 &  0.0000 &   0.0000 &  0.0000 &  0.0000 &  0.0000 & $ 3s^{ 2.00} \, 3p^{ 6.00} $ \\
      \mr
      K  & 19 & 2 &  -599.1190  & 51.2038 &  0.0000 &  0.0000 &   0.0000 &  0.0000 &  0.0000 &  0.0000 & $ 4s^{ 1.00} $ \\
      Ca & 20 & 1 &  -676.7079  & 56.6004 &  0.0000 &  0.0000 &   0.0000 &  0.0000 &  0.0000 &  0.0000 & $ 4s^{ 2.00} $ \\
      Sc & 21 & 2 &  -759.6763  & 53.1310 & +0.2094 & +0.2094 &  -0.4189 & +0.0031 & +0.0031 & -0.0062 & $ 4s^{ 1.99} \, 3d^{ 1.00} $ \\
      Ti & 22 & 3 &  -848.3215  & 50.2245 & -0.1626 & -0.1626 &  +0.3250 & -0.0026 & -0.0026 & +0.0051 & $ 4s^{ 1.99} \, 3d^{ 2.00} $ \\
      V  & 23 & 4 &  -942.7941  & 47.6799 &  0.0000 &  0.0000 &   0.0000 &  0.0000 &  0.0000 &  0.0000 & $ 4s^{ 1.99} \, 3d^{ 3.00} \, 5s^{ 0.01} $ \\
      Cr & 24 & 7 & -1043.2424  & 37.9685 &  0.0000 &  0.0000 &   0.0000 &  0.0000 &  0.0000 &  0.0000 & $ 4s^{ 1.00} \, 3d^{ 5.00} $ \\
      Mn & 25 & 6 & -1149.6148  & 37.0228 & +0.7490 & +0.7490 &  -1.4982 & +0.0160 & +0.0160 & -0.0321 & $ 4s^{ 1.00} \, 3d^{ 5.98} \, 4d^{ 0.02} $ \\
      Fe & 26 & 5 & -1262.2043  & 35.7532 & -0.5400 & -0.5400 &  +1.0797 & -0.0120 & -0.0120 & +0.0239 & $ 4s^{ 1.00} \, 3d^{ 6.99} \, 4d^{ 0.01} $ \\
      Co & 27 & 4 & -1381.1643  & 34.5419 &  0.0000 &  0.0000 &   0.0000 &  0.0000 &  0.0000 &  0.0000 & $ 4s^{ 1.00} \, 3d^{ 8.00} $ \\
      Ni & 28 & 3 & -1506.5944  & 33.4315 & -0.4075 & -0.4075 &  +0.8148 & -0.0097 & -0.0097 & +0.0193 & $ 4s^{ 1.00} \, 3d^{ 9.00} $ \\
      Cu & 29 & 2 & -1638.6938  & 32.2616 &  0.0000 &  0.0000 &   0.0000 &  0.0000 &  0.0000 &  0.0000 & $ 4s^{ 1.00} \, 3d^{10.00} $ \\
      Zn & 30 & 1 & -1777.5332  & 35.0221 &  0.0000 &  0.0000 &   0.0000 &  0.0000 &  0.0000 &  0.0000 & $ 4s^{ 2.00} \, 3d^{10.00} $ \\
      Ga & 31 & 2 & -1921.1399  & 41.4244 & -5.5787 & -5.5787 & +11.1575 & -0.1068 & -0.1068 & +0.2135 & $ 4s^{ 2.00} \, 4p^{ 1.00} $ \\
      Ge & 32 & 3 & -2073.0829  & 42.1351 & +3.9175 & +3.9175 &  -7.8353 & +0.0737 & +0.0737 & -0.1474 & $ 4s^{ 2.00} \, 4p^{ 2.00} $ \\
      As & 33 & 4 & -2231.8086  & 41.4771 &  0.0000 &  0.0000 &   0.0000 &  0.0000 &  0.0000 &  0.0000 & $ 4s^{ 2.00} \, 4p^{ 3.00} $ \\
      Se & 34 & 3 & -2397.3677  & 41.6200 & -2.8141 & -2.8141 &  +5.6285 & -0.0536 & -0.0536 & +0.1072 & $ 4s^{ 2.00} \, 4p^{ 4.00} $ \\
      Br & 35 & 2 & -2569.7593  & 40.7722 & +2.2376 & +2.2376 &  -4.4751 & +0.0435 & +0.0435 & -0.0870 & $ 4s^{ 2.00} \, 4p^{ 5.00} $ \\
      Kr & 36 & 1 & -2749.2009  & 39.7020 &  0.0000 &  0.0000 &   0.0000 &  0.0000 &  0.0000 &  0.0000 & $ 4s^{ 2.00} \, 4p^{ 6.00} $ \\
      \br
    \end{tabular}}
\end{table}
\begin{table}[!htb]
  \centering
  \caption{
    Same as Table \ref{tab:Gaussian_UHF_631pG_all}, but with the dAug-CC-pV5Z basis.}
  \label{tab:Gaussian_UHF_dCCpV_all}
  {\tiny
    \begin{tabular}{lrcD{.}{.}{4}D{.}{.}{4}D{.}{.}{4}D{.}{.}{4}D{.}{.}{4}D{.}{.}{4}D{.}{.}{4}D{.}{.}{4}l}
      \br
      \multicolumn{1}{c}{Atom} & \multicolumn{1}{c}{Z} & \multicolumn{1}{c}{Mult.} & \multicolumn{1}{c}{Energy} & \multicolumn{1}{c}{$ Z \avr{r^2} $} & \multicolumn{1}{c}{$ Q_x $} & \multicolumn{1}{c}{$ Q_y $} & \multicolumn{1}{c}{$ Q_z $} & \multicolumn{1}{c}{$ \beta_x $} & \multicolumn{1}{c}{$ \beta_y $} & \multicolumn{1}{c}{$ \beta_z $} & \multicolumn{1}{c}{Config.} \\
      \mr
      Li &  3 & 2 &    -7.4327  & 18.6252 &  0.0000 &  0.0000 &   0.0000 &  0.0000 &  0.0000 &  0.0000 & $ 2s^{ 1.00} $ \\
      Be &  4 & 1 &   -14.5730  & 17.3210 &  0.0000 &  0.0000 &   0.0000 &  0.0000 &  0.0000 &  0.0000 & $ 2s^{ 2.00} $ \\
      B  &  5 & 2 &   -24.5331  & 15.7731 & -2.4912 & -2.4912 &  +4.9823 & -0.1252 & -0.1252 & +0.2504 & $ 2s^{ 1.99} \, 2p^{ 1.00} \, 3d^{ 0.01} $ \\
      C  &  6 & 3 &   -37.6937  & 13.7574 & +1.4846 & +1.4846 &  -2.9689 & +0.0855 & +0.0855 & -0.1711 & $ 2s^{ 1.99} \, 2p^{ 2.00} $ \\
      N  &  7 & 4 &   -54.4045  & 12.0782 &  0.0000 &  0.0000 &   0.0000 &  0.0000 &  0.0000 &  0.0000 & $ 2s^{ 2.00} \, 2p^{ 3.00} $ \\
      O  &  8 & 3 &   -74.8188  & 11.2130 & -0.9517 & -0.9517 &  +1.9037 & -0.0673 & -0.0673 & +0.1346 & $ 2s^{ 2.00} \, 2p^{ 3.99} \, 3p^{ 0.01} $ \\
      F  &  9 & 2 &   -99.4161  & 10.2470 & +0.6738 & +0.6738 &  -1.3476 & +0.0521 & +0.0521 & -0.1042 & $ 2s^{ 2.00} \, 2p^{ 5.00} $ \\
      Ne & 10 & 1 &  -128.5468  &  9.3741 &  0.0000 &  0.0000 &   0.0000 &  0.0000 &  0.0000 &  0.0000 & $ 2s^{ 2.00} \, 2p^{ 6.00} $ \\
      \mr
      Na & 11 & 2 &  -161.8587  & 27.1413 &  0.0000 &  0.0000 &   0.0000 &  0.0000 &  0.0000 &  0.0000 & $ 3s^{ 1.00} $ \\
      Al & 13 & 2 &  -241.8807  & 33.1885 & -5.7634 & -5.7634 & +11.5268 & -0.1377 & -0.1377 & +0.2753 & $ 3s^{ 1.99} \, 3p^{ 1.00} \, 3d^{ 0.01} $ \\
      Si & 14 & 3 &  -288.8588  & 32.1436 & +3.6302 & +3.6302 &  -7.2605 & +0.0895 & +0.0895 & -0.1790 & $ 3s^{ 1.99} \, 3p^{ 2.00} \, 3d^{ 0.01} $ \\
      P  & 15 & 4 &  -340.7192  & 30.2655 &  0.0000 &  0.0000 &   0.0000 &  0.0000 &  0.0000 &  0.0000 & $ 3s^{ 2.00} \, 3p^{ 3.00} $ \\
      S  & 16 & 3 &  -397.5132  & 29.1710 & -2.1695 & -2.1695 &  +4.3391 & -0.0590 & -0.0590 & +0.1179 & $ 3s^{ 1.99} \, 3p^{ 4.00} \, 3d^{ 0.01} $ \\
      Cl & 17 & 2 &  -459.4897  & 27.6256 & +1.6258 & +1.6258 &  -3.2517 & +0.0466 & +0.0466 & -0.0933 & $ 3s^{ 2.00} \, 3p^{ 5.00} $ \\
      Ar & 18 & 1 &  -526.8173  & 26.0376 &  0.0000 &  0.0000 &   0.0000 &  0.0000 &  0.0000 &  0.0000 & $ 3s^{ 2.00} \, 3p^{ 6.00} $ \\
      \mr
      Sc & 21 & 2 &  -759.7408  & 53.1077 & +0.1976 & +0.1976 &  -0.3955 & +0.0029 & +0.0029 & -0.0059 & $ 4s^{ 1.99} \, 3d^{ 1.00} $ \\
      Ti & 22 & 3 &  -848.4026  & 50.2023 & -0.1508 & -0.1508 &  +0.3016 & -0.0024 & -0.0024 & +0.0048 & $ 4s^{ 2.00} \, 3d^{ 2.00} $ \\
      V  & 23 & 4 &  -942.8935  & 47.6556 &  0.0000 &  0.0000 &   0.0000 &  0.0000 &  0.0000 &  0.0000 & $ 4s^{ 1.99} \, 3d^{ 3.00} \, 5s^{ 0.01} $ \\
      Cr & 24 & 7 & -1043.3567  & 37.6585 &  0.0000 &  0.0000 &   0.0000 &  0.0000 &  0.0000 &  0.0000 & $ 4s^{ 1.00} \, 3d^{ 5.00} $ \\
      Mn & 25 & 6 & -1149.7533  & 36.7555 & +0.6276 & +0.6276 &  -1.2551 & +0.0135 & +0.0135 & -0.0271 & $ 4s^{ 1.00} \, 3d^{ 5.99} \, 4d^{ 0.01} $ \\
      Fe & 26 & 5 & -1262.3743  & 35.4942 & -0.4318 & -0.4318 &  +0.8636 & -0.0096 & -0.0096 & +0.0193 & $ 4s^{ 1.00} \, 3d^{ 6.99} \, 4d^{ 0.01} $ \\
      Co & 27 & 4 & -1381.3674  & 34.2853 &  0.0000 &  0.0000 &   0.0000 &  0.0000 &  0.0000 &  0.0000 & $ 4s^{ 1.00} \, 3d^{ 8.00} $ \\
      Ni & 28 & 3 & -1506.8305  & 33.1827 & +0.2806 & +0.3348 &  -0.6154 & +0.0067 & +0.0080 & -0.0147 & $ 4s^{ 1.00} \, 3d^{ 9.00} $ \\
      Cu & 29 & 2 & -1638.9574  & 36.3225 & -0.0156 & -0.2304 &  +0.2462 &  0.0003 & -0.0050 & +0.0054 & $ 4s^{ 2.00} \, 3d^{ 9.00} $ \\
      Zn & 30 & 1 & -1777.8480  & 34.9792 &  0.0000 &  0.0000 &   0.0000 &  0.0000 &  0.0000 &  0.0000 & $ 4s^{ 2.00} \, 3d^{10.00} $ \\
      Ga & 31 & 2 & -1923.2644  & 40.7328 & -5.6952 & -5.6952 &  +11.3903 & -0.1108 & -0.1108 & +0.2217 & $ 4s^{ 1.99} \, 4p^{ 1.00} \, 4d^{ 0.01} $ \\
      Ge & 32 & 3 & -2075.3638  & 41.5004 & +3.9233 & +3.9233 &  -7.8466 & +0.0749 & +0.0749 & -0.1499 & $ 4s^{ 1.99} \, 4p^{ 2.00} \, 4d^{ 0.01} $ \\
      As & 33 & 4 & -2234.2398  & 41.0402 &  0.0000 &  0.0000 &   0.0000 &  0.0000 &  0.0000 &  0.0000 & $ 4s^{ 2.00} \, 4p^{ 3.00} $ \\
      Se & 34 & 3 & -2399.8753  & 41.1769 & -2.5759 & -2.5759 &  +5.1518 & -0.0496 & -0.0496 & +0.0992 & $ 4s^{ 2.00} \, 4p^{ 4.00} $ \\
      Br & 35 & 2 & -2572.4483  & 40.5058 & +2.0362 & +2.0362 &  -4.0723 & +0.0398 & +0.0398 & -0.0797 & $ 4s^{ 2.00} \, 4p^{ 5.00} $ \\
      Kr & 36 & 1 & -2752.0549  & 39.5314 &  0.0000 &  0.0000 &   0.0000 &  0.0000 &  0.0000 &  0.0000 & $ 4s^{ 2.00} \, 4p^{ 6.00} $ \\
      \br
    \end{tabular}}
\end{table}
\begin{table}[!htb]
  \centering
  \caption{
    Same as Table \ref{tab:Gaussian_UHF_631pG_all}, but with the STO-3G basis.}
  \label{tab:Gaussian_UHF_STO_all}
  {\tiny
    \begin{tabular}{lrcD{.}{.}{4}D{.}{.}{4}D{.}{.}{4}D{.}{.}{4}D{.}{.}{4}D{.}{.}{4}D{.}{.}{4}D{.}{.}{4}l}
      \br
      \multicolumn{1}{c}{Atom} & \multicolumn{1}{c}{Z} & \multicolumn{1}{c}{Mult.} & \multicolumn{1}{c}{Energy} & \multicolumn{1}{c}{$ Z \avr{r^2} $} & \multicolumn{1}{c}{$ Q_x $} & \multicolumn{1}{c}{$ Q_y $} & \multicolumn{1}{c}{$ Q_z $} & \multicolumn{1}{c}{$ \beta_x $} & \multicolumn{1}{c}{$ \beta_y $} & \multicolumn{1}{c}{$ \beta_z $} & \multicolumn{1}{c}{Config.} \\
      \mr
      Li &  3 & 2 &    -7.3155   & 13.0955 &  0.0000 &  0.0000 &  0.0000 &  0.0000 &  0.0000 &  0.0000 & $ 2s^{ 1.00} $ \\
      Be &  4 & 1 &   -14.3519   & 12.3657 &  0.0000 &  0.0000 &  0.0000 &  0.0000 &  0.0000 &  0.0000 & $ 2s^{ 2.00} $ \\
      B  &  5 & 2 &   -24.1490   & 10.6497 & -1.3369 & -1.3369 & +2.6738 & -0.0995 & -0.0995 & +0.1990 & $ 2s^{ 2.00} \, 2p^{ 1.00} $ \\
      C  &  6 & 3 &   -37.1984   & 10.5788 & +1.0169 & +1.0169 & -2.0335 & +0.0762 & +0.0762 & -0.1524 & $ 2s^{ 2.00} \, 2p^{ 2.00} $ \\
      N  &  7 & 4 &   -53.7190   & 10.1785 &  0.0000 &  0.0000 &  0.0000 &  0.0000 &  0.0000 &  0.0000 & $ 2s^{ 2.00} \, 2p^{ 3.00} $ \\
      O  &  8 & 3 &   -73.8042   &  9.1334 & -0.5942 & -0.5942 & +1.1884 & -0.0516 & -0.0516 & +0.1031 & $ 2s^{ 2.00} \, 2p^{ 4.00} $ \\
      F  &  9 & 2 &   -97.9865   &  8.2689 & +0.4626 & +0.4626 & -0.9252 & +0.0443 & +0.0443 & -0.0887 & $ 2s^{ 2.00} \, 2p^{ 5.00} $ \\
      Ne & 10 & 1 &  -126.6045   &  7.3940 &  0.0000 &  0.0000 &  0.0000 &  0.0000 &  0.0000 &  0.0000 & $ 2s^{ 2.00} \, 2p^{ 6.00} $ \\
      \mr
      Na & 11 & 2 &  -159.6684   & 12.1291 &  0.0000 &  0.0000 &  0.0000 &  0.0000 &  0.0000 &  0.0000 & $ 3s^{ 1.00} $ \\
      Mg & 12 & 1 &  -197.0074   & 15.6258 &  0.0000 &  0.0000 &  0.0000 &  0.0000 &  0.0000 &  0.0000 & $ 3s^{ 2.00} $ \\
      Al & 13 & 2 &  -238.8584   & 19.1823 & -2.0286 & -2.0286 & +4.0571 & -0.0838 & -0.0838 & +0.1677 & $ 3s^{ 2.00} \, 3p^{ 1.00} $ \\
      Si & 14 & 3 &  -285.4662   & 22.0780 & +1.9050 & +1.9050 & -3.8102 & +0.0684 & +0.0684 & -0.1368 & $ 3s^{ 2.00} \, 3p^{ 2.00} $ \\
      P  & 15 & 4 &  -336.8688   & 22.6495 &  0.0000 &  0.0000 &  0.0000 &  0.0000 &  0.0000 &  0.0000 & $ 3s^{ 2.00} \, 3p^{ 3.00} $ \\
      S  & 16 & 3 &  -393.1302   & 22.8523 & -1.3907 & -1.3907 & +2.7811 & -0.0482 & -0.0482 & +0.0965 & $ 3s^{ 2.00} \, 3p^{ 4.00} $ \\
      Cl & 17 & 2 &  -454.5422   & 24.7260 & +1.3180 & +1.3180 & -2.6361 & +0.0423 & +0.0423 & -0.0845 & $ 3s^{ 2.00} \, 3p^{ 5.00} $ \\
      Ar & 18 & 1 &  -521.2229   & 22.9344 &  0.0000 &  0.0000 &  0.0000 &  0.0000 &  0.0000 &  0.0000 & $ 3s^{ 2.00} \, 3p^{ 6.00} $ \\
      \mr
      K  & 19 & 2 &  -593.0778   & 33.1098 &  0.0000 &  0.0000 &  0.0000 &  0.0000 &  0.0000 &  0.0000 & $ 4s^{ 1.00} $ \\
      Ca & 20 & 1 &  -669.9889   & 42.2174 &  0.0000 &  0.0000 &  0.0000 &  0.0000 &  0.0000 &  0.0000 & $ 4s^{ 2.00} $ \\
      Sc & 21 & 2 &  -751.9936   & 44.3275 & +2.4323 & +2.4323 & -4.8643 & +0.0435 & +0.0435 & -0.0870 & $ 4s^{ 1.98} \, 3d^{ 1.02} $ \\
      Ti & 22 & 3 &  -839.5552   & 36.4583 & -0.8418 & -0.8418 & +1.6833 & -0.0183 & -0.0183 & +0.0366 & $ 4s^{ 1.99} \, 3d^{ 2.01} $ \\
      V  & 23 & 4 &  -932.3822   & 51.8131 &  0.0000 &  0.0000 &  0.0000 &  0.0000 &  0.0000 &  0.0000 & $ 4s^{ 2.00} \, 4p^{ 3.00} $ \\
      Cr & 24 & 7 & -1031.9037  & 43.3117 & -0.4164 & -0.4164 & +0.8326 & -0.0076 & -0.0076 & +0.0152 & $ 4s^{ 1.00} \, 3d^{ 2.00} \, 4p^{ 3.00} $ \\
      Mn & 25 & 6 & -1137.6484  & 31.6207 &  0.0000 &  0.0000 &  0.0000 &  0.0000 &  0.0000 &  0.0000 & $ 4s^{ 2.00} \, 3d^{ 5.00} $ \\
      Fe & 26 & 5 & -1249.0414  & 41.1316 & -3.8513 & -3.8513 & +7.7028 & -0.0742 & -0.0742 & +0.1484 & $ 4s^{ 2.00} \, 3d^{ 5.00} \, 4p^{ 1.00} $ \\
      Co & 27 & 4 & -1366.3895  & 48.9607 & +3.8432 & +3.8432 & -7.6865 & +0.0622 & +0.0622 & -0.1244 & $ 4s^{ 2.00} \, 3d^{ 5.00} \, 4p^{ 2.00} $ \\
      Ni & 28 & 3 & -1490.1299  & 46.3092 & +3.8145 & +3.8145 & -7.6292 & +0.0653 & +0.0653 & -0.1306 & $ 4s^{ 2.00} \, 3d^{ 6.00} \, 4p^{ 2.00} $ \\
      Cu & 29 & 2 & -1620.3878  & 37.6504 & -3.5702 & -3.5702 & +7.1405 & -0.0752 & -0.0752 & +0.1503 & $ 4s^{ 2.00} \, 3d^{ 8.00} \, 4p^{ 1.00} $ \\
      Zn & 30 & 1 & -1757.1763  & 35.4563 & -4.8110 & -4.8110 & +9.6221 & -0.1076 & -0.1076 & +0.2151 & $ 4s^{ 2.00} \, 3d^{ 8.00} \, 4p^{ 2.00} $ \\
      Ga & 31 & 2 & -1900.7285  & 31.3321 & -2.8420 & -2.8420 & +5.6842 & -0.0719 & -0.0719 & +0.1438 & $ 4s^{ 2.00} \, 4p^{ 1.00} $ \\
      Ge & 32 & 3 & -2051.6363  & 32.0788 & +2.3143 & +2.3143 & -4.6288 & +0.0572 & +0.0572 & -0.1144 & $ 4s^{ 2.00} \, 4p^{ 2.00} $ \\
      As & 33 & 4 & -2209.2637  & 33.7744 &  0.0000 &  0.0000 &  0.0000 &  0.0000 &  0.0000 &  0.0000 & $ 4s^{ 2.00} \, 4p^{ 3.00} $ \\
      Se & 34 & 3 & -2373.5273  & 35.4117 & -1.8814 & -1.8814 & +3.7627 & -0.0421 & -0.0421 & +0.0842 & $ 4s^{ 2.00} \, 4p^{ 4.00} $ \\
      Br & 35 & 2 & -2544.6368  & 35.2509 & +1.6414 & +1.6414 & -3.2825 & +0.0369 & +0.0369 & -0.0738 & $ 4s^{ 2.00} \, 4p^{ 5.00} $ \\
      Kr & 36 & 1 & -2722.7060  & 34.8481 &  0.0000 &  0.0000 &  0.0000 &  0.0000 &  0.0000 &  0.0000 & $ 4s^{ 2.00} \, 4p^{ 6.00} $ \\
      \mr
      Rb & 37 & 2 & -2907.6043  & 44.3734 &  0.0000 &  0.0000 &  0.0000 &  0.0000 &  0.0000 &  0.0000 & $ 5s^{ 1.00} $ \\
      Sr & 38 & 1 & -3099.1167  & 51.3762 &  0.0000 &  0.0000 &  0.0000 &  0.0000 &  0.0000 &  0.0000 & $ 5s^{ 2.00} $ \\
      Y  & 39 & 2 & -3297.3418  & 58.4961 & +0.9769 & +0.9769 & -1.9539 & +0.0132 & +0.0132 & -0.0265 & $ 5s^{ 1.72} \, 4d^{ 1.28} $ \\
      Zr & 40 & 3 & -3503.0360  & 53.2014 & -1.2644 & -1.2644 & +2.5289 & -0.0188 & -0.0188 & +0.0377 & $ 5s^{ 1.81} \, 4d^{ 2.19} $ \\
      Nb & 41 & 6 & -3715.8900  & 45.9028 & +1.1380 & +1.1380 & -2.2759 & +0.0197 & +0.0197 & -0.0393 & $ 5s^{ 1.00} \, 4d^{ 4.00} $ \\
      Mo & 42 & 7 & -3935.8966  & 43.4772 &  0.0000 &  0.0000 &  0.0000 &  0.0000 &  0.0000 &  0.0000 & $ 5s^{ 1.00} \, 4d^{ 5.00} $ \\
      Tc & 43 & 6 & -4163.2052  & 49.5440 &  0.0000 &  0.0000 &  0.0000 &  0.0000 &  0.0000 &  0.0000 & $ 5s^{ 2.00} \, 4d^{ 5.00} $ \\
      Ru & 44 & 5 & -4397.7215  & 50.7468 & +0.6444 & +0.6444 & -1.2887 & +0.0101 & +0.0101 & -0.0201 & $ 5s^{ 2.00} \, 4d^{ 6.00} $ \\
      Rh & 45 & 4 & -4639.7207  & 49.6014 & -0.5554 & -0.5554 & +1.1108 & -0.0089 & -0.0089 & +0.0178 & $ 5s^{ 2.00} \, 4d^{ 7.00} $ \\
      Pd & 46 & 1 & -4889.1921  & 48.0030 & +0.9705 & +0.9705 & -1.9409 & +0.0160 & +0.0160 & -0.0321 & $ 5s^{ 2.00} \, 4d^{ 8.00} $ \\
      Ag & 47 & 2 & -5146.5464  & 47.5080 & -0.6013 & -0.3270 & +0.9283 & -0.0100 & -0.0055 & +0.0155 & $ 5s^{ 2.00} \, 4d^{ 9.00} $ \\
      Cd & 48 & 1 & -5411.5327  & 41.7214 &  0.0000 &  0.0000 &  0.0000 &  0.0000 &  0.0000 &  0.0000 & $ 5s^{ 2.00} \, 4d^{10.00} $ \\
      In & 49 & 2 & -5682.7774  & 47.0137 & -3.2899 & -3.2899 & +6.5798 & -0.0555 & -0.0555 & +0.1109 & $ 5s^{ 2.00} \, 5p^{ 1.00} $ \\
      Sn & 50 & 3 & -5963.2061  & 50.2820 & +2.9919 & +2.9919 & -5.9838 & +0.0472 & +0.0472 & -0.0943 & $ 5s^{ 2.00} \, 5p^{ 2.00} $ \\
      Sb & 51 & 4 & -6251.3625  & 54.2873 &  0.0000 &  0.0000 &  0.0000 &  0.0000 &  0.0000 &  0.0000 & $ 5s^{ 2.00} \, 5p^{ 3.00} $ \\
      Te & 52 & 3 & -6547.1224  & 56.9144 & -2.6502 & -2.6502 & +5.3004 & -0.0369 & -0.0369 & +0.0738 & $ 5s^{ 2.00} \, 5p^{ 4.00} $ \\
      I  & 53 & 2 & -6850.6762  & 57.2039 & +2.3607 & +2.3607 & -4.7214 & +0.0327 & +0.0327 & -0.0654 & $ 5s^{ 2.00} \, 5p^{ 5.00} $ \\
      Xe & 54 & 1 & -7162.1042  & 56.8048 &  0.0000 &  0.0000 &  0.0000 &  0.0000 &  0.0000 &  0.0000 & $ 5s^{ 2.00} \, 5p^{ 6.00} $ \\
      \br
    \end{tabular}}
\end{table}
\begin{landscape}
  \begin{table}[!htb]
    \centering
    \caption{
      The total energies, the root-mean radii, $ Z \avr{r^2} $, the quadrupole moments, $ Q $,
      and the deformation parameters, $ \beta $, 
      for $ \mathrm{Al} $, $ \mathrm{Cu} $, and $ \mathrm{Ga} $ 
      calculated with the several many-body calculation methods with the 6-31+G basis.
      Configurations of the valence electrons calculated by the natural orbital analysis are also shown.
      The Hartree atomic unit is used for total energies, $ Z \avr{r^2} $, and $ Q $.}
    \label{tab:Gaussian_631pG_Al_Cu_Ga}
    \begin{indented}
    \item[]
      {\tiny
        \begin{tabular}{llD{.}{.}{4}D{.}{.}{4}D{.}{.}{4}D{.}{.}{4}D{.}{.}{4}D{.}{.}{4}D{.}{.}{4}D{.}{.}{4}l}
          \br
          \multicolumn{1}{c}{Atom} & \multicolumn{1}{c}{Method} & \multicolumn{1}{c}{Energy} & \multicolumn{1}{c}{$ Z \avr{r^2} $} & \multicolumn{1}{c}{$ Q_x $} & \multicolumn{1}{c}{$ Q_y $} & \multicolumn{1}{c}{$ Q_z $} & \multicolumn{1}{c}{$ \beta_x $} & \multicolumn{1}{c}{$ \beta_y $} & \multicolumn{1}{c}{$ \beta_z $} & \multicolumn{1}{c}{Config.} \\
          \mr
          Al & UHF     &  -241.8545   & 33.4055 & -5.5694 & -5.5694 & +11.1387 & -0.1322 & -0.1322 & +0.2643 & $ 3s^{2.00} \, 3p^{1.00} \, 4s^{0.00} \, 4p^{0.00} $ \\
          Al & CCD     &  -241.8839   & 33.8409 & -5.7148 & -5.7148 & +11.4296 & -0.1339 & -0.1339 & +0.2677 & $ 3s^{1.91} \, 3p^{1.07} \, 4s^{0.01} \, 4p^{0.02} $ \\
          Al & CCSD    &  -241.8841   & 33.8935 & -5.7422 & -5.7422 & +11.4842 & -0.1343 & -0.1343 & +0.2686 & $ 3s^{1.91} \, 3p^{1.07} \, 4s^{0.01} \, 4p^{0.02} $ \\
          Al & CID     &  -241.8839   & 33.8409 & -5.7148 & -5.7148 & +11.4296 & -0.1339 & -0.1339 & +0.2677 & $ 3s^{1.91} \, 3p^{1.07} \, 4s^{0.01} \, 4p^{0.02} $ \\
          Al & CISD    &  -241.8840   & 33.8814 & -5.7367 & -5.7367 & +11.4733 & -0.1342 & -0.1342 & +0.2684 & $ 3s^{1.91} \, 3p^{1.07} \, 4s^{0.01} \, 4p^{0.02} $ \\
          Al & MP2     &  -241.8717   & 33.7371 & -5.7126 & -5.7126 & +11.4254 & -0.1342 & -0.1342 & +0.2684 & $ 3s^{1.97} \, 3p^{1.02} \, 4s^{0.00} \, 4p^{0.01} $ \\
          Al & MP3     &  -241.8791   & 33.8428 & -5.7476 & -5.7476 & +11.4952 & -0.1346 & -0.1346 & +0.2692 & $ 3s^{1.94} \, 3p^{1.04} \, 4s^{0.00} \, 4p^{0.01} $ \\
          Al & MP4     &  -241.8822   & 33.8727 & -5.7491 & -5.7491 & +11.4983 & -0.1345 & -0.1345 & +0.2691 & $ 3s^{1.92} \, 3p^{1.06} \, 4s^{0.01} \, 4p^{0.01} $ \\
          Al & LDA     &  -241.2799   & 32.8660 & -5.5932 & -5.5932 & +11.1867 & -0.1349 & -0.1349 & +0.2698 & $ 3s^{2.00} \, 3p^{1.00} \, 4s^{0.00} \, 4p^{0.00} $ \\
          \mr
          Cu & UHF     & -1638.6938  & 32.2616 &  0.0000 &  0.0000 &   0.0000 &  0.0000 &  0.0000 &  0.0000 & $ 3d^{10.00} \, 4s^{1.00} \, 4p^{0.00} \, 4d^{0.00} $ \\
          Cu & CCD     & -1638.9170  & 32.2339 &  0.0000 &  0.0000 &   0.0000 &  0.0000 &  0.0000 &  0.0000 & $ 3d^{ 9.91} \, 4s^{0.99} \, 4p^{0.02} \, 4d^{0.07} $ \\
          Cu & CCSD    & -1638.9226  & 32.3642 &  0.0000 &  0.0000 &   0.0000 &  0.0000 &  0.0000 &  0.0000 & $ 3d^{ 9.90} \, 4s^{0.99} \, 4p^{0.02} \, 4d^{0.08} $ \\
          Cu & CID     & -1638.9035  & 32.2233 &  0.0000 &  0.0000 &   0.0000 &  0.0000 &  0.0000 &  0.0000 & $ 3d^{ 9.93} \, 4s^{0.99} \, 4p^{0.01} \, 4d^{0.06} $ \\
          Cu & CISD    & -1638.9101  & 32.2921 &  0.0000 &  0.0000 &   0.0000 &  0.0000 &  0.0000 &  0.0000 & $ 3d^{ 9.91} \, 4s^{0.99} \, 4p^{0.02} \, 4d^{0.07} $ \\
          Cu & MP2     & -1638.9657  & 32.4330 &  0.0000 &  0.0000 &   0.0000 &  0.0000 &  0.0000 &  0.0000 & $ 3d^{ 9.88} \, 4s^{1.00} \, 4p^{0.01} \, 4d^{0.12} $ \\
          Cu & MP3     & -1638.8891  & 32.1425 &  0.0000 &  0.0000 &   0.0000 &  0.0000 &  0.0000 &  0.0000 & $ 3d^{ 9.95} \, 4s^{0.99} \, 4p^{0.01} \, 4d^{0.04} $ \\
          Cu & MP4     & -1638.9490  & 32.5433 &  0.0000 &  0.0000 &   0.0000 &  0.0000 &  0.0000 &  0.0000 & $ 3d^{ 9.84} \, 4s^{1.00} \, 4p^{0.02} \, 4d^{0.14} $ \\
          Cu & LDA     & -1637.4831  & 31.5039 &  0.0000 &  0.0000 &   0.0000 &  0.0000 &  0.0000 &  0.0000 & $ 3d^{10.00} \, 4s^{1.00} \, 4p^{0.00} \, 4d^{0.00} $ \\
          \mr
          Ga & UHF     & -1921.1399  & 41.4244 & -5.5787 & -5.5787 & +11.1575 & -0.1068 & -0.1068 & +0.2135 & $ 4s^{2.00} \, 4p^{1.00} \, 5s^{0.00} \, 5p^{0.00} $ \\
          Ga & CCD     & -1921.1686  & 41.7576 & -5.6909 & -5.6909 & +11.3819 & -0.1080 & -0.1080 & +0.2161 & $ 4s^{1.92} \, 4p^{1.06} \, 5s^{0.01} \, 5p^{0.01} $ \\
          Ga & CCSD    & -1921.1688  & 41.8071 & -5.7146 & -5.7146 & +11.4292 & -0.1083 & -0.1083 & +0.2167 & $ 4s^{1.92} \, 4p^{1.06} \, 5s^{0.01} \, 5p^{0.02} $ \\
          Ga & CID     & -1921.1686  & 41.7576 & -5.6909 & -5.6909 & +11.3819 & -0.1080 & -0.1080 & +0.2161 & $ 4s^{1.92} \, 4p^{1.06} \, 5s^{0.01} \, 5p^{0.01} $ \\
          Ga & CISD    & -1921.1688  & 41.7947 & -5.7092 & -5.7092 & +11.4187 & -0.1083 & -0.1083 & +0.2166 & $ 4s^{1.92} \, 4p^{1.06} \, 5s^{0.01} \, 5p^{0.02} $ \\
          Ga & MP2     & -1921.1576  & 41.6788 & -5.6963 & -5.6963 & +11.3923 & -0.1083 & -0.1083 & +0.2167 & $ 4s^{1.97} \, 4p^{1.02} \, 5s^{0.00} \, 5p^{0.01} $ \\
          Ga & MP3     & -1921.1646  & 41.7557 & -5.7190 & -5.7190 & +11.4381 & -0.1086 & -0.1086 & +0.2171 & $ 4s^{1.95} \, 4p^{1.04} \, 5s^{0.00} \, 5p^{0.01} $ \\
          Ga & MP4     & -1921.1673  & 41.7795 & -5.7177 & -5.7177 & +11.4356 & -0.1085 & -0.1085 & +0.2170 & $ 4s^{1.93} \, 4p^{1.05} \, 5s^{0.01} \, 5p^{0.01} $ \\
          Ga & LDA     & -1919.7297  & 40.5292 & -5.6550 & -5.6550 & +11.3100 & -0.1106 & -0.1106 & +0.2212 &  $ 4s^{2.00} \, 4p^{1.00} \, 5s^{0.00} \, 5p^{0.00} $ \\
          \br
        \end{tabular}}
    \end{indented}
  \end{table}
  \begin{table}[!htb]
    \centering
    \caption{
      Same as Table \ref{tab:Gaussian_631pG_Al_Cu_Ga}, but with the dAug-CC-pV5Z basis.}
    \label{tab:Gaussian_dCCpV_Al_Cu_Ga}
    \begin{indented}
    \item[]
      {\tiny
        \begin{tabular}{llD{.}{.}{4}D{.}{.}{4}D{.}{.}{4}D{.}{.}{4}D{.}{.}{4}D{.}{.}{4}D{.}{.}{4}D{.}{.}{4}l}
          \br
          \multicolumn{1}{c}{Atom} & \multicolumn{1}{c}{Method} & \multicolumn{1}{c}{Energy} & \multicolumn{1}{c}{$ Z \avr{r^2} $} & \multicolumn{1}{c}{$ Q_x $} & \multicolumn{1}{c}{$ Q_y $} & \multicolumn{1}{c}{$ Q_z $} & \multicolumn{1}{c}{$ \beta_x $} & \multicolumn{1}{c}{$ \beta_y $} & \multicolumn{1}{c}{$ \beta_z $} & \multicolumn{1}{c}{Config.} \\
          \mr
          Al & UHF     &  -241.8807  & 33.1885 & -5.7634 & -5.7634 & +11.5268 & -0.1377 & -0.1377 & +0.2753 & $ 3s^{1.99} \, 3p^{1.00} \, 3d^{0.01} \, 4p^{0.00} $ \\
          Al & CCD     &  -241.9324  & 32.5801 & -5.2419 & -5.2419 & +10.4841 & -0.1275 & -0.1275 & +0.2551 & $ 3s^{1.90} \, 3p^{1.05} \, 3d^{0.03} \, 4p^{0.01} $ \\
          Al & CCSD    &  -241.9332  & 32.5620 & -5.2529 & -5.2529 & +10.5055 & -0.1279 & -0.1279 & +0.2557 & $ 3s^{1.89} \, 3p^{1.05} \, 3d^{0.04} \, 4p^{0.01} $ \\
          Al & CID     &  -241.9324  & 32.5801 & -5.2419 & -5.2419 & +10.4841 & -0.1275 & -0.1275 & +0.2551 & $ 3s^{1.90} \, 3p^{1.05} \, 3d^{0.03} \, 4p^{0.01} $ \\
          Al & CISD    &  -241.9333  & 32.5568 & -5.2513 & -5.2513 & +10.5029 & -0.1279 & -0.1279 & +0.2557 & $ 3s^{1.89} \, 3p^{1.06} \, 3d^{0.04} \, 4p^{0.01} $ \\
          Al & MP2     &  -241.9199  & 32.8683 & -5.4066 & -5.4066 & +10.8129 & -0.1304 & -0.1304 & +0.2608 & $ 3s^{1.95} \, 3p^{1.02} \, 3d^{0.02} \, 4p^{0.01} $ \\
          Al & MP3     &  -241.9283  & 32.7040 & -5.2941 & -5.2941 & +10.5885 & -0.1283 & -0.1283 & +0.2566 & $ 3s^{1.92} \, 3p^{1.03} \, 3d^{0.03} \, 4p^{0.01} $ \\
          Al & MP4     &  -241.9314  & 32.6312 & -5.2596 & -5.2596 & +10.5194 & -0.1278 & -0.1278 & +0.2555 & $ 3s^{1.91} \, 3p^{1.04} \, 3d^{0.03} \, 4p^{0.01} $ \\
          Al & LDA     &  -241.3109  & 32.9046 & -5.4817 & -5.4817 & +10.9637 & -0.1321 & -0.1321 & +0.2641 & $ 3s^{2.00} \, 3p^{1.00} \, 3d^{0.00} \, 4p^{0.00} $ \\
          \mr
          Cu & UHF     & -1638.9574  & 36.3225 & -0.0156 & -0.2304 &  +0.2462 & -0.0003 & -0.0050 & +0.0054 & $ 3d^{ 9.00} \, 4s^{2.00} \, 4p^{0.00} \, 4d^{0.00} \, 4f^{0.00} \, 5s^{0.00} \, 5p^{0.00} $ \\
          Cu & CCD     & -1639.3940  & 34.4852 & +0.1256 & -0.2824 &  +0.1568 & +0.0029 & -0.0065 & +0.0036 & $ 3d^{ 8.91} \, 4s^{1.91} \, 4p^{0.09} \, 4d^{0.05} \, 4f^{0.02} \, 5s^{0.00} \, 5p^{0.01} $ \\
          Cu & CCSD    & -1639.3988  & 34.7050 & +0.0187 & -0.3167 &  +0.2980 & +0.0004 & -0.0072 & +0.0068 & $ 3d^{ 8.91} \, 4s^{1.90} \, 4p^{0.09} \, 4d^{0.05} \, 4f^{0.02} \, 5s^{0.01} \, 5p^{0.01} $ \\
          Cu & CID     & -1639.3686  & 34.6404 & +0.0560 & -0.2844 &  +0.2284 & +0.0013 & -0.0065 & +0.0052 & $ 3d^{ 8.93} \, 4s^{1.96} \, 4p^{0.04} \, 4d^{0.04} \, 4f^{0.02} \, 5s^{0.00} \, 5p^{0.00} $ \\
          Cu & CISD    & -1639.3717  & 34.6775 & +0.0999 & -0.2904 &  +0.1905 & +0.0023 & -0.0066 & +0.0044 & $ 3d^{ 8.92} \, 4s^{1.96} \, 4p^{0.04} \, 4d^{0.04} \, 4f^{0.02} \, 5s^{0.00} \, 5p^{0.00} $ \\
          Cu & MP2     & -1639.4302  & 34.3392 & -0.1485 & -0.2480 &  +0.3968 & -0.0034 & -0.0057 & +0.0092 & $ 3d^{ 8.89} \, 4s^{1.95} \, 4p^{0.05} \, 4d^{0.07} \, 4f^{0.02} \, 5s^{0.00} \, 5p^{0.00} $ \\
          Cu & MP3     & -1639.3866  & 34.4844 & -0.0161 & -0.2404 &  +0.2565 & -0.0004 & -0.0055 & +0.0059 & $ 3d^{ 8.92} \, 4s^{1.92} \, 4p^{0.08} \, 4d^{0.04} \, 4f^{0.02} \, 5s^{0.00} \, 5p^{0.01} $ \\
          Cu & MP4     & -1639.4025  & 34.5798 & +0.0410 & -0.2632 &  +0.2224 & +0.0009 & -0.0060 & +0.0051 & $ 3d^{ 8.90} \, 4s^{1.91} \, 4p^{0.09} \, 4d^{0.06} \, 4f^{0.02} \, 5s^{0.00} \, 5p^{0.01} $ \\
          Cu & LDA     & -1637.7692  & 31.2608 &  0.0000 &  0.0000 &  0.0000 &  0.0000 &  0.0000 &  0.0000 & $ 3d^{10.00} \, 4s^{1.00} \, 4p^{0.00} \, 4d^{0.00} \, 4f^{0.00} \, 5s^{0.00} \, 5p^{0.00} $ \\
          \mr
          Ga & UHF     & -1923.2644  & 40.7328 & -5.6952 & -5.6952 & +11.3903 & -0.1108 & -0.1108 & +0.2217 & $ 4s^{1.99} \, 4p^{1.00} \, 4d^{0.01} \, 4f^{0.00} \, 5s^{0.00} \, 5p^{0.00} \, 5d^{0.00} $ \\
          Ga & CCD     & -1923.5235  & 39.7317 & -5.1260 & -5.1260 & +10.2519 & -0.1023 & -0.1023 & +0.2045 & $ 4s^{1.92} \, 4p^{1.04} \, 4d^{0.04} \, 4f^{0.01} \, 5s^{0.00} \, 5p^{0.01} \, 5d^{0.01} $ \\
          Ga & CCSD    & -1923.5288  & 39.7447 & -5.1447 & -5.1447 & +10.2892 & -0.1026 & -0.1026 & +0.2052 & $ 4s^{1.92} \, 4p^{1.04} \, 4d^{0.05} \, 4f^{0.01} \, 5s^{0.01} \, 5p^{0.02} \, 5d^{0.01} $ \\
          Ga & CID     & -1923.5088  & 39.7560 & -5.2203 & -5.2203 & +10.4408 & -0.1041 & -0.1041 & +0.2082 & $ 4s^{1.95} \, 4p^{1.03} \, 4d^{0.04} \, 4f^{0.01} \, 5s^{0.00} \, 5p^{0.01} \, 5d^{0.01} $ \\
          Ga & CISD    & -1923.5141  & 39.7442 & -5.2196 & -5.2196 & +10.4391 & -0.1041 & -0.1041 & +0.2082 & $ 4s^{1.94} \, 4p^{1.03} \, 4d^{0.04} \, 4f^{0.01} \, 5s^{0.00} \, 5p^{0.01} \, 5d^{0.01} $ \\
          Ga & MP2     & -1923.5387  & 39.7807 & -5.1518 & -5.1518 & +10.3039 & -0.1027 & -0.1027 & +0.2053 & $ 4s^{1.95} \, 4p^{1.02} \, 4d^{0.05} \, 4f^{0.01} \, 5s^{0.00} \, 5p^{0.01} \, 5d^{0.01} $ \\
          Ga & MP3     & -1923.5151  & 39.8152 & -5.1556 & -5.1556 & +10.3115 & -0.1026 & -0.1026 & +0.2053 & $ 4s^{1.93} \, 4p^{1.03} \, 4d^{0.04} \, 4f^{0.01} \, 5s^{0.00} \, 5p^{0.01} \, 5d^{0.01} $ \\
          Ga & MP4     & -1923.5338  & 39.7092 & -5.1079 & -5.1079 & +10.2156 & -0.1020 & -0.1020 & +0.2039 & $ 4s^{1.92} \, 4p^{1.04} \, 4d^{0.06} \, 4f^{0.01} \, 5s^{0.00} \, 5p^{0.02} \, 5d^{0.01} $ \\
          Ga & LDA     & -1921.8127  & 39.9893 & -5.4132 & -5.4132 & +10.8265 & -0.1073 & -0.1073 & +0.2146 & $ 4s^{2.00} \, 4p^{1.00} \, 4d^{0.00} \, 4f^{0.00} \, 5s^{0.00} \, 5p^{0.00} \, 5d^{0.00} $ \\
          \br
        \end{tabular}}
    \end{indented}      
  \end{table}
\end{landscape}
\begin{table}[!htb]
  \centering
  \caption{
    The single-particle quadrupole moments $ q^{\urm{(sp)}}_i $ 
    and the single-particle root-mean-square radii $ \avr{r^2}_{nl} $ 
    for the occupied orbitals in the $ \mathrm{Al} $ atom calculated 
    by the unrestricted Hartree-Fock method with the 6-31+G basis.
    The single-particle energy of each orbital, $ \epsilon_j $, and 
    the corresponding orbital are also shown.
    The Hartree atomic unit is used for $ \epsilon_j $, $ \avr{r^2}_{nl} $, and $ q_z^{\urm{(sp)}} $.}
  \label{tab:Q_Al_MO}
  \begin{indented}
  \item[]
    \begin{tabular}{rlD{.}{.}{4}D{.}{.}{4}D{.}{.}{4}D{.}{.}{4}D{.}{.}{4}}
      \br
      Spin & Orbital & \multicolumn{1}{c}{$ \epsilon_j $} & \multicolumn{1}{c}{$ \avr{r^2}_{nl} $} & \multicolumn{1}{c}{$ q_x^{\urm{(sp)}} $} & \multicolumn{1}{c}{$ q_y^{\urm{(sp)}} $} & \multicolumn{1}{c}{$ q_z^{\urm{(sp)}} $} \\
      \mr
      $ \alpha $ & $ 1s $   & -58.4930 &  0.0195 &  0.0000 &  0.0000 &   0.0000 \\
      $ \alpha $ & $ 2s $   &  -4.9115 &  0.4591 &  0.0000 &  0.0000 &   0.0000 \\
      $ \alpha $ & $ 2p_z $ &  -3.2244 &  0.4535 & -0.1814 & -0.1814 &  +0.3628 \\
      $ \alpha $ & $ 2p_x $ &  -3.2186 &  0.4561 & +0.3648 & -0.1824 &  -0.1824 \\
      $ \alpha $ & $ 2p_y $ &  -3.2186 &  0.4561 & -0.1824 & +0.3648 &  -0.1824 \\
      $ \alpha $ & $ 3s $   &  -0.4238 &  7.9466 &  0.0000 &  0.0000 &   0.0000 \\
      $ \alpha $ & $ 3p_z $ &  -0.2099 & 13.9269 & -5.5708 & -5.5708 & +11.1416 \\
      \mr
      $ \beta $  & $ 1s $   & -58.4903 &  0.0195 &  0.0000 &  0.0000 &   0.0000 \\
      $ \beta $  & $ 2s $   &  -4.9081 &  0.4588 &  0.0000 &  0.0000 &   0.0000 \\
      $ \beta $  & $ 2p_x $ &  -3.2173 &  0.4557 & +0.3646 & -0.1823 &  -0.1823 \\
      $ \beta $  & $ 2p_y $ &  -3.2173 &  0.4557 & -0.1823 & +0.3646 &  -0.1823 \\
      $ \beta $  & $ 2p_z $ &  -3.2080 &  0.4547 & -0.1819 & -0.1819 &  +0.3637 \\
      $ \beta $  & $ 3s $   &  -0.3628 &  7.8434 &  0.0000 &  0.0000 &   0.0000 \\
      \mr
      \multicolumn{3}{l}{Total}        & 33.4055 & -5.5693 & -5.5693 & +11.1387 \\
      \br
    \end{tabular}
  \end{indented}
\end{table}
\begin{table}[!htb]
  \centering
  \caption{
    Same as Table \ref{tab:Q_Al_MO}, but for the $ \mathrm{Sc} $ atom.}
  \label{tab:Q_Sc_MO}
  \begin{indented}
  \item[]
    \begin{tabular}{rlD{.}{.}{4}D{.}{.}{4}D{.}{.}{4}D{.}{.}{4}D{.}{.}{4}}
      \br
      Spin & Orbital & \multicolumn{1}{c}{$ \epsilon_j $} & \multicolumn{1}{c}{$ \avr{r^2}_{nl} $} & \multicolumn{1}{c}{$ q_x^{\urm{(sp)}} $} & \multicolumn{1}{c}{$ q_y^{\urm{(sp)}} $} & \multicolumn{1}{c}{$ q_z^{\urm{(sp)}} $} \\
      \mr
      $ \alpha $ & $ 1s $      & -165.8952 &  0.0073 &  0.0000 &  0.0000 &  0.0000 \\
      $ \alpha $ & $ 2s $      &  -19.0956 &  0.1393 &  0.0000 &  0.0000 &  0.0000 \\
      $ \alpha $ & $ 2p_z $    &  -15.6861 &  0.1162 & -0.0465 & -0.0465 & +0.0930 \\
      $ \alpha $ & $ 2p_x $    &  -15.6830 &  0.1164 & +0.0931 & -0.0466 & -0.0466 \\
      $ \alpha $ & $ 2p_y $    &  -15.6830 &  0.1164 & -0.0466 & +0.0931 & -0.0466 \\
      $ \alpha $ & $ 3s $      &   -2.5988 &  1.3396 & +0.0016 & +0.0016 & -0.0031 \\
      $ \alpha $ & $ 3p_x $    &   -1.6232 &  1.6265 & +1.3012 & -0.6506 & -0.6506 \\
      $ \alpha $ & $ 3p_y $    &   -1.6232 &  1.6265 & -0.6506 & +1.3012 & -0.6506 \\
      $ \alpha $ & $ 3p_z $    &   -1.5893 &  1.6107 & -0.6443 & -0.6443 & +1.2886 \\
      $ \alpha $ & $ 3d_{xy} $ &   -0.3375 &  3.7873 & +1.0821 & +1.0821 & -2.1642 \\
      $ \alpha $ & $ 4s $      &   -0.2171 & 17.4198 & +0.1076 & +0.1076 & -0.2152 \\
      \mr
      $ \beta $  & $ 1s $      & -165.8951 &  0.0073 &  0.0000 &  0.0000 &  0.0000 \\
      $ \beta $  & $ 2s $      &  -19.0832 &  0.1391 & -0.0001 & -0.0001 & +0.0001 \\
      $ \beta $  & $ 2p_z $    &  -15.6837 &  0.1162 & -0.0465 & -0.0465 & +0.0929 \\
      $ \beta $  & $ 2p_x $    &  -15.6657 &  0.1161 & +0.0929 & -0.0464 & -0.0464 \\
      $ \beta $  & $ 2p_y $    &  -15.6657 &  0.1161 & -0.0464 & +0.0929 & -0.0464 \\
      $ \beta $  & $ 3s $      &   -2.5451 &  1.3370 & -0.0194 & -0.0194 & +0.0387 \\
      $ \beta $  & $ 3p_z $    &   -1.5798 &  1.6156 & -0.6462 & -0.6462 & +1.2925 \\
      $ \beta $  & $ 3p_x $    &   -1.5273 &  1.6196 & +1.2957 & -0.6478 & -0.6478 \\
      $ \beta $  & $ 3p_y $    &   -1.5273 &  1.6196 & -0.6478 & +1.2957 & -0.6478 \\
      $ \beta $  & $ 4s $      &   -0.2051 & 18.5388 & -0.9703 & -0.9703 & +1.9406 \\
      \mr
      \multicolumn{3}{l}{Total}            & 53.1312 & +0.2095 & +0.2095 & -0.4189 \\
      \br
    \end{tabular}    
  \end{indented}
\end{table}
\begin{table}[!htb]
  \centering
  \caption{
    Same as Table \ref{tab:Q_Al_MO}, but for the $ \mathrm{Ti} $ atom.}
  \label{tab:Q_Ti_MO}
  \begin{indented}
  \item[]
    \begin{tabular}{rlD{.}{.}{4}D{.}{.}{4}D{.}{.}{4}D{.}{.}{4}D{.}{.}{4}}
      \br
      Spin & Orbital & \multicolumn{1}{c}{$ \epsilon_j $} & \multicolumn{1}{c}{$ \avr{r^2}_{nl} $} & \multicolumn{1}{c}{$ q_x^{\urm{(sp)}} $} & \multicolumn{1}{c}{$ q_y^{\urm{(sp)}} $} & \multicolumn{1}{c}{$ q_z^{\urm{(sp)}} $} \\
      \mr
      $ \alpha $ & $ 1s $      & -183.2759 &  0.0066 &  0.0000 &  0.0000 &  0.0000 \\
      $ \alpha $ & $ 2s $      &  -21.4572 &  0.1250 &  0.0000 &  0.0000 &  0.0000 \\
      $ \alpha $ & $ 2p_x $    &  -17.8273 &  0.1034 & +0.0827 & -0.0414 & -0.0414 \\
      $ \alpha $ & $ 2p_y $    &  -17.8273 &  0.1034 & -0.0414 & +0.0827 & -0.0414 \\
      $ \alpha $ & $ 2p_z $    &  -17.8244 &  0.1036 & -0.0414 & -0.0414 & +0.0829 \\
      $ \alpha $ & $ 3s $      &   -2.9463 &  1.1792 & -0.0016 & -0.0016 & +0.0032 \\
      $ \alpha $ & $ 3p_z $    &   -1.9071 &  1.4095 & -0.5638 & -0.5638 & +1.1276 \\
      $ \alpha $ & $ 3p_x $    &   -1.8680 &  1.4017 & +1.1213 & -0.5607 & -0.5607 \\
      $ \alpha $ & $ 3p_y $    &   -1.8680 &  1.4017 & -0.5607 & +1.1213 & -0.5607 \\
      $ \alpha $ & $ 3d_{yz} $ &   -0.4248 &  2.8534 & -1.6305 & +0.8153 & +0.8153 \\
      $ \alpha $ & $ 3d_{zx} $ &   -0.4248 &  2.8534 & +0.8153 & -1.6305 & +0.8153 \\
      $ \alpha $ & $ 4s $      &   -0.2338 & 15.4686 & +0.0616 & +0.0616 & -0.1232 \\
      \mr
      $ \beta $  & $ 1s $      & -183.2754 &  0.0066 &  0.0000 &  0.0000 &  0.0000 \\
      $ \beta $  & $ 2s $      &  -21.4247 &  0.1247 & +0.0001 & +0.0001 & -0.0001 \\
      $ \beta $  & $ 2p_x $    &  -17.8015 &  0.1031 & +0.0825 & -0.0412 & -0.0412 \\
      $ \beta $  & $ 2p_y $    &  -17.8015 &  0.1031 & -0.0412 & +0.0825 & -0.0412 \\
      $ \beta $  & $ 2p_z $    &  -17.7791 &  0.1030 & -0.0412 & -0.0412 & +0.0824 \\
      $ \beta $  & $ 3s $      &   -2.8211 &  1.1778 & +0.0172 & +0.0172 & -0.0344 \\
      $ \beta $  & $ 3p_x $    &   -1.7472 &  1.4036 & +1.1229 & -0.5614 & -0.5614 \\
      $ \beta $  & $ 3p_y $    &   -1.7472 &  1.4036 & -0.5614 & +1.1229 & -0.5614 \\
      $ \beta $  & $ 3p_z $    &   -1.6867 &  1.4105 & -0.5642 & -0.5642 & +1.1284 \\
      $ \beta $  & $ 4s $      &   -0.2112 & 17.3797 & +0.5815 & +0.5815 & -1.1629 \\
      \mr
      \multicolumn{3}{l}{Total}            & 50.2249 & -0.1625 & -0.1625 & +0.3249 \\
      \br
    \end{tabular}    
  \end{indented}
\end{table}
\begin{table}[!htb]
  \centering
  \caption{
    Same as Table \ref{tab:Q_Al_MO}, but for the $ \mathrm{Ni} $ atom.}
  \label{tab:Q_Ni_MO}
  \begin{indented}
  \item[]
    \begin{tabular}{rlD{.}{.}{4}D{.}{.}{4}D{.}{.}{4}D{.}{.}{4}D{.}{.}{4}}
      \br
      Spin & Orbital & \multicolumn{1}{c}{$ \epsilon_j $} & \multicolumn{1}{c}{$ \avr{r^2}_{nl} $} & \multicolumn{1}{c}{$ q_x^{\urm{(sp)}} $} & \multicolumn{1}{c}{$ q_y^{\urm{(sp)}} $} & \multicolumn{1}{c}{$ q_z^{\urm{(sp)}} $} \\
      \mr
      $ \alpha $ & $ 1s $           & -305.4079 &  0.0040 &  0.0000 &  0.0000 &  0.0000 \\
      $ \alpha $ & $ 2s $           &  -37.7247 &  0.0720 &  0.0000 &  0.0000 & -0.0001 \\
      $ \alpha $ & $ 2p_x $         &  -32.7657 &  0.0574 & +0.0459 & -0.0230 & -0.0230 \\
      $ \alpha $ & $ 2p_y $         &  -32.7657 &  0.0574 & -0.0230 & +0.0459 & -0.0230 \\
      $ \alpha $ & $ 2p_z $         &  -32.7281 &  0.0574 & -0.0229 & -0.0229 & +0.0459 \\
      $ \alpha $ & $ 3s $           &   -4.7265 &  0.6527 & +0.0073 & +0.0073 & -0.0147 \\
      $ \alpha $ & $ 3p_x $         &   -3.1649 &  0.7371 & +0.5897 & -0.2948 & -0.2948 \\
      $ \alpha $ & $ 3p_y $         &   -3.1649 &  0.7371 & -0.2948 & +0.5897 & -0.2948 \\
      $ \alpha $ & $ 3p_z $         &   -3.0843 &  0.7415 & -0.2966 & -0.2966 & +0.5932 \\
      $ \alpha $ & $ 3d_{x^2-y^2} $ &   -0.5446 &  1.4257 & +0.4073 & +0.4073 & -0.8147 \\
      $ \alpha $ & $ 3d_{xy}      $ &   -0.4962 &  1.5120 & +0.4320 & +0.4320 & -0.8640 \\
      $ \alpha $ & $ 3d_{yz}      $ &   -0.4616 &  1.5739 & -0.8994 & +0.4497 & +0.4497 \\
      $ \alpha $ & $ 3d_{zx}      $ &   -0.4616 &  1.5739 & +0.4497 & -0.8994 & +0.4497 \\
      $ \alpha $ & $ 3d_{z^2}     $ &   -0.4493 &  1.5996 & -0.4427 & -0.4427 & +0.8854 \\
      $ \alpha $ & $ 4s $           &   -0.2395 & 13.2134 & +0.1360 & +0.1360 & -0.2720 \\
      \mr
      $ \beta $  & $ 1s $           & -305.4054 &  0.0040 &  0.0000 &  0.0000 &  0.0000 \\
      $ \beta $  & $ 2s $           &  -37.6883 &  0.0719 &  0.0000 &  0.0000 &  0.0000 \\
      $ \beta $  & $ 2p_x $         &  -32.7198 &  0.0572 & +0.0458 & -0.0229 & -0.0229 \\
      $ \beta $  & $ 2p_y $         &  -32.7198 &  0.0572 & -0.0229 & +0.0458 & -0.0229 \\
      $ \beta $  & $ 2p_z $         &  -32.7194 &  0.0573 & -0.0229 & -0.0229 & +0.0459 \\
      $ \beta $  & $ 3s $           &   -4.6262 &  0.6549 & -0.0005 & -0.0005 & +0.0010 \\
      $ \beta $  & $ 3p_z $         &   -3.0570 &  0.7407 & -0.2963 & -0.2963 & +0.5926 \\
      $ \beta $  & $ 3p_x $         &   -3.0065 &  0.7383 & +0.5907 & -0.2953 & -0.2953 \\
      $ \beta $  & $ 3p_y $         &   -3.0065 &  0.7383 & -0.2953 & +0.5907 & -0.2953 \\
      $ \beta $  & $ 3d_{xy} $      &   -0.4702 &  1.4877 & +0.4251 & +0.4251 & -0.8501 \\
      $ \beta $  & $ 3d_{yz} $      &   -0.4160 &  1.5895 & -0.9083 & +0.4541 & +0.4541 \\
      $ \beta $  & $ 3d_{zx} $      &   -0.4160 &  1.5895 & +0.4541 & -0.9083 & +0.4541 \\
      $ \beta $  & $ 3d_{z^2} $     &   -0.3977 &  1.6301 & -0.4655 & -0.4655 & +0.9309 \\
      \mr
      \multicolumn{3}{l}{Total}                 & 33.4318 & -0.4074 & -0.4074 & +0.8148 \\
      \br
    \end{tabular}
  \end{indented}
\end{table}
\begin{table}[!htb]
  \centering
  \caption{
    Same as Table \ref{tab:Q_Al_MO}, but for the $ \mathrm{Ga} $ atom.}
  \label{tab:Q_Ga_MO}
  \begin{indented}
  \item[]
    \begin{tabular}{rlD{.}{.}{4}D{.}{.}{4}D{.}{.}{4}D{.}{.}{4}D{.}{.}{4}}
      \br
      Spin & Orbital & \multicolumn{1}{c}{$ \epsilon_j $} & \multicolumn{1}{c}{$ \avr{r^2}_{nl} $} & \multicolumn{1}{c}{$ q_x^{\urm{(sp)}} $} & \multicolumn{1}{c}{$ q_y^{\urm{(sp)}} $} & \multicolumn{1}{c}{$ q_z^{\urm{(sp)}} $} \\
      \mr
      $ \alpha $ & $ 1s $           & -378.6802 &  0.0033 &  0.0000 &  0.0000 &   0.0000 \\
      $ \alpha $ & $ 2s $           &  -47.5198 &  0.0623 &  0.0000 &  0.0000 &   0.0000 \\
      $ \alpha $ & $ 2p_z $         &  -42.4832 &  0.0452 & -0.0181 & -0.0181 &  +0.0362 \\
      $ \alpha $ & $ 2p_x $         &  -42.4826 &  0.0452 & +0.0362 & -0.0181 &  -0.0181 \\
      $ \alpha $ & $ 2p_y $         &  -42.4826 &  0.0452 & -0.0181 & +0.0362 &  -0.0181 \\
      $ \alpha $ & $ 3s $           &   -6.2558 &  0.5350 & +0.0004 & +0.0004 &  -0.0008 \\
      $ \alpha $ & $ 3p_x $         &   -4.4690 &  0.5485 & +0.4388 & -0.2194 &  -0.2194 \\
      $ \alpha $ & $ 3p_y $         &   -4.4690 &  0.5485 & -0.2194 & +0.4388 &  -0.2194 \\
      $ \alpha $ & $ 3p_z $         &   -4.4678 &  0.5474 & -0.2189 & -0.2189 &  +0.4379 \\
      $ \alpha $ & $ 3d_{x^2-y^2} $ &   -1.1813 &  0.7753 & +0.2215 & +0.2215 &  -0.4430 \\
      $ \alpha $ & $ 3d_{xy} $      &   -1.1813 &  0.7753 & +0.2215 & +0.2215 &  -0.4430 \\
      $ \alpha $ & $ 3d_{yz} $      &   -1.1761 &  0.7753 & -0.4430 & +0.2215 &  +0.2215 \\
      $ \alpha $ & $ 3d_{zx} $      &   -1.1761 &  0.7753 & +0.2215 & -0.4430 &  +0.2215 \\
      $ \alpha $ & $ 3d_{z^2} $     &   -1.1743 &  0.7757 & -0.2239 & -0.2239 &  +0.4478 \\
      $ \alpha $ & $ 4s $           &   -0.4471 &  7.5754 & +0.0036 & +0.0036 &  -0.0072 \\
      $ \alpha $ & $ 4p_z $         &   -0.2086 & 13.9574 & -5.5830 & -5.5830 & +11.1659 \\
      \mr
      $ \beta $  & $ 1s $           & -378.6792 &  0.0033 &  0.0000 &  0.0000 &   0.0000 \\
      $ \beta $  & $ 2s $           &  -47.5192 &  0.0623 &  0.0000 &  0.0000 &   0.0000 \\
      $ \beta $  & $ 2p_x $         &  -42.4823 &  0.0452 & +0.0362 & -0.0181 &  -0.0181 \\
      $ \beta $  & $ 2p_y $         &  -42.4823 &  0.0452 & -0.0181 & +0.0362 &  -0.0181 \\
      $ \beta $  & $ 2p_z $         &  -42.4796 &  0.0452 & -0.0181 & -0.0181 &  +0.0362 \\
      $ \beta $  & $ 3s $           &   -6.2535 &  0.5349 & +0.0007 & +0.0007 &  -0.0013 \\
      $ \beta $  & $ 3p_x $         &   -4.4682 &  0.5483 & +0.4386 & -0.2193 &  -0.2193 \\
      $ \beta $  & $ 3p_y $         &   -4.4682 &  0.5483 & -0.2193 & +0.4386 &  -0.2193 \\
      $ \beta $  & $ 3p_z $         &   -4.4585 &  0.5477 & -0.2191 & -0.2191 &  +0.4382 \\
      $ \beta $  & $ 3d_{x^2-y^2} $ &   -1.1808 &  0.7753 & +0.2215 & +0.2215 &  -0.4430 \\
      $ \beta $  & $ 3d_{xy} $      &   -1.1808 &  0.7753 & +0.2215 & +0.2215 &  -0.4430 \\
      $ \beta $  & $ 3d_{yz} $      &   -1.1720 &  0.7753 & -0.4430 & +0.2215 &  +0.2215 \\
      $ \beta $  & $ 3d_{zx} $      &   -1.1720 &  0.7753 & +0.2215 & -0.4430 &  +0.2215 \\
      $ \beta $  & $ 3d_{z^2} $     &   -1.1691 &  0.7756 & -0.2133 & -0.2133 &  +0.4267 \\
      $ \beta $  & $ 4s $           &   -0.3877 &  7.3770 & -0.0070 & -0.0070 &  +0.0139 \\
      \mr
      \multicolumn{3}{l}{Total}                 & 41.4245 & -5.5788 & -5.5788 & +11.1576 \\
      \br
    \end{tabular}
  \end{indented}
\end{table}

\clearpage
% 
%%%%%%%%%%%%%%%%%%%%%%%%%%%%%%%%%%%%%%%%%%%%%%%%%% 
\providecommand{\newblock}{}

\end{document}